\documentclass[%
 reprint,
superscriptaddress,
nofootinbib,
 amsmath,amssymb,
 aps,
prc,
floatfix,
]{revtex4-2}

\usepackage{graphicx}
\usepackage{dcolumn}
\usepackage{bm}
\usepackage{xcolor}
\usepackage[pdftex,colorlinks,linkcolor = blue,urlcolor  = blue,citecolor = blue]{hyperref}

\begin{document}


\title{New experimental constraint on the $^{185}$W($n,\gamma$)$^{186}$W cross section}
\author{A.~C.~Larsen}%
\email{a.c.larsen@fys.uio.no}
\affiliation{Department of Physics, University of Oslo, N-0316 Oslo, Norway}

\author{G.~M.~Tveten}
\email{gry@xal.no}
\affiliation{Department of Physics, University of Oslo, N-0316 Oslo, Norway}
\affiliation{Expert Analytics AS, N-0179 Oslo, Norway}

\author{T.~Renstr\o m}
\affiliation{Department of Physics, University of Oslo, N-0316 Oslo, Norway}
\affiliation{Expert Analytics AS, N-0179 Oslo, Norway}

\author{H.~Utsunomiya}
\affiliation{Department of Physics, Konan University, Okamoto 8-9-1, Higashinada, Kobe 658-8501, Japan}
\affiliation{Shanghai Advanced Research Institute, Chinese Academy of Sciences, Shanghai 201210, China}

\author{E.~Algin}
\affiliation{Department of Metallurgical and Materials Engineering, Pamukkale University, 20160 Denizli, Turkey}

\author{T.~Ari-izumi}
\affiliation{Department of Physics, Konan University, Okamoto 8-9-1, Higashinada, Kobe 658-8501, Japan}

\author{K.~O.~Ay}
\affiliation{Department of Physics, Eskisehir Osmangazi University, 26480 Eskisehir, Turkey}

\author{F.~L.~Bello~Garrote}
\affiliation{Department of Physics, University of Oslo, N-0316 Oslo, Norway}

\author{L.~Crespo~Campo}
\affiliation{Department of Physics, University of Oslo, N-0316 Oslo, Norway}

\author{F.~Furmyr}
\affiliation{Department of Physics, University of Oslo, N-0316 Oslo, Norway}

\author{S.~Goriely}
\affiliation{Institut d’Astronomie et d’Astrophysique, Universit\'{e} Libre de Bruxelles,
Campus de la Plaine, CP-226, 1050 Brussels, Belgium}

\author{A.~G\"orgen}
\affiliation{Department of Physics, University of Oslo, N-0316 Oslo, Norway}

\author{M.~Guttormsen}
\affiliation{Department of Physics, University of Oslo, N-0316 Oslo, Norway}

\author{V.~W.~Ingeberg}
\affiliation{Department of Physics, University of Oslo, N-0316 Oslo, Norway}

\author{B.~V.~Kheswa}
\affiliation{%
iThemba LABS, 
P.O. Box 722, 
7129 Somerset West,
South Africa
}%
\affiliation{
Department of Applied Physics and Engineering Mathematics,
University of Johannesburg, Johannesburg, 2028, 
South Africa
}

\author{I.~K.~B.~Kullmann}
\affiliation{Institute d'Astronomie et d'Astrophysique, Universit\'{e} Libre de Bruxelles, Belgium}

\author{T.~Laplace}
\affiliation{Department of Nuclear Engineering, University of California, Berkeley, 94720, USA}

\author{E.~Lima}
\affiliation{Department of Physics, University of Oslo, N-0316 Oslo, Norway}

\author{M.~Markova}
\affiliation{Department of Physics, University of Oslo, N-0316 Oslo, Norway}

\author{J.~E.~Midtb\o}
\affiliation{Department of Physics, University of Oslo, N-0316 Oslo, Norway}

\author{S.~Miyamoto}
\affiliation{Laboratory of Advanced Science and Technology for Industry, University of Hyogo, 3-1-2 Kouto, Kamigori, Ako-gun, Hyogo 678-1205, Japan}

\author{A.~H.~Mj\o s}
\affiliation{Department of Physics, University of Oslo, N-0316 Oslo, Norway}

\author{V.~Modamio}
\affiliation{Department of Physics, University of Oslo, N-0316 Oslo, Norway}

\author{M.~Ozgur}
\affiliation{Department of Physics, Eskisehir Osmangazi University, 26480 Eskisehir, Turkey}

\author{F.~Pogliano}
\affiliation{Department of Physics, University of Oslo, N-0316 Oslo, Norway}

\author{S.~Riemer-S\o rensen}
\affiliation{Institute of Theoretical Astrophysics, University of Oslo, N-0316 Oslo, Norway}
\affiliation{Department of Mathematics and Cybernetics, SINTEF Digital, N-0314 Oslo, Norway}

\author{E.~Sahin}
\affiliation{Department of Physics, University of Oslo, N-0316 Oslo, Norway}

\author{S.~Shen}
\affiliation{Institute of Theoretical Astrophysics, University of Oslo, N-0316 Oslo, Norway}

\author{S.~Siem}
\affiliation{Department of Physics, University of Oslo, N-0316 Oslo, Norway}

\author{A.~Spyrou}
\affiliation{Physics Department, Michigan State University, East Lansing, Michigan 48824, USA}
\affiliation{National Superconducting Cyclotron Laboratory, Michigan State University, East Lansing, Michigan 48824, USA}
\affiliation{Joint Institute for Nuclear Astrophysics Center for the Evolution of the Elements, University of Notre Dame, Notre Dame, Indiana 46556, USA}

\author{M.~Wiedeking}
\affiliation{%
iThemba LABS, 
P.O. Box 722, 
7129 Somerset West,
South Africa
}
\affiliation{School of Physics, University of the Witwatersrand, Johannesburg 2050, South Africa}

\date{\today}

\begin{abstract}
In this work, we present new data on the $^{182,183,184}$W($\gamma,n$) cross sections, utilizing a quasi-monochromatic photon beam produced at the NewSUBARU synchrotron radiation facility.
Further, we have extracted the nuclear level density and $\gamma$-ray strength function of $^{186}$W from data on the $^{186}$W($\alpha,\alpha^\prime\gamma$)$^{186}$W reaction measured at the Oslo Cyclotron Laboratory.
Combining previous measurements on the $^{186}$W($\gamma,n$) cross section with our new $^{182,183,184}$W($\gamma,n$) and ($\alpha,\alpha^\prime\gamma$)$^{186}$W data sets, we have deduced the $^{186}$W $\gamma$-ray strength function in the range of $1 < E_\gamma < 6$ MeV and $7 < E_\gamma < 14$ MeV.

Our data are used to extract the level density and $\gamma$-ray strength  functions needed as input to the nuclear-reaction code \textsf{TALYS}, providing an indirect, experimental constraint for the $^{185}$W($n,\gamma$)$^{186}$W cross section and reaction rate. 
Compared to the recommended Maxwellian-averaged cross section (MACS) in the KADoNiS-1.0 data base, our results are on average lower for the relevant energy range $k_B T \in [5,100]$ keV, and we provide a smaller uncertainty for the MACS. 
The theoretical values of Bao \textit{et al.} and the cross section experimentally constrained on photoneutron data of Sonnabend \textit{et al.} are significantly higher than our result. 
The lower value by Mohr \textit{et al.} is in very good agreement with our deduced MACS. 
Our new results could have implications for the $s$-process and in particular the predicted $s$-process production of $^{186,187}$Os nuclei.

\end{abstract}

\maketitle


\section{Introduction}
\label{sec:intro}
\begin{figure*}[htb]
\center
\includegraphics[width=1.6\columnwidth]{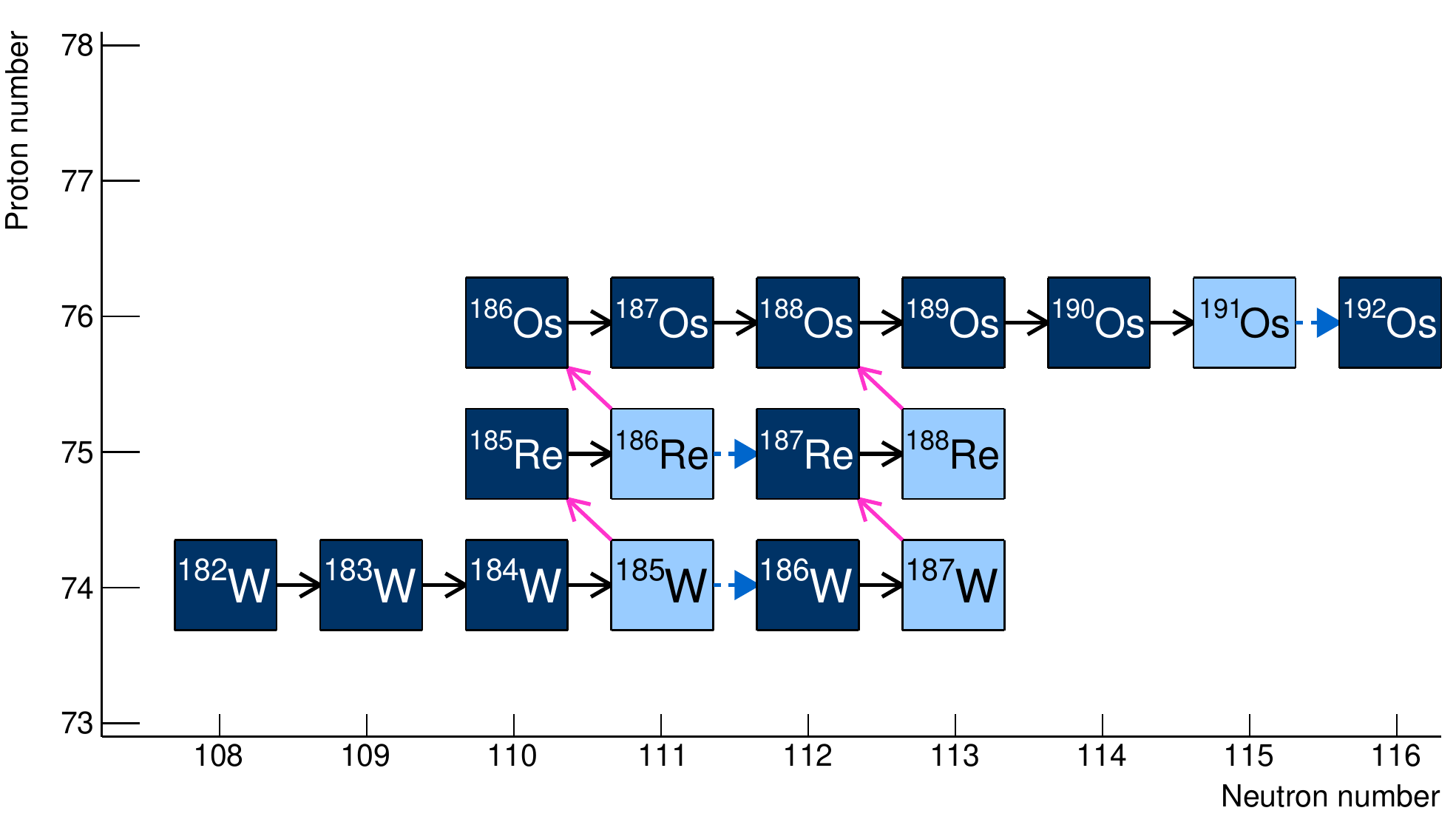}
\caption{(Color online) Schematic illustration of the nuclear chart in the W-Re-Os region. The black arrows indicate $(n,\gamma)$ reactions on stable or near-stable isotopes, the blue dashed arrows show the possible $(n,\gamma)$ branch on the long-lived W, Re and Os isotopes, while the pink arrows display the $\beta^-$ decay branch. }
\label{fig:chart}
\end{figure*}
Neutron-capture reactions are known to be the main producers of elements heavier than iron in our Universe~\cite{burbidge1957,cameron1957}. 
The rapid ($r$) and the slow ($s$) neutron-capture processes are traditionally believed to account for almost 100\% of the Solar-system heavy-element abundances~\cite{arnould2007,cowan2021}. 
The $r$ process takes place in an environment with an extremely high neutron density typicallly larger than $10^{24}$ neutrons/cm$^3$, which produces very neutron-rich nuclei within a  short time window ($\approx$1s).  
In contrast, the $s$ process is, as the name implies, a slow process; the neutron density is comparatively low ($\sim 10^{6}-10^{8}$ neutrons/cm$^3$ in asymptotic giant branch stars~\cite{kappeler2011}) and it can take from days to thousands of years between each neutron-capture reaction. 
Consequently, the $s$-process ``path'' in the nuclear chart remains close to the valley of stability, as the $\beta$-decay rates are typically much faster than the $(n,\gamma)$ rates when an unstable nucleus is reached. 

However, this is not true for some particular nuclei along the $s$-process path. 
At the \textit{branch points}~\cite{ward1976} the $\beta$-decay rate is comparable to the ($n,\gamma$) rate, so that there is a non-negligible possibility for the nucleus to \textit{either} undergo $\beta$-decay \textit{or} capture another neutron. 
On the one hand, such branch points could complicate the $s$-process nucleosynthesis calculation significantly; on the other hand, they may provide valuable information about the neutron density and/or  temperature at the astrophysical site for which the $s$ process operates~\cite{kappeler1991,sonnabend2003,mohr2004}.

In this work, we focus on the branch-point nucleus $^{185}$W, with a laboratory half-life of 75.1(3) days~\cite{Emery1972}.
This nucleus is of interest for the Re/Os cosmochronology first discussed by Clayton~\cite{Clayton1964}. 
The main idea behind the Re/Os cosmochronology is the following:
the matter from which the Solar system  was formed, contained a given amount of $^{187}$Re and $^{187}$Os. 
Further, $^{187}$Re is usually assigned a pure $r$-process origin, while $^{187}$Os is produced only in the $s$ process.
As $^{187}$Re has a very long half-life of more than $4\cdot 10^{10}$ years~\cite{galeazzi2000}, Clayton suggested to use the solar-system amount of $^{187}$Re and $^{187}$Os as a ``clock'', which would display the time span for which nucleosynthesis events produced various elements up to the time of the formation of our Solar system. 
Provided that the $^{187}$Os amount stemming from the $s$ process can be reliably calculated, the extra amount of $^{187}$Os originates from the $^{187}$Re decay. 
Thus, at least in principle, the abundances of the parent/child pair $^{187}$Re/$^{187}$Os  can be used as a cosmochronometer, although not without complications~\cite{yokoi1983re,arnould1984validity,Mosconi2010}.
As discussed in Refs.~\cite{arnould1984validity,kappeler1991,shizuma2005,Mosconi2010}, the branchings at $^{185}$W and $^{186}$Re (see Fig.~\ref{fig:chart}) could well have a non-negligible impact on this cosmochronometer. 
Moreover, several authors~\cite{kappeler1991,sonnabend2003,mohr2004,Humayun_2007} have discussed the $^{185}$W and $^{186}$Re branchings as a ``neutron dosimeter'' for the effective $s$-process neutron density; this application again depends on the radiative neutron-capture cross sections of $^{185}$W and $^{186}$Re. 
No direct measurement of the neutron-capture cross section is possible on these target nuclei, and only constraints on the electromagnetic decay of the compound system have been obtained through photoneutron experiments at relatively high photon energies~\cite{mohr2004,sonnabend2003}.

Here we present new photoneutron data on $^{182,183,184}$W that complete the ($\gamma,n$) measurements on the W isotopes (Sec.~\ref{sec:expNEWSubaru}). 
Moreover, in Sec.~\ref{sec:expOslo}, we present the $^{186}$W($\alpha,\alpha' \gamma$) data taken at the Oslo Cyclotron Laboratory, and the data analysis with the resulting level density and $\gamma$ strength function of $^{186}$W.
Using our new data to constrain the input to the nuclear reaction code \textsf{TALYS-1.9}~\cite{TALYS} we estimate the $^{185}$W($n,\gamma$)$^{186}$W Maxwellian-averaged cross section and reaction rate, and compare to previous measurements and evaluations in Sec.~\ref{subsec:astro}.
Finally, we give a summary and outlook in Sec.~\ref{sec:summary}.

\section{The ($\gamma,n$) experiments}
\label{sec:expNEWSubaru}
\subsection{Experimental details}

\begin{figure*}[t]
\center
\includegraphics[width=1.7\columnwidth]{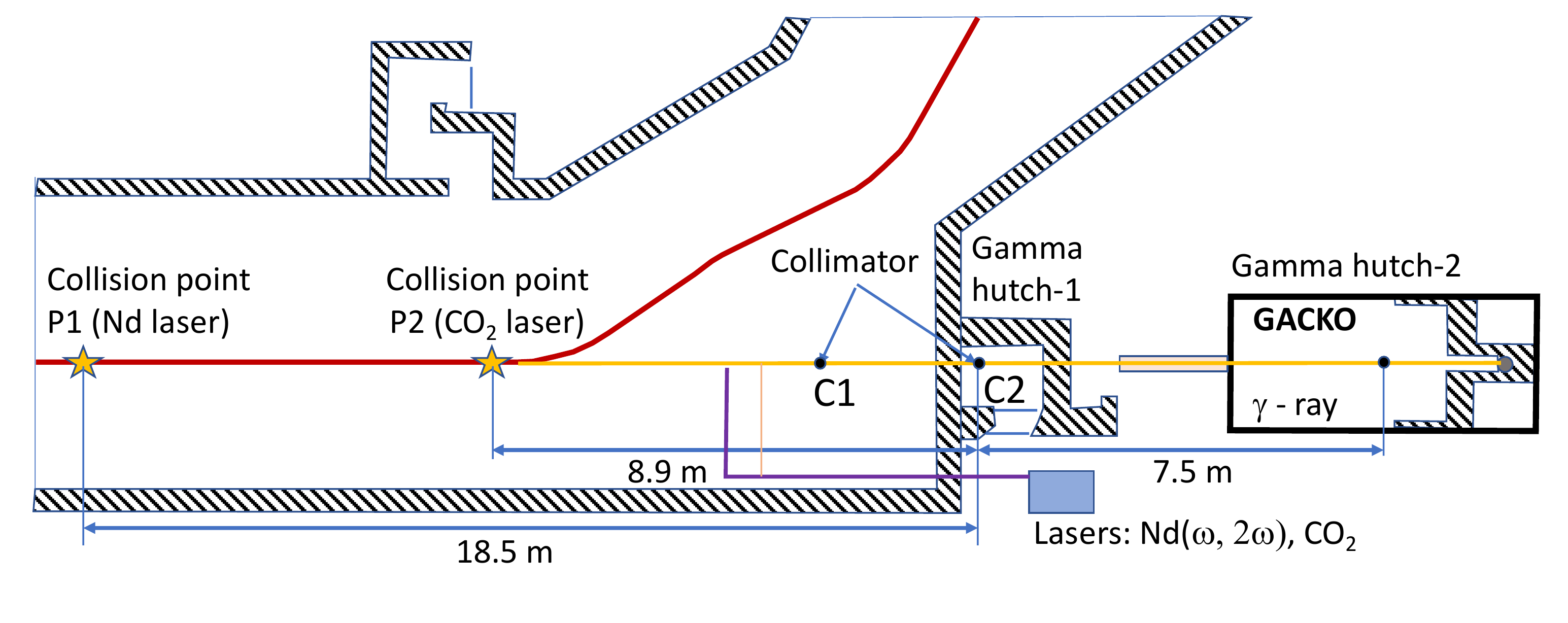}
\caption{(Color online) A schematic illustration of the experimental set up at NewSUBARU used in the ($\gamma, n$) cross-section measurements.}
\label{fig:setup}
\end{figure*}
The photo-neutron measurements on $^{182,183,184}$W took place at the NewSUBARU synchrotron radiation facility.
Figure~\ref{fig:setup} shows a schematic illustration of the $\gamma$-ray beam line and experimental setup.
Beams of $\gamma$ rays  were produced through laser Compton scattering (LCS) of 1064 nm photons in head-on collisions with relativistic electrons at the most-efficient collision point P1. The $\gamma$ beams were collimated using the Pb collimators C1 and C2, each 10 cm long, with 3 mm and 2 mm apertures, respectively.  
The beam profile on target nearly follows the geometrical aperture of the collimator C2 with respect to the collision point P1, thus avoiding any interaction between beam and other materials than the target. Throughout the experiment, the laser was periodically on for 80 ms and off for 20 ms, in order to measure background neutrons and $\gamma$-rays. 
In this experiment, the beams produced had an energy resolution ranging from 0.6 MeV to 0.9 MeV (full-width at half maximum, FWHM).

The electrons were injected from a linear accelerator into the NewSUBARU storage ring with an initial energy of 974 MeV, then subsequently decelerated to nominal energies ranging from 608 MeV to 849 MeV, providing LCS $\gamma$-ray beams of energies up to 13 MeV and down to the neutron separation energies of the W isotopes (thus varied for each individual case). 
The maximum $\gamma$-ray energy of the beams was increased in steps of 0.25 MeV. 
The electron beam energy has been calibrated with the accuracy on the order of 10$^{-5}$ \cite{calofns}.  
The energy is reproduced in every injection of an electron beam from a linear accelerator to the storage ring. 
The reproducibility of the electron energy is assured in the deceleration down to 0.5 GeV by an automated control of the electron beam-optics parameters.

The energy profiles of the produced $\gamma$-ray beams were measured with a $3.5$\textit{in.}$\times 4.0$\textit{in.} LaBr$_3$(Ce) (LaBr$_3$) detector. 
The measured LaBr$_3$ spectra were reproduced by a \textsf{Geant4} code~\cite{thesisioana, geant1, geant2, geant3} that incorporated the kinematics of the LCS process, including the beam emittance and the interactions between the LCS beam and the LaBr$_3$ detector. 
In this way we were routinely able to simulate the energy profile of the incoming $\gamma$ beams with the maximum energies accurately determined by the calibrated electron beam energy by best reproducing the LaBr$_3$ spectra \cite{Utsu15, Fili14}.

The W targets were made from isotopically enriched tungsten as metallic powder. 
The material was pressed together and enclosed in an aluminium cylinder with a thin cap. 
The targets had areal densities of $0.7421$~g/cm$^2$  ($^{182}$W), $0.754$~g/cm$^2$  ($^{183}$W), and $1.7925$~g/cm$^2$ ($^{184}$W). 
Due to the presence of the Al cap, we limited the $\gamma$-ray beam energy maximum 13 MeV to avoid getting contaminant neutrons from $^{27}$Al ($S_n$=13.056 MeV). 

To measure the emitted neutrons, a high-efficiency $4\pi$ detector was used, consisting of 20 $^3\rm{He}$ proportional counters, arranged in three concentric rings and embedded in a 36 $\times$ 36 $\times$ 50 cm$^3$ polyethylene neutron moderator~\cite{itoh_2011}. 
The ring ratio technique, originally developed by Berman and Fultz~\cite{Berman_ring_ratio}, was used to determine the average energy of the neutrons from the ($\gamma,n$) reactions. 
The efficiency of the neutron detector varies with the average neutron energy. 
The efficiency was measured with a calibrated $^{252}$Cf source with the emission rate of 2.27 $\cdot$ 10$^4$ s$^{-1}$ with 2.2\% uncertainty, and the energy dependence was determined by Monte Carlo simulations \cite{Nyhu15}. 
The efficiency of the neutron detector was simulated using isotropically distributed, mono-energetic neutrons. 
Once the neutron detection efficiency for a given beam energy has been determined, we were able to deduce the number of ($\gamma, n$) reactions that took place during each run. 

The LCS $\gamma$-ray flux was monitored by a $8$\textit{in.}$\times 12$\textit{in.} NaI(Tl) (NaI)
detector during neutron measurement runs with 100$\%$ detection efficiency for the beam energies used in this experiment. The number of incoming $\gamma$ rays per measurement was determined using the pile-up and Poisson-fitting technique described in Refs.~\cite{KONDO2011462,UTSUNOMIYA2018103}.

\subsection{Analysis}

\begin{figure}[t]
\includegraphics[width=0.5\textwidth]{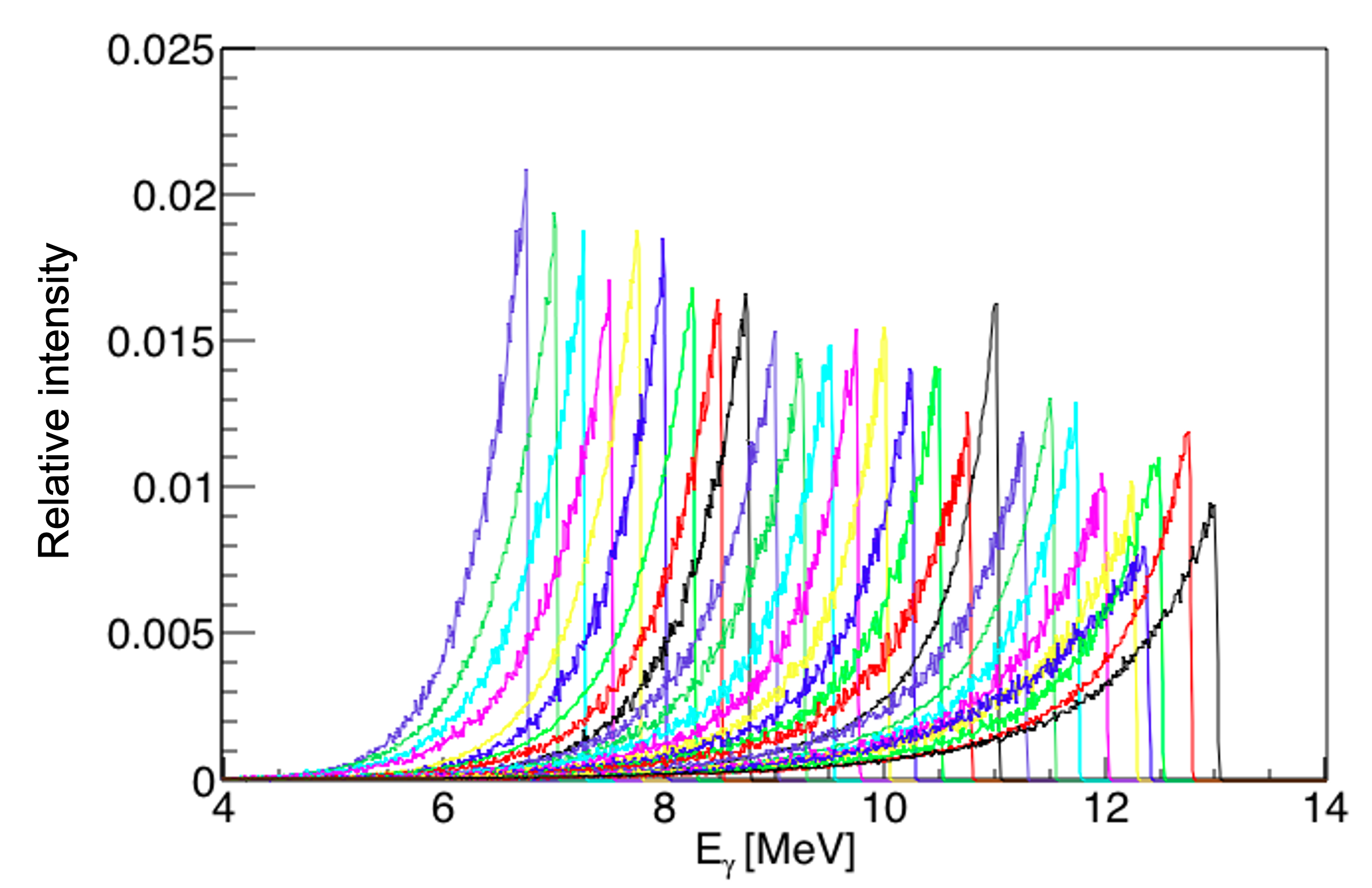}
\caption{(Color online) The simulated energy profiles for the $\gamma$ beams used. The distributions (integrated over all $E_\gamma$) are normalized to unity. }
\label{fig:GammaProfile}
\end{figure}

The measured photo-neutron cross section for an incoming beam with maximum $\gamma$ energy $E_{\rm max}$ is given by the convoluted cross section,
\begin{equation}
\sigma^{E_{\rm max}}_{\rm exp}=\int_{S_n}^{E_{\rm max}}D^{E_{\rm max}}(E_{\gamma})\sigma(E_{\gamma})dE_{\gamma}=\frac{N_n}{N_tN_{\gamma}\xi\epsilon_n g}.
\label{eq:cross1}
\end{equation}
Here, $D^{E_{\rm max}}$ is the normalized energy distribution ($\int_{S_n}^{E_{\rm max}} D^{E_{\rm max}}dE_{\gamma}= 1$) of
the $\gamma$-ray beam obtained from \textsf{Geant4} simulations. 
Examples of the simulated $\gamma$-beam profiles, $D^{E_{\rm max}}$, 
are shown in Fig.~\ref{fig:GammaProfile}. Furthermore, $\sigma(E_{\gamma})$ is the true photo-neutron cross section as a function of energy. 
The quantity $N_n$ represents the number of neutrons detected, $N_t$ gives the number of target nuclei per unit area, $N_{\gamma}$ is the number of $\gamma$ rays incident on target, $\epsilon_n$ represents the neutron detection efficiency, and finally $\xi=(1-e^{-\mu t})/(\mu t)$ gives a correction factor for self-attenuation in the target. 
The factor $g$ represents the fraction of the $\gamma$ flux above $S_n$. 
\begin{figure}[tb]
\includegraphics[width=0.48\textwidth]{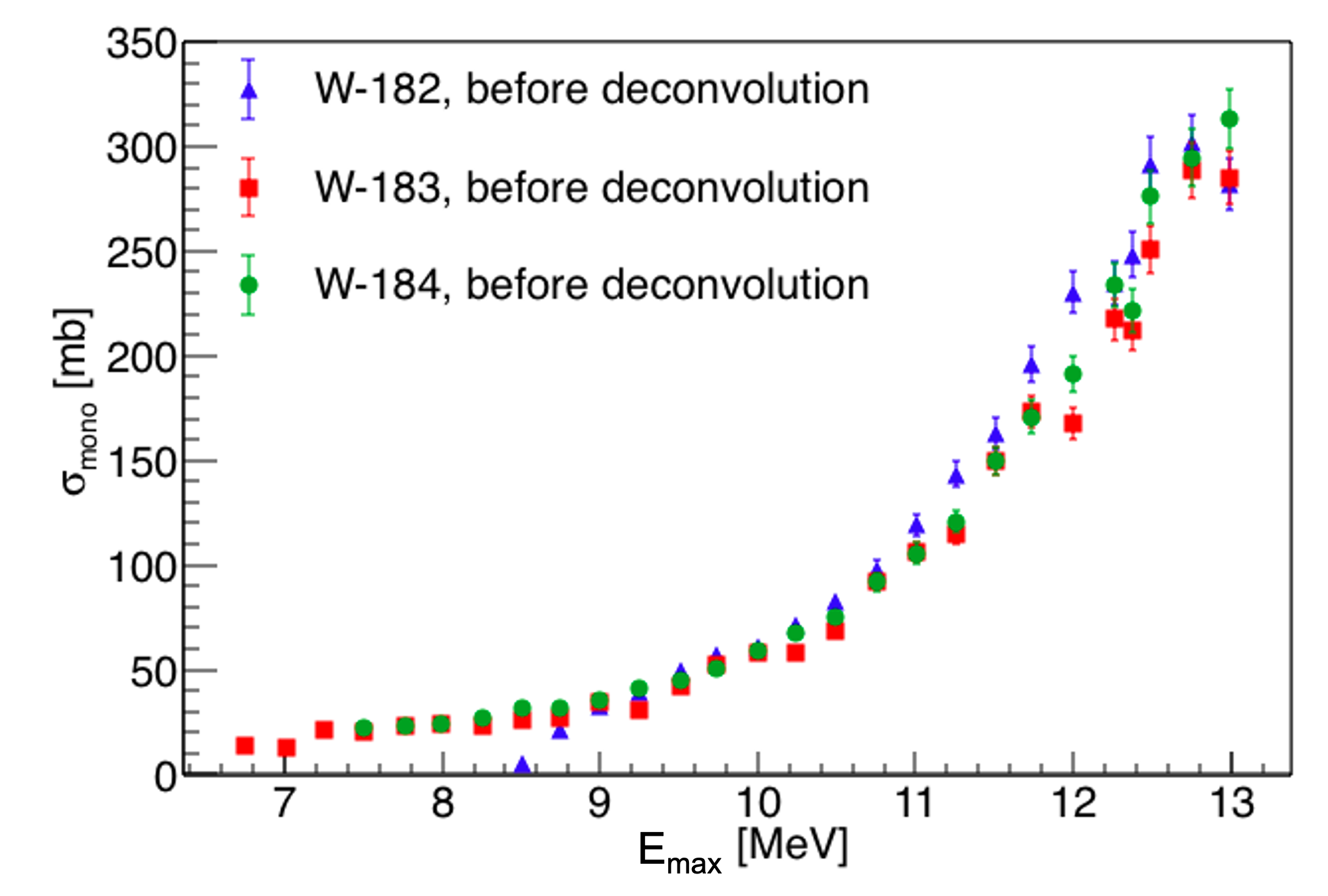}
\caption{(Color online) Monochromatic cross sections of $^{182,183,184}\rm{W}$. The error bars contain statistical uncertainties from the number of detected neutrons, the uncertainty in the efficiency of the neutron detector and the uncertainly in the pile-up method used to determine the integrated $\gamma$-flux on target.}
\label{fig:Monocross}
\end{figure}

We have determined the convoluted cross sections $\sigma^{E_{\rm max}}_{\rm exp}$ given by Eq.~(\ref{eq:cross1}) for $\gamma$ beams with maximum energies in the range $S_{n}\leq E_{\rm max} \leq$ 13 MeV. 
The convoluted cross sections $\sigma^{E_{\rm max}}_{\rm exp}$ are not connected to a specific $E_{\gamma}$, and we choose to plot them as a function of $E_{\rm max}$. 
The convoluted cross sections of the three W isotopes, which are often called monochromatic cross sections, are shown in Fig.~\ref{fig:Monocross}. 
The error bars in Fig.~\ref{fig:Monocross} represent the total uncertainty in the quantities comprising Eq.~(\ref{eq:cross1}), and consists of $\sim 3.2\%$ from the efficiency determination of the neutron detector, $\sim 1\%$ from the pile-up method that gives the number of $\gamma$ rays, and the statistical uncertainty in the number of detected neutrons \cite{UTSUNOMIYA2018103}. 
The statistical uncertainty ranges between $\sim$ 5.0 $\%$ close to neutron threshold and 4.4 $\%$ for the highest maximum $\gamma$-ray beam energies. The systematic error is dominated by the uncertainty from the pile-up method and from the simulated efficiency of the neutron detector. 
For the total uncertainty, we have added these uncorrelated errors in quadrature.

By approximating the integral in Eq.~(\ref{eq:cross1}) with a sum for each $\gamma$-beam profile, we are able to express the unfolding problem as a set of linear equations
\begin{equation}
\sigma_{\rm f }=\bf{D}\sigma,
\end{equation}
where $\sigma_{\rm f}$ is the cross section folded with the beam profile {\bf D}.  
The indexes $i$ and $j$ of the matrix element $D_{ij}$ corresponds to $E_{\rm max}$ and $E_{\gamma}$, respectively.
The set of equations is given by
\begin{equation}
\begin{pmatrix}\sigma_{\rm{1}}\\\sigma_{\rm{2}}\\ \vdots \\ \sigma_N \end{pmatrix}_{\rm f}\\\mbox{}=
\begin{pmatrix}D_{ 11} & D_{ 12} & \cdots &\cdots &D_{ 1M} \\ D_{ 21} & D_{ 22} &
\cdots & \cdots & D_{ 2M} \\ \vdots & \vdots & \vdots & \vdots & \vdots \\ D_{ N1} & D_{ N2}& \cdots & \cdots &D_{ NM}\end{pmatrix}
\begin{pmatrix}\sigma_{1}\\\sigma_{2}\\ \vdots \\ \vdots \\\sigma_{M} \end{pmatrix}.
\label{eq:matrise_unfolding}
\end{equation}
Each row of $\bf{D}$ corresponds to a \textsf{Geant4} simulated $\gamma$
beam profile belonging to a specific measurement characterized by $E_{\rm max}$   
(see Fig.~\ref{fig:GammaProfile} for a visual representation of some of the rows in the response matrix $\bf{D}$). 
It is clear that $\bf{D}$ is highly asymmetrical.

The number of $\gamma$-ray beam energies used to study the cross section is much lower than the bin size (10 keV) of the simulated beam profiles above $S_n$. As the system of linear equations in Eq.~(\ref{eq:matrise_unfolding}) is under-determined, the true $\sigma$ vector cannot be extracted by matrix inversion. 
In order to find $\sigma$, we utilize a folding iteration method. The main features of this method are as follows \cite{renstrom2018_Dy}:

\begin{itemize}

\item [1)] As a starting point, we choose for the 0th iteration, a constant trial function $\sigma^0$.
This initial vector is multiplied with $\bf{D}$, and we get the 0th folded vector $\sigma^0_{\rm f}= {\bf D} \sigma^{0}$.
\item[2)] The next trial input function, $\sigma^1$, can be established by adding the difference of the experimentally measured spectrum, $\sigma_{\rm{exp}}$, and the folded spectrum, $\sigma^0 _{\rm f}$,
to $\sigma^0$. In order to be able to add the folded and the input vector together, we first perform a  Piecewise Cubic Hermite Interpolating Polynomial (pchip) interpolation on the folded vector so that the two vectors have equal dimensions. Our new input vector is:

\begin{equation}
\sigma^1 = \sigma^0 + (\sigma_{\rm{exp}}-\sigma^0 _{\rm f}).
\end{equation} 

\item[3)] The steps 1) and 2) are iterated $i$ times giving
\begin{eqnarray}
\sigma^i_{\rm f} &=& {\bf D} \sigma^{i}
\\
\sigma^{i+1}     &=& \sigma^i + (\sigma_{\rm{exp}}-\sigma^i _{\rm f})
\end{eqnarray}
until convergence is achieved. 
This means that
$\sigma^{i+1}_{\rm f} \approx \sigma_{\rm exp}$ within the statistical errors.
In order to quantitatively check convergence, we calculate the reduced $\chi^2$ of $\sigma^{i+1}_{\rm f}$ and
$\sigma_{\rm{exp}}$ after each iteration.
Approximately four iterations are usually enough for convergence, which is defined when the reduced $\chi^2$ value approaches $\approx 1$.
\end{itemize}

We stopped iterating when the $\chi^2$ became lower than unity. In principle, the iteration could continue until the reduced $\chi^2$ approaches zero,
but that results in large unrealistic fluctuations in $\sigma^i$ due to over-fitting to the measured points $\sigma_{\rm exp}$.

\begin{figure}[tb]
\includegraphics[clip,width=1.0\columnwidth]{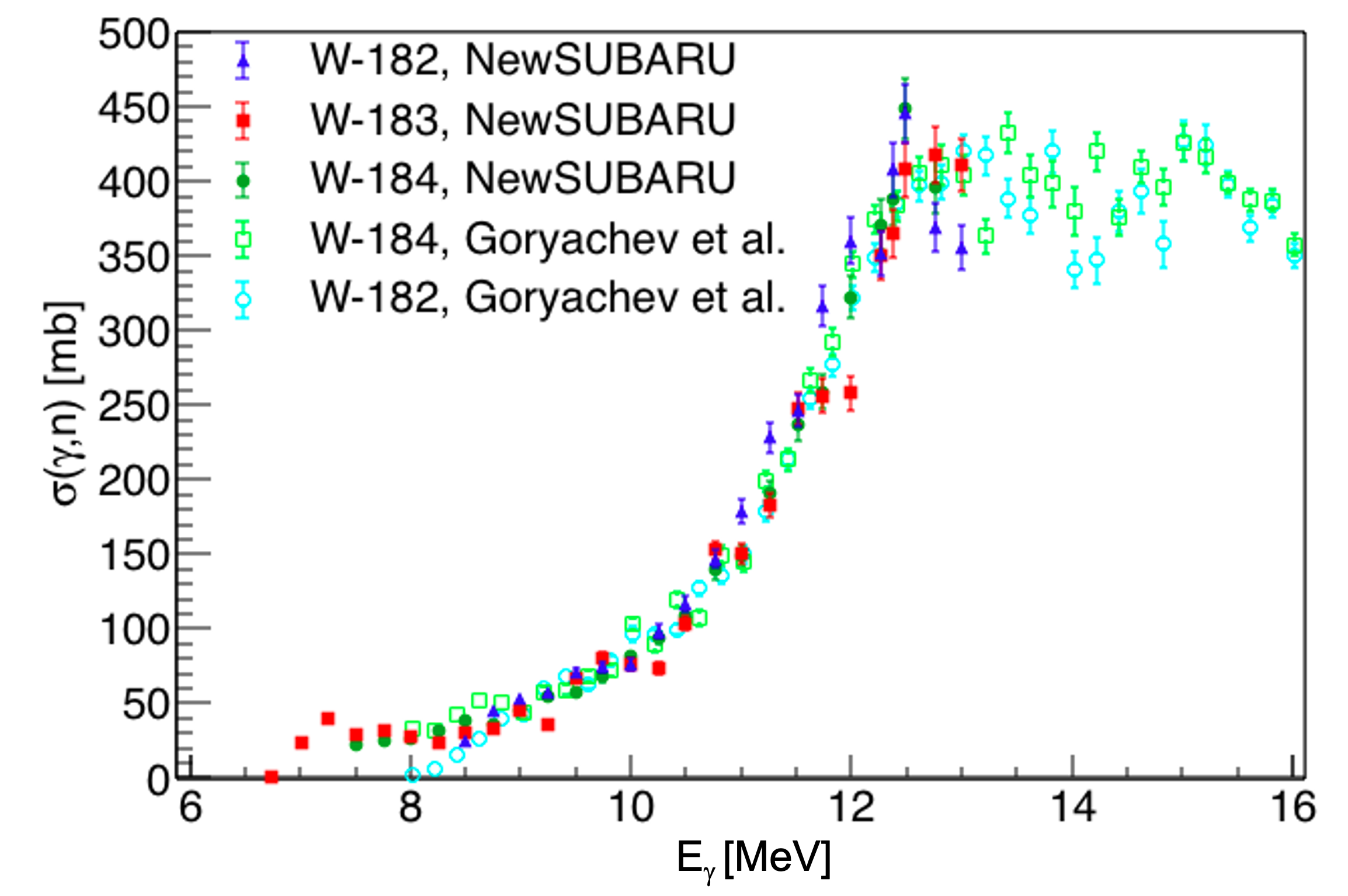}
\caption{(Color online) Cross sections of $^{182,183,184}\rm{W}$ obtained after deconvolution. Also shown are cross sections of $^{182,184}$W from Goryachev \textit{et al.}~\cite{Goryachev1978}.}
\label{fig:Unfolded}
\end{figure}

We estimate the total uncertainty in the unfolded cross sections by calculating an upper limit of the monochromatic cross sections from Fig.~\ref{fig:Monocross} by adding and subtracting the errors to the measured cross section values. These upper and lower limits are then unfolded separately, resulting in the unfolded cross sections shown in Fig.~\ref{fig:Unfolded}.

In Fig.~\ref{fig:Unfolded}, the unfolded cross sections for $^{182,183,184}$W are evaluated at the maximum energies of the incoming $\gamma$ beams. 
The error bars represent the statistical errors and the systematic error due to the uncertainty in the absolute efficiency calibration of the neutron detector. 
The results are compared to data on $^{182,184}$W from Goryachev \textit{et al.}~\cite{Goryachev1978}, and the agreement is overall quite reasonable although some local discrepancies can be observed. 
These discrepancies are sometimes not within the given uncertainties, and could be due to unknown systematic errors.


\section{The Oslo experiment}
\label{sec:expOslo}
\subsection{Experimental details}
The $^{186}$W($\alpha,\alpha^\prime\gamma$) inelastic-scattering experiment was performed at the Oslo Cyclotron Laboratory. A fully-ionized 30-MeV $\alpha$ beam was delivered by the MC-35 Scanditronix cyclotron and directed to the $^{186}$W target. The radio frequency was set to 23.76 MHz, giving a beam burst every 42.09 ns.
The experiment was run for about eight days with typical beam intensities of $1.5-2.2$ enA. 
The target was mounted on a 24-$\mu$m carbon backing, and the target thickness was 0.31 mg/cm$^2$ with enrichment $>98\%$ in $^{186}$W.

To detect the outgoing charged particles, we used the Silicon Ring (SiRi)~\cite{Guttormsen_NIMA_2011} placed in backward angles with respect to the beam direction.
SiRi is a $\Delta E$-$E$ telescope array consisting of eight 1550-$\mu$m thick back ($E$) detectors, each of which has a 130-$\mu$m thick front ($\Delta E$) detector divided in eight strips. 
A 10.5-$\mu$m thick Al foil was placed in front of SiRi to reduce the amount $\delta$ electrons from the target.
SiRi covers about $6\%$ of $4\pi$ and the strips have an angular resolution of about 2$^\circ$, where the center of the strip is at $126-140^\circ$ (in steps of 2$^\circ$); measured from the center of the front detector (at 133$^\circ$), the distance of SiRi from the center of the target was 5 cm.

\begin{figure*}[tb]
\includegraphics[width=2.1\columnwidth]{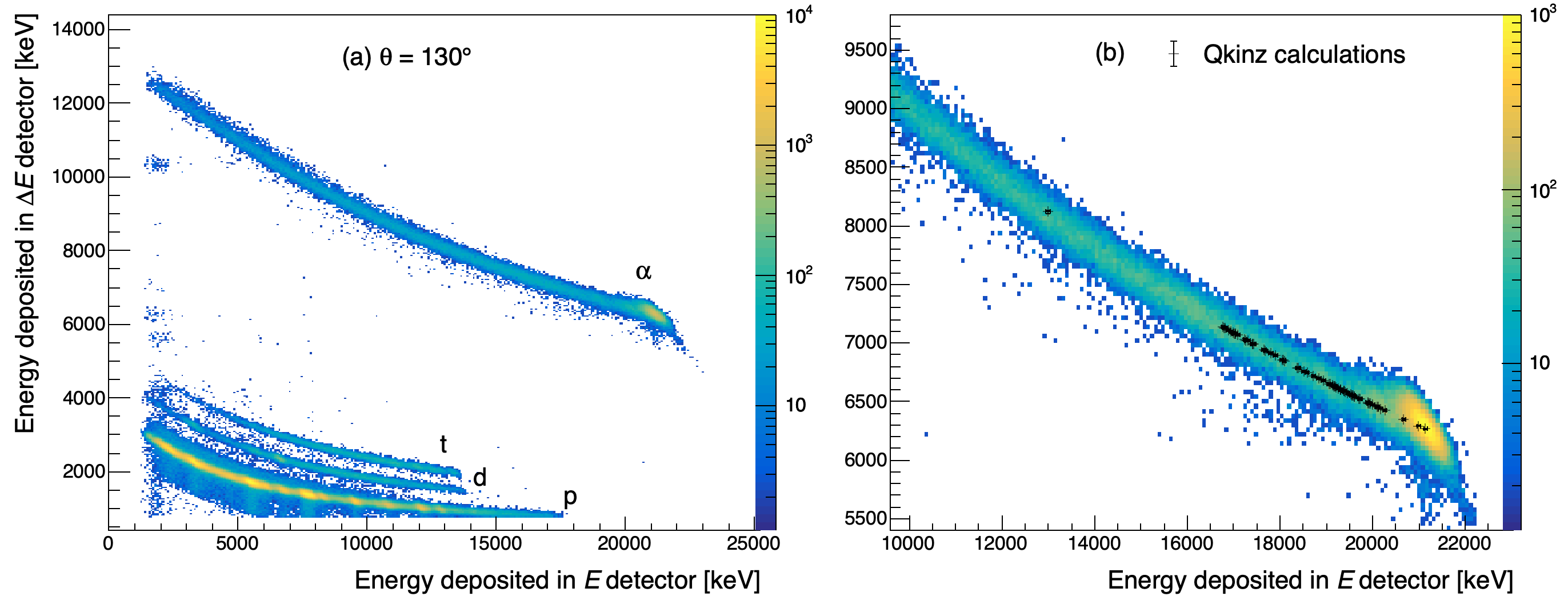}
\caption{(Color online) (a) Particle-identification spectrum for one of the front strips at 130$^\circ$ with its corresponding back detector ($\Delta E$--$E$ banana plot); (b) a zoom on the $\alpha$-particle banana with the \textsf{Qkinz} calculations used for calibration (crosses).}
\label{fig:siribananas}
\end{figure*}

The $\Delta E$-$E$ telescopes allow for separating different charged-particle species. 
Figure~\ref{fig:siribananas}a shows the measured protons, deuterons, tritons, and $\alpha$ particles for a strip at 130${^\circ}$. 
To select the $^{186}$W($\alpha,\alpha^\prime$) events, a gate was set on the ``banana'' corresponding to the $\alpha$ particles. 
To calibrate the SiRi front and back detectors, we used range calculations for our setup with the \textsf{Qkinz} code~\cite{Qkinz}, see Fig.~\ref{fig:siribananas}b.

The resolution of the $\alpha$ particles was measured to be $330$--$360$ keV FWHM for the peak of the elastically-scattered $\alpha$ particles. 
The relatively poor resolution is mainly due to a rather elongated beam spot on the target ($\approx 3$--$4$ mm in diameter in the vertical direction, and $\approx 1$ mm in the horizontal direction).
The master-gate signal for the data acquisition system was a logical signal of 2$\mu$s generated when an $E$ detector gave a signal above threshold, which was set to $\approx 200$ mV. 

Using the CACTUS array~\cite{guttormsen_CACTUS_1990}, we measured $\gamma$ rays in coincidence with the inelastic scattered $\alpha$ particles. 
In the configuration used for this experiment, CACTUS consisted of 26 NaI(Tl) crystals of cylindrical shape (5\textit{in.}$\times$ 5\textit{in.}). 
All crystals were collimated with lead collimators and had 2-mm thick Cu shields in front to attenuate X-rays.
The NaI(Tl) detectors were mounted on the spherical CACTUS frame, so that the front end of each crystal was positioned 22 cm from the center of the target.
The efficiency of CACTUS (for 26 NaI(Tl) detectors) is 14.1(2)\% as measured with a $^{60}$Co source, and with a resolution of $\approx 6.8$\% FWHM for $E_\gamma = 1.33$ MeV.
Using analog electronics, we obtained a lower threshold of about 350 keV for the NaI(Tl) detectors. 

The CACTUS detectors were calibrated in energy by gating on the protons in SiRi. 
As the target had a significant contamination of carbon (from the backing) and oxygen, we used peaks in the proton spectrum from the $^{12}$C($\alpha,p\gamma$)$^{15}$N and $^{16}$O($\alpha,p\gamma$)$^{19}$F reactions to further identify $\gamma$ rays for calibration. 
In particular, we used the 5.269-MeV transition from the 5/2$^+$ first-excited level in $^{15}$N together with the 1.868-MeV transition from the 13/2$^+$ level at $E_x = 4.648$ MeV in $^{19}$F.
Then we cross-checked the obtained calibration with the 1235-keV and 2583-keV lines of $^{19}$F, in addition to the 511-keV $\gamma$ ray from positron annihilation. 

To obtain $\alpha$--$\gamma$ coincident events, we applied a gate on the time-to-digital converter (TDC) spectra for the prompt peak, and subtracting randomly correlated events. 
The start of the TDCs is given by the master gate, and the stop signal is generated from the NaI(Tl) detectors (each NaI(Tl) has an individual TDC), with a built-in delay from the Mesytec shapers of $\approx 400$ ns. 
The range of the TDCs was 1.2 $\mu$s.
The gate on the prompt peak was set to $\Delta t = 0 \pm 20$ ns, while the gate for the background subtraction was set to $\Delta t = 135 \pm 20$ ns. 

\begin{figure*}
\center
\includegraphics[width=2.0\columnwidth]{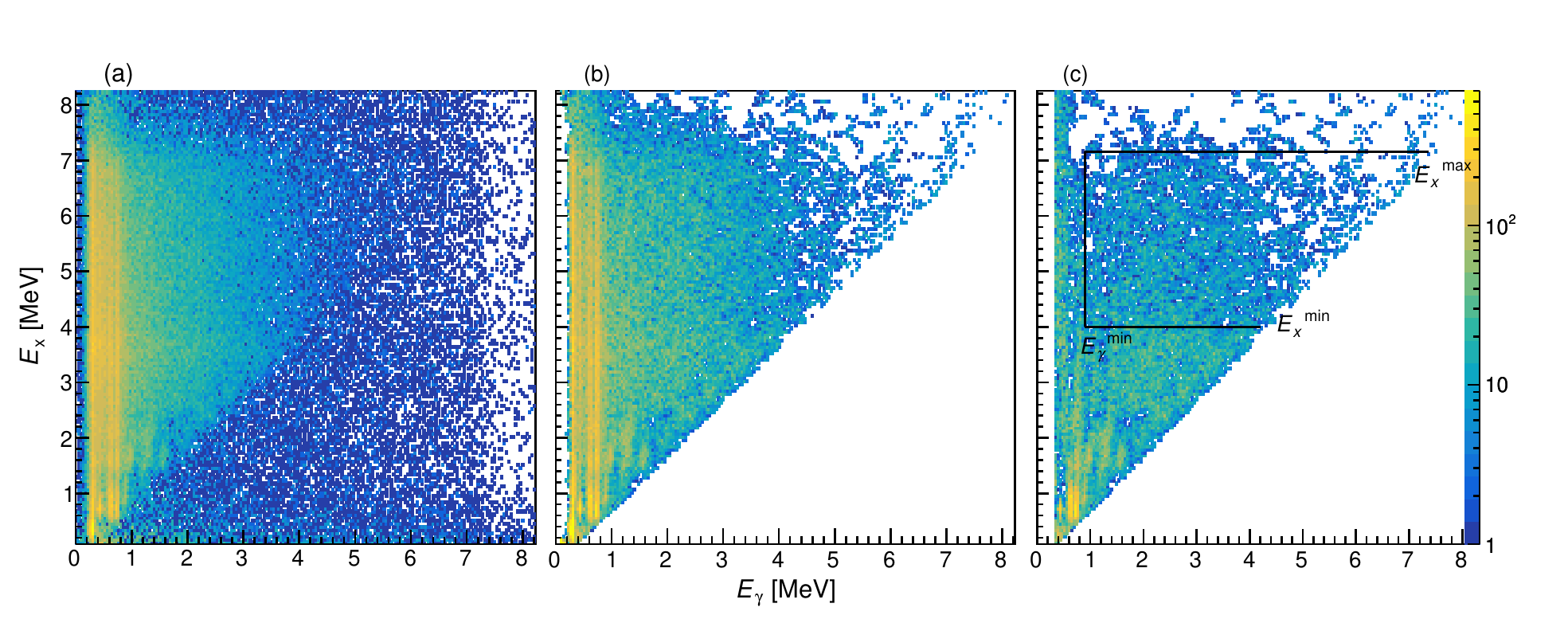}
\caption{(Color online) Excitation-energy vs. $\gamma$-ray energy matrices of $^{186}$W. (a) Background-subtracted data; (b) unfolded $\gamma$-ray spectra; (c) first-generation $\gamma$-ray spectra. The lines indicate the limits set for the further analysis.}
\label{fig:matrices}
\end{figure*}

Using the reaction kinematics, we determined the initial excitation energy of the residual nucleus from the deposited energy of the $\alpha$ particles in SiRi. 
Applying the time gates for the $\gamma$ rays, we obtained excitation-energy tagged, background-subtracted $\gamma$-ray spectra as shown in Fig.~\ref{fig:matrices}a. 

The $\gamma$-ray spectra needed to be corrected for the CACTUS detector response. 
For this purpose, we applied the iterative unfolding method of Ref.~\cite{guttormsen_unfolding_1996} available in the Oslo-method software package~\cite{Oslosoftware}. 
This method takes the raw $\gamma$-ray spectrum as a starting point for the unfolded (``true'') spectrum. 
This trial spectrum is folded with the known detector response, and then compared with the raw spectrum. 
By taking the difference between the folded spectrum and the raw spectrum, a new, improved trial spectrum is made. 
This process is repeated until the folded spectrum is approximately equal to the raw spectrum, within the experimental uncertainties. 
To preserve the experimental statistical fluctuations, and not introduce artificial, spurious ones, the Compton subtraction method is also applied. 
This takes advantage of the fact that the Compton distribution is very smooth.
For more details, see Ref.~\cite{guttormsen_unfolding_1996}.
The unfolded $\gamma$-ray spectra for each $E_x$ bin are shown in Fig.~\ref{fig:matrices}b.

After unfolding, the first-generation $\gamma$ rays were extracted from the data by applying an iterative subtraction method~\cite{guttormsen_fg_1987}.
The first-generation $\gamma$ rays are the ones that are emitted first in the decay cascades, and their distribution represents the branching ratios for the various $\gamma$ transitions at a given $E_x$ bin. 
The principle behind the subtraction method is as follows.
For a given $E_x$ bin, say, at $E_x = 4$ MeV, the unfolded spectrum contains \textit{all} the $\gamma$ rays from \textit{all} the possible decay cascades originating from the levels populated in that $E_x$ bin. 
If we now consider the $E_x$ bins \textit{below} $E_x = 4$ MeV, they will contain all the same $\gamma$ rays as the $E_x = 4$ MeV bin, \textit{except} the first-generation $\gamma$s at $E_x = 4$ MeV. 
This is true if the $E_x$ bins have the same decay cascades whether the levels in the bin were populated directly through the nuclear reaction, or if they were populated from $\gamma$ decay of above-lying levels.
We refer the reader to Ref.~\cite{larsen_syst_2011} for a more in-depth discussion on the assumptions behind the first-generation method.
The first-generation $\gamma$ spectra are displayed in Fig.~\ref{fig:matrices}c.

\subsection{Extraction of level density and $\gamma$-ray transmission coefficient}
\label{subsec:rhosigchi}
We now exploit the fact that the first-generation $\gamma$ spectra represent the (relative) branching ratios for a given initial excitation-energy bin, and that we have many such branching ratios available for a large $E_x$ region. 
In the spirit of Fermi's Golden Rule~\cite{Dirac1927,Fermi1950}, where the decay rate is proportional to the level density at the final excitation energy and the reduced transition probability for decay between a given initial and final level, we use the following ansatz~\cite{schiller2000}:
\begin{equation}
    P(E_\gamma,E_x) \propto \rho(E_x - E_\gamma)\cdot \mathcal{T}(E_\gamma),
\end{equation}
where $P(E_\gamma,E_x)$ is the matrix of first-generation $\gamma$ rays (Fig.~\ref{fig:matrices}c), $\rho(E_x - E_\gamma)$ is the level density at the excitation energy where the $\gamma$ transition ``lands'' and $\mathcal{T}(E_\gamma)$ is the $\gamma$-ray transmission coefficient.
Note that $\mathcal{T}(E_\gamma)$ is only a function of $E_\gamma$, which means that the Brink-Axel hypothesis~\cite{brink1955,axel1962} is invoked. 
Brink stated that
\begin{quote}
    ``...we assume that the energy dependence of the photo effect is independent of the detailed structure of the initial state so that, if it were possible to perform the photo effect on an excited state, the cross section for absorption of a photon of energy $E$ would still have an energy dependence given by (15).'' 
\end{quote}
where ``(15)'' is referring to the equation describing the Giant  Dipole Resonance (GDR) with a Lorentzian function that only depends on the $\gamma$-transition energy. 
Brink's original formulation (as well as Axel's application of Brink's hypothesis) concerned only $E1$ transitions, and there is a wealth of recent works in the literature discussing the validity and/or violation of the hypothesis; see, \textit{e.g.}, Refs.~\cite{Misch_PhysRevC.90.065808,JOHNSON201572,Hung_PhysRevLett.118.022502,Guttormsen_PhysRevLett.116.012502,Martin_PhysRevLett.119.182503,CrespoCampo_PhysRevC.98.054303,Scholz_PhysRevC.101.045806,Angell_PhysRevC.86.051302,ISAAK2019225}. 
\begin{figure*}[h!t]
\center
\includegraphics[clip,width=2.0\columnwidth]{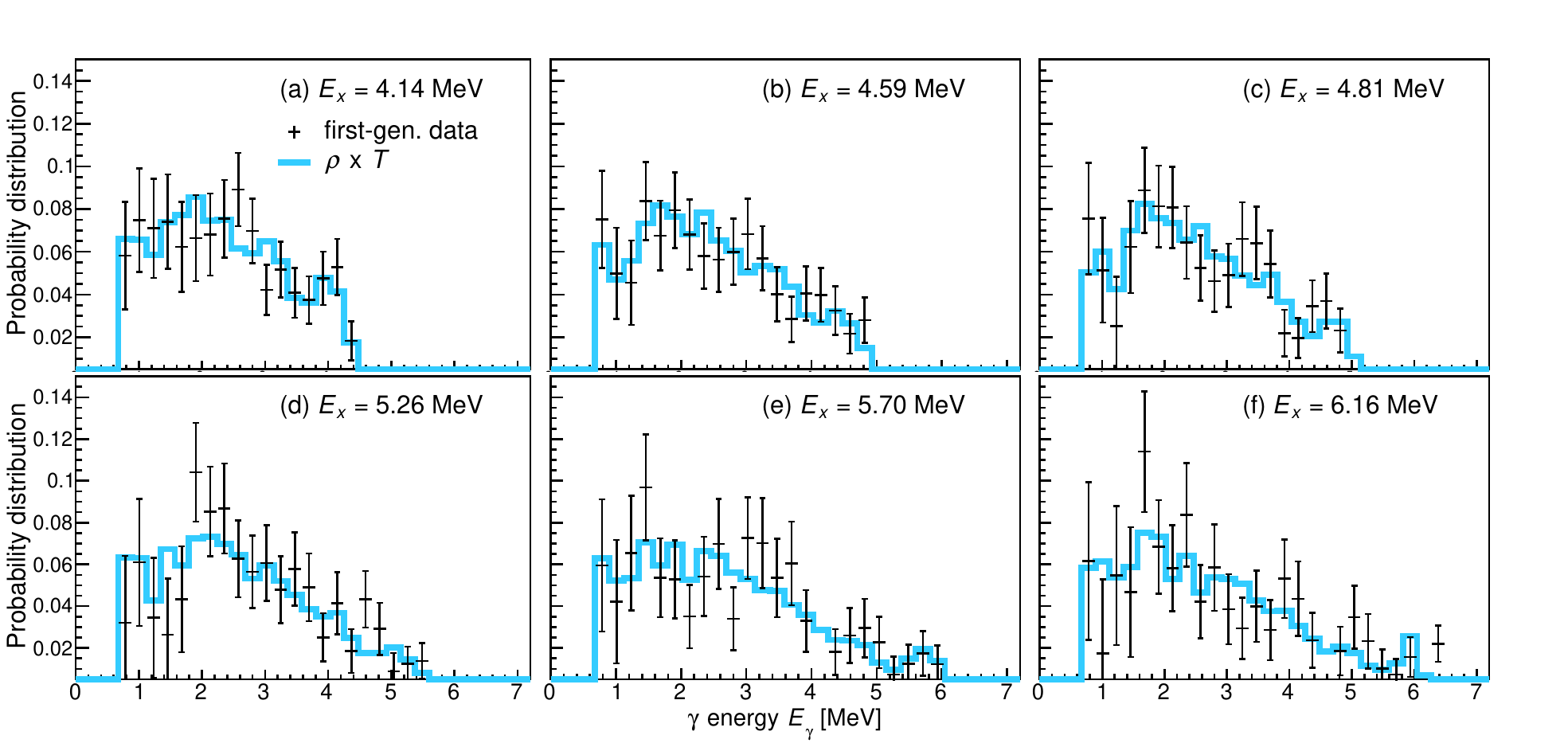}
\caption{(Color online) Experimental first-generation spectra (black crosses) compared to the predicted ones using the extracted level density and $\gamma$-transmission coefficient (blue line) for various excitation-energy bins (224-keV wide).}
\label{fig:doesitwork}
\end{figure*}

A necessary condition for the Oslo method is that the Brink hypothesis is at least approximately true for the specific excitation-energy region used for extracting the level density and $\gamma$-ray transmission coefficient. 
We have performed tests of this assumption for the application in the Oslo method in Ref.~\cite{larsen_syst_2011}.
When the Brink hypothesis is applicable, we can fit the data of the first-generation $\gamma$ rays to obtain a reliable estimate of the level density and the $\gamma$-ray transmission coefficient through an iterative optimization using a least-squares fit: 
\begin{equation}
    \chi^2_{\mathrm{red}} = \frac{1}{N_{\mathrm{free}}}\sum_{E_i=E_x^{\mathrm{min}}}^{E_x^{\mathrm{max}}} \sum_{E_\gamma = E_\gamma^{\mathrm{min}}}^{E_i} \frac{\left[P(E_\gamma,E_i)-P_{\mathrm{th}}(E_\gamma,E_i)\right]^2}{\left[\Delta P(E_\gamma,E_i) \right]^2}.
\end{equation}
Here, $P(E_\gamma,E_i)$ is the experimental matrix of first-generation $\gamma$ rays where each row is normalized to unity: 
\begin{equation}
\sum_{E_i = E_\gamma^{\mathrm{min}}}^{E_i} P(E_\gamma,E_i) = 1,
\end{equation}
and $\Delta P(E_\gamma,E_i)$ is the uncertainties in the first-generation matrix (including statistical errors and an estimate for systematic uncertainties due to unfolding and the first-generation method, see Ref.~\cite{schiller2000}).
Moreover, $N_\mathrm{free}$ is the number of degrees of freedom and $P_{\mathrm{th}}(E_\gamma,E_i)$ is the approximation for the theoretical first-generation matrix~\cite{schiller2000}: 
\begin{equation}
    P_{\mathrm{th}}(E_\gamma,E_i) = \frac{\rho(E_i - E_\gamma)\mathcal{T}(E_\gamma)}{\sum_{E_\gamma = E_\gamma^{\mathrm{min}}}^{E_i} \rho(E_i - E_\gamma)\mathcal{T}(E_\gamma) }.
\end{equation}
The number of degrees of freedom, $N_\mathrm{free}$, is given by $N_\mathrm{free} = N_{\mathrm{ch}}(P) - N_{\mathrm{ch}}(\rho) - N_{\mathrm{ch}}(\mathcal{T})$. 
For the present data set, we have used $E_\gamma^{\mathrm{min}} = 0.90$ MeV, $E_x^{\mathrm{min}} = 4.0$ MeV, and  $E_x^{\mathrm{max}} = 7.2$ MeV as shown in Fig.~\ref{fig:matrices}c.
Note that the neutron separation energy $S_n$ of $^{186}$W is 7.1920(12) MeV~\cite{NNDC}, and as we have no way of discriminating against neutrons, the Oslo method is usually limited to a maximum excitation energy (close to) $S_n$. 
With bin size of 224 keV, and the limits applied as shown in Fig.~\ref{fig:matrices}c, we have the number of pixels in the first-generation matrix $N_{\mathrm{ch}}(P) = 330$, while the number of elements in the vectors of $\rho$ and $\mathcal{T}$ is $N_{\mathrm{ch}}(\rho) = N_{\mathrm{ch}}(\mathcal{T}) = 39$, giving $N_\mathrm{free} = 252$. 
It is important to note that the number of data points in the first-generation matrix, $N_{\mathrm{ch}}(P)$, is much bigger than the number of points to be estimated, which is $2 \times 39$ points; this is why the method usually converges very well. 
When convergence is reached, the extracted $\rho(E_x - E_\gamma)$ and $\mathcal{T}(E_\gamma)$ are the ones that best describe the experimental $P(E_\gamma,E_i)$ matrix.  
For this case, we obtain  $\chi^2_{\mathrm{red}} = 0.85$ after 20 iterations.

As a visual illustration of the fit, Fig.~\ref{fig:doesitwork} shows some of the experimental first-generation spectra together with the spectra obtained for $P_{\mathrm{th}}$. 
Overall, the agreement is quite good, although we remark that the experimental errors are rather large. 
Note that the fit is performed on \textit{all} the first-generation spectra (for 15 excitation-energy bins), and so the fit is still well constrained. 

Schiller \textit{et al.} showed~\cite{schiller2000} that the $\chi^2$ minimization obtains a unique solution for the relative variation of neighboring points in the functions $\rho$ and $\mathcal{T}$; however, an equally good fit to the experimental $P$ matrix is given by the transformation
\begin{eqnarray}
\tilde{\rho}(E_i-E_\gamma)&=&\mathcal{A}\exp[\alpha(E_i-E_\gamma)]\,\rho(E_i-E_\gamma),
\label{eq:array1}\\
\tilde{{\mathcal{T}}}(E_\gamma)&=&B\exp(\alpha E_\gamma){\mathcal{T}} (E_\gamma).
\label{eq:array2}
\end{eqnarray}
Here, $\alpha$ is the common slope adjustment of $\rho$ and $\mathcal{T}$, while $\mathcal{A}$ and $B$ gives the absolute scaling of $\rho$ and $\mathcal{T}$, respectively. 
These parameters must be determined from external data, as described in the following sections.

\subsection{Normalization of level density}
\label{subsec:nldnorm}
To normalize the level density by determining the $\alpha$ and $\mathcal{A}$ parameters, we make use of discrete levels~\cite{NNDC} at low $E_x$ and data on $s$-wave neutron resonance spacings~\cite{Mughabghab2018} at the neutron separation energy $S_n$. 
The average $s$-wave neutron resonance spacing $D_0 = 9.3(16)$ eV~\cite{Mughabghab2018} represents the spacing of levels with $J^\pi = 1^-, 2^-$ as the target nucleus $^{185}$W has ground-state spin/parity $I_t^\pi = \frac{3}{2}^-$.
To obtain the \textit{total} level density at $S_n$, we need to apply a model for the spin distribution, in particular the \textit{spin cutoff parameter} $\sigma_J(E_x)$.
Here, we use as a starting point the model of von Egidy and Bucurescu~\cite{Egidy_PhysRevC.72.044311,Egidy_PhysRevC.73.049901} employing the rigid-body moment of inertia. 
However, as shown by Uhrenholt \textit{et al.}~\cite{UHRENHOLT2013127}, at excitation energies around 7$-$8 MeV for heavy nuclei, a full rigid-body moment of inertia might not be reached yet: in Fig.~10 of Ref.~\cite{UHRENHOLT2013127}, the effective moment of inertia is $\approx 85$\% of the rigid-body moment of inertia at $E_x \approx 8$ MeV. 
We take this as the reference value for which we will vary the spin cutoff parameter to obtain an estimate for the systematic uncertainty connected to the spin distribution, with the effective moment of inertia ranging from 70\%$-$100\% of the rigid-body moment of inertia:
\begin{equation}
    \sigma_J^2(S_n) = \eta \, 0.0146 A^{5/3} \frac{1+\sqrt{1+4a(S_n-E_1)}}{2a},
\end{equation}
where $\eta$ is the reduction factor set to 0.85(15), $A$ is the mass number of the nucleus (here 186), $a$ is the level-density parameter and $E_1$ is an excitation-energy shift taken from the global systematics of von Egidy and Bucurescu~\cite{Egidy_PhysRevC.72.044311,Egidy_PhysRevC.73.049901} calculated with the \textsf{robin.c} code in the Oslo-method software package (see Table~\ref{tab:table_nld}).
This gives us a range of values for the estimated $\rho(S_n)$, which is then calculated as~\cite{schiller2000,larsen_syst_2011}
\begin{equation}
    \rho(S_n) =  \frac{2\sigma_J^2}{D_0\left[(I_t+1)e^{-(I_t+1)^2/2\sigma_J^2} + I_te^{-I_t^2/2\sigma_J^2}\right]},
\label{eq:d2rho}
\end{equation}
assuming an equal parity distribution for all spins at the neutron separation energy.
Uncertainties in the $D_0$ value and the spin cutoff parameter are propagated (for a derivation, see Appendix~\ref{appendix1}). 
All the applied parameters are given in Table~\ref{tab:table_nld}.

Moreover, due to the argument in the level density function being $E_i - E_\gamma$, we get an upper limit for the extracted level density given by $E_x^{\mathrm{max}}- E_\gamma^{\mathrm{min}}$. 
Therefore, we need to make an extrapolation from our data points up to $\rho(S_n)$. 
Here, we use the constant-temperature (CT) model of Ericson~\cite{Ericson1959}:
\begin{equation}
    \rho_{\mathrm{CT}}(E_x) = \frac{1}{T}\exp{\frac{E_x-E_0}{T}},
\end{equation}
where $T$ denotes the nuclear ``temperature'' and $E_0$ is a shift; both parameters are usually obtained from fits to discrete data and to neutron resonance spacings. 
The parameters used for $^{186}$W are shown in Table~\ref{tab:table_nld}.

\begin{table*}[htb]
\caption{Parameters used for the normalization of the level density and $\gamma$-ray transmission coefficient. Note that the $E_0$ parameter is adjusted to make $\rho_{\mathrm{CT}}(S_n)$ match with $\rho(S_n) = 26.5\cdot 10^5$ MeV$^{-1}$. The uncertainty in $\left< \Gamma_{\gamma0} \right>$ from Mughabghab~\cite{Mughabghab2018} is given as $5$ meV; however, this uncertainty seems too small based on the experimental errors in the radiative width for other W isotopes, and we have chosen a more conservative uncertainty in line with the experimental errors of $^{182,183,184,186}$W.   } 
\begin{tabular}{cccccccccccc}
\hline
\hline
$S_n$  & $I_t^{\pi}$ & $D_0$ & $\sigma^2_J(S_n)$ & $a$          & $E_1$ & $\rho(S_n)$ & $T$ & $E_0$ & $\left< \Gamma_{\gamma0} \right>$ & $\sigma_d^2$ & $E_d$\\
(MeV)  &             & (eV)  &                   & (MeV$^{-1}$) & (MeV) &  10$^5$ (MeV$^{-1}$) & (MeV)   & (MeV) & (meV)  & & (MeV)    \\
\hline
7.192 & $3/2^{-}$	 & 9.3(16)  & 47(8)       &  19.38       & 0.28  & 26.5(64) & 0.51(1) & -0.0077 & 60$^{+13}_{-9}$ & 7.3(13) & 0.86(19)  \\

\hline
\hline
\end{tabular}
\label{tab:table_nld}
\end{table*}
\begin{figure}
\center
\includegraphics[width=1.0\columnwidth]{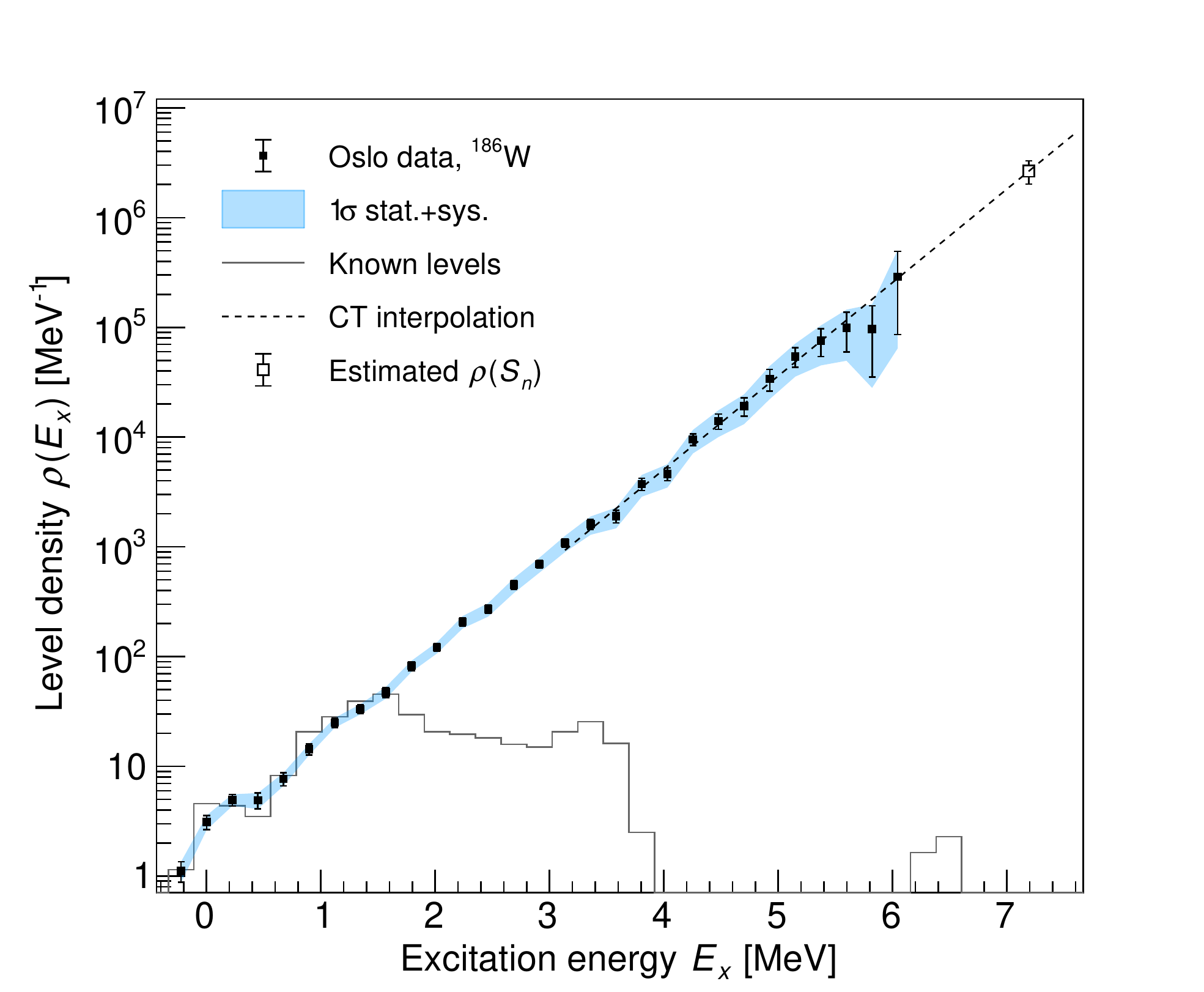}
\caption{(Color online) Normalized level density of $^{186}$W. The discrete levels~\cite{NNDC} are binned with the same bin size as our data (224 keV/channel). The dashed line shows the CT-model interpolation between our data and $\rho(S_n)$. The black error bars represent statistical uncertainties from the experiment and systematic errors connected to the unfolding procedure and the first-generation method. The blue band includes also systematic errors from the normalization procedure (see text).   }
\label{fig:leveldensity}
\end{figure}

From the Oslo-method software, statistical uncertainties and an estimate of systematic errors due to the unfolding procedure and the first-generation method are calculated as described in Ref.~\cite{schiller2000}. 
We also include systematic errors from the normalization procedure, accounting for the uncertainty in the experimental $D_0$ value as well as the uncertainty in the moment of inertia and thus the spin cutoff parameter as described above.
We estimate the uncertainty (approximately one standard deviation) including all these factors as 
\begin{equation}
    \delta \rho = \rho_{\rm rec} \sqrt{ \left( \frac{\delta D_0}{D_0}\right)^2 + \left( \frac{\delta \sigma_J}{\sigma_J}\right)^2 + \left( \frac{\Delta \rho_{\rm rec}}{\rho_{\rm rec}}\right)^2},
\end{equation}
where $\rho_{\rm rec}$ is the central value (``recommended'' normalization), and $\Delta \rho_{\rm rec}$ represents statistical uncertainties and systematic errors from unfolding and the first-generation method. 
The resulting normalized level density is shown in Fig.~\ref{fig:leveldensity}.

\subsection{Normalization of $\gamma$-ray strength}
\label{subsec:gsfnorm}
\begin{figure}
\center
\includegraphics[width=1.0\columnwidth]{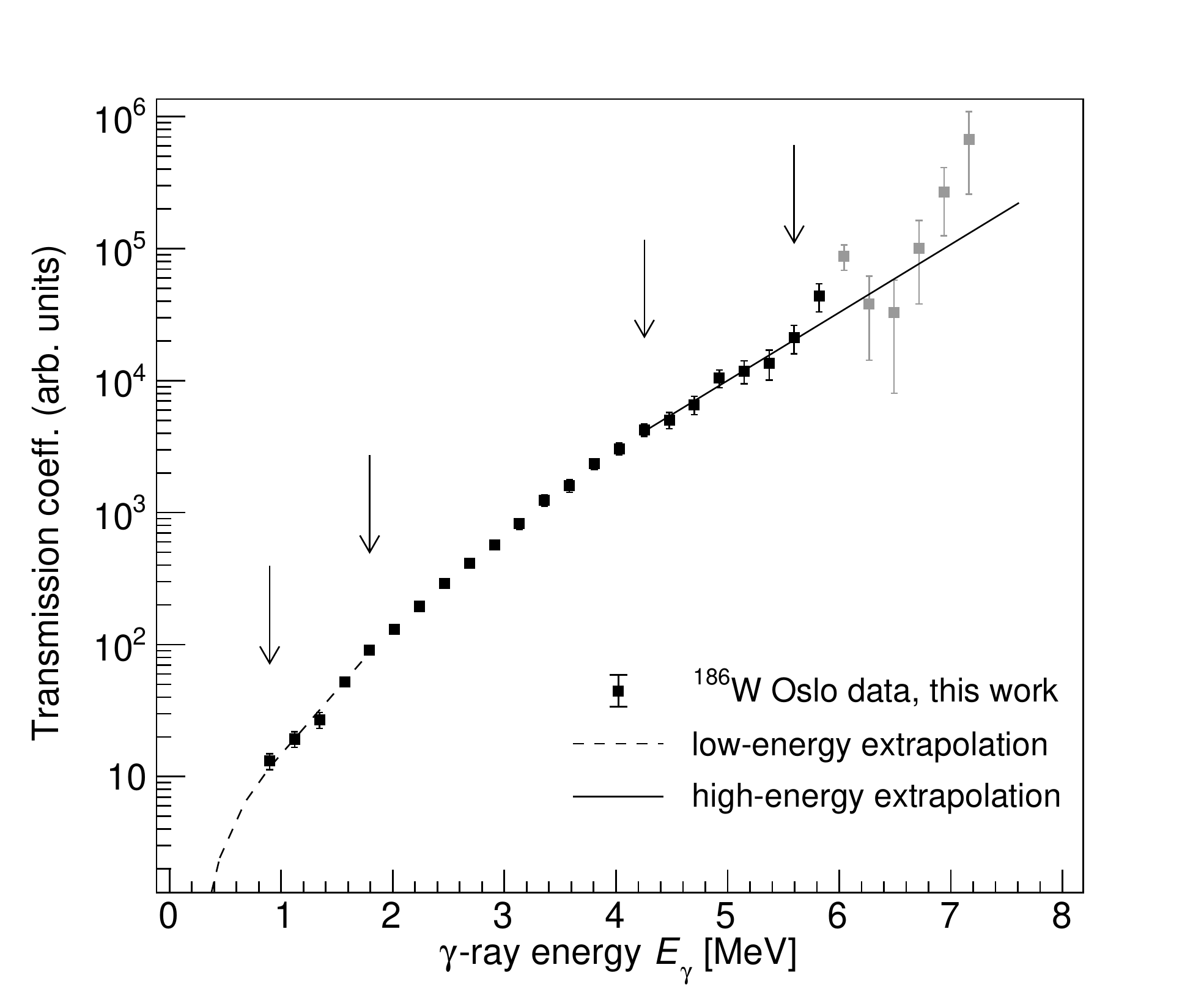}
\caption{Gamma-ray transmission coefficient of $^{186}$W before normalization. The arrows indicate the fit regions used for determining the extrapolations (see text). The gray data points are not considered further in the analysis due to very low statistics in the first-generation matrix for these $\gamma$ energies.}
\label{fig:gammatransmission}
\end{figure}
Having the normalized level density at hand, we proceed to normalizing the $\gamma$-ray transmission coefficient $\mathcal{T}(E_\gamma)$ by determining the scaling parameter $B$ in Eq.~(\ref{eq:array2}). 
Here we make use of the relation between the average, total radiative width $\left< \Gamma_{\gamma 0} \right>$ deduced from $s$-wave neutron resonances, the level density and the transmission coefficient~\cite{Voinov2001,larsen_syst_2011}:
\begin{align}
\langle \Gamma_{\gamma0}\rangle & =
 \frac{B D_0}{4\pi}\int_{E_{\gamma}=0}^{S_n}dE_{\gamma}\mathcal{T}(E_{\gamma})\rho(S_n-E_{\gamma}) \times \nonumber \\ 
 \sum_{J= -1}^{1} &\left[ g(S_{n}-E_{\gamma},I_{t}-1/2+J) + g(S_{n}-E_{\gamma},I_{t}+1/2+J)\right],
\label{eq:width}
\end{align}
where $g$ is the spin distribution~\cite{Bethe1936,Ericson1958}:
\begin{equation}
    g(E_x,J) \simeq \frac{2J+1}{2\sigma_J^2}\exp\left[-(J+1/2)^2/2\sigma_J^2\right].
\label{eq:spindist}
\end{equation}
As we need the spin distribution for the excitation-energy range $E_x \in [0,S_n]$, we make use of the spin cutoff parameter in the general form~\cite{Capote2009} 
\begin{equation}
    \sigma_J^2(E_x) = \sigma_d^2 + \frac{E_x-E_d}{S_n-E_d}\left( \sigma_J^2(S_n) - \sigma_d^2  \right),
\end{equation}
which is motivated also from microscopic calculations (e.g., shell-model calculations~\cite{renstrom2018_Ni} and the work of Uhrenholt \textit{et al.}~\cite{UHRENHOLT2013127}).
Here, $\sigma_d^2$ represents the spin cutoff parameter at the low excitation energy $E_d$, where the levels are still resolved and with firm spin/parity assignments~\cite{NNDC}, see Table~\ref{tab:table_nld}. 

We need to estimate the $\gamma$-ray transmission coefficient for $E_\gamma < E_\gamma^{\mathrm{min}}$, i.e., where we do not have experimental data, in order to calculate the integral in Eq.~(\ref{eq:width}). 
Therefore, we extrapolate with a fit to the low-energy data points using the functional form $E_\gamma^3\exp(p_1 E_\gamma+p_2)$, where $p_1$ and $p_2$ are free parameters\footnote{This functional form is motivated by shell-model calculations of the low-energy $\gamma$ strength, e.g. Refs.~\cite{Schwengner2013,Brown2014}.}. 
Moreover, the statistics is very low at high $\gamma$-ray energies, and so we make use of an extrapolation here as well, here using a simple exponential, $\exp(p_3 E_\gamma+p_4)$, where  $p_3$ and $p_4$ are again free parameters. 
The fit regions and the extrapolation functions are shown in Fig.~\ref{fig:gammatransmission}. 
The  data points in gray color ($E_\gamma > 6$ MeV) are from a region in the first-generation matrix with very low statistics (see Fig.~\ref{fig:matrices}c), and we therefore choose to exclude those data points from the further analysis. 

\begin{figure}
\center
\includegraphics[width=1.0\columnwidth]{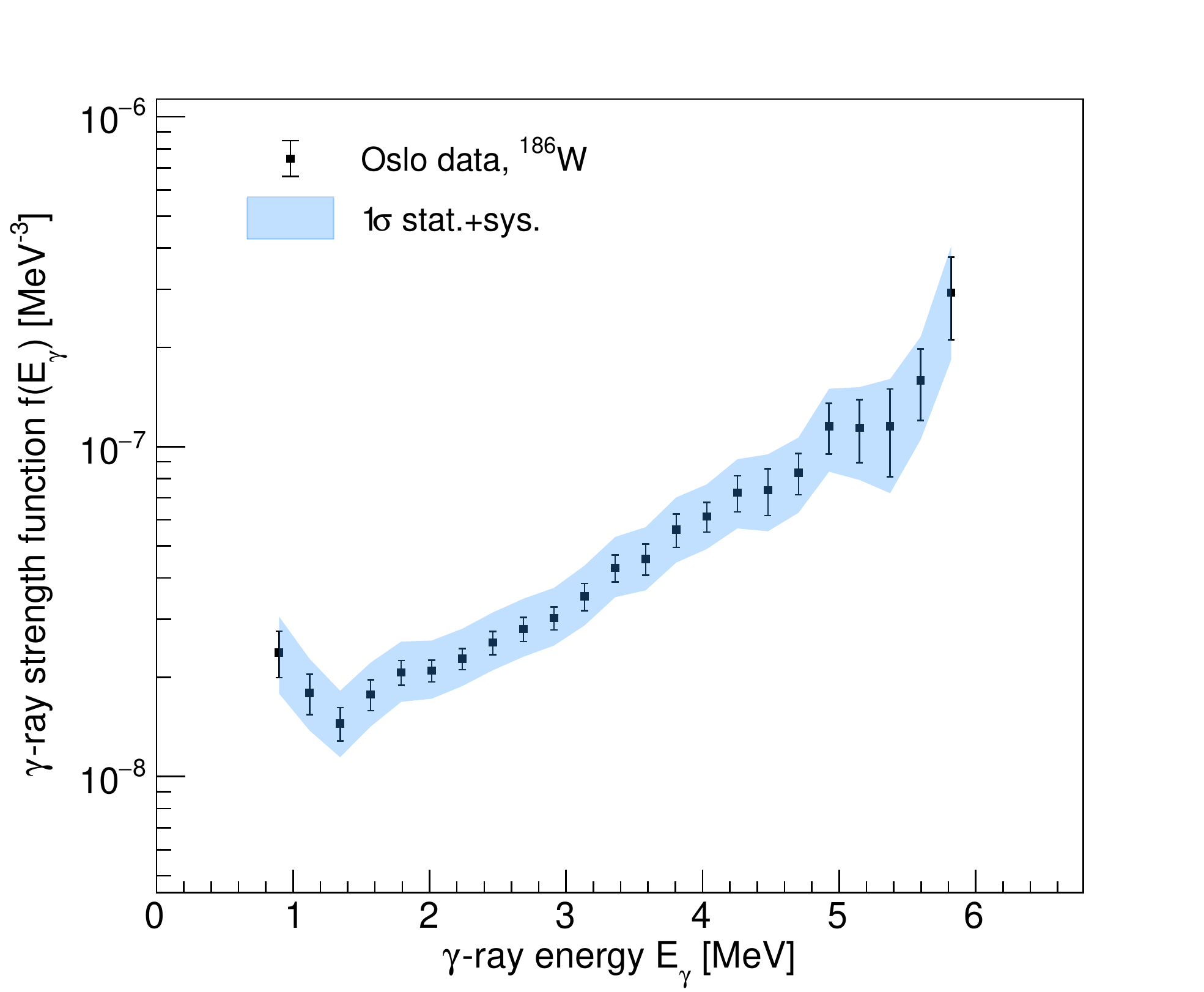}
\caption{(Color online) Gamma-ray strength function of $^{186}$W. The black error bars represent statistical uncertainties from the experiment and systematic errors connected to the unfolding procedure and the first-generation method. The blue band includes also systematic errors from the normalization procedure (see text).}
\label{fig:gammastrength}
\end{figure}

To obtain the $\gamma$-ray strength function, we use the fact that $\gamma$ decay at high excitation energies is largely dominated by dipole transitions (see, e.g., Refs.~\cite{Kopecky1990,Larsen2013,Jones2018}). 
As our experimental data in principle contain transitions of both electric and magnetic character, we get the total dipole strength function $f(E_\gamma)$ through
\begin{equation}
    f(E_\gamma) = \frac{\mathcal{T}(E_\gamma)}{2\pi E_\gamma ^3}.
\end{equation}
In accordance with the approach for the level density, we estimate the uncertainty in the $\gamma$-ray strength function through
\begin{equation}
    \delta f = f_{\rm rec} \sqrt{\left( \frac{\delta D_0}{D_0}\right)^2 + \left( \frac{\delta \sigma_J}{\sigma_J}\right)^2 + \left( \frac{\delta \Gamma_{\gamma 0}}{\Gamma_{\gamma 0}}\right)^2 +  \left( \frac{\Delta f_{\rm rec}}{f_{\rm rec}}\right)^2},
\end{equation}
where $\Delta f_{\rm rec}$ is again the central value (``recommended'' normalization), and $\Delta f_{\rm rec}$ represents statistical uncertainties and systematic errors from unfolding and the first-generation method.
The resulting, normalized $\gamma$-ray strength function is shown in Fig.~\ref{fig:gammastrength}.


\section{Results and discussion}
\label{sec:results}
\subsection{Comparison to other data and models}
\begin{figure}[ht]
\center
\includegraphics[width=1.0\columnwidth]{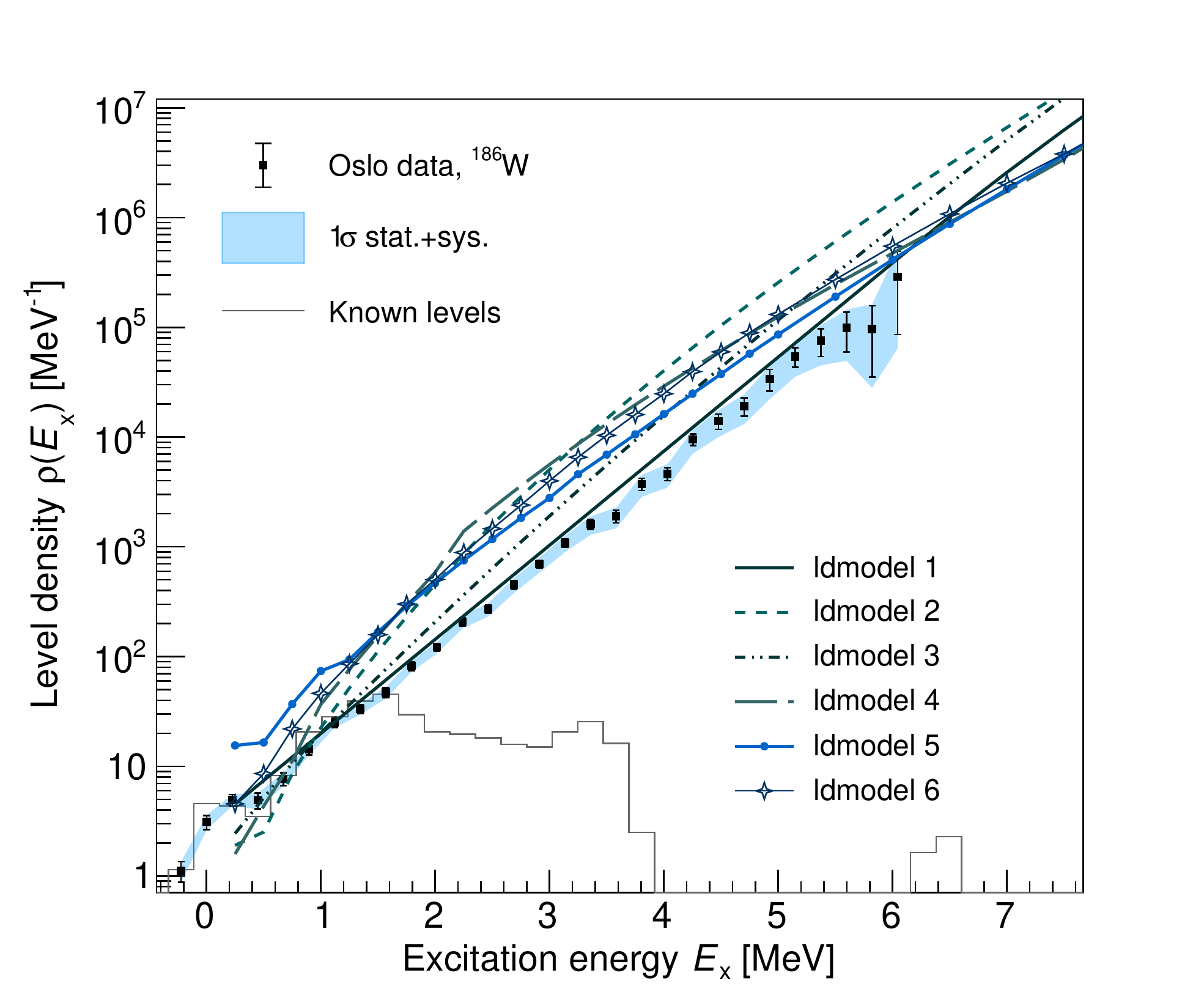}
\caption{(Color online) Comparison of the level-density data from this work with models included in the \textsf{TALYS} code (see text).}
\label{fig:nldmodels}
\end{figure}

The level-density data are compared to various models available in the \textsf{TALYS-1.9} code~\cite{TALYS}, see Fig.~\ref{fig:nldmodels}.
The models are: \textit{ldmodel 1}, the composite formula of Gilbert and Cameron~\cite{gilbert1965}; \textit{ldmodel 2}, the back-shifted Fermi gas model~\cite{Huizenga1969}; \textit{ldmodel 3}, the generalized superfluid model~\cite{Ignatyuk1993}; \textit{ldmodel 4}, calculated within the Hartree-Fock-BCS approach~\cite{Demetriou2001}; \textit{ldmodel 5}, the combinatorial-plus-Hartree-Fock-Bogoliubov approach~\cite{Goriely2008}; and \textit{ldmodel 6}, the combinatorial model combined with a temperature-dependent Hartree-Fock-Bogoliubov calculation~\cite{Hilaire2012}.

From a first look, none of the models seem to be in good agreement with the data, and we remark that the \textsf{TALYS} level densities have not been normalized to the $D_0$ value from Ref.~\cite{Mughabghab2018}. 
In adition, we take notice of two important issues:
(\textit{i}) the spin cutoff parameter we have used in our normalization procedure might not be representative of the corresponding spin distribution in the \textsf{TALYS} models; (\textit{ii}) our data can be re-normalized more coherently for each model by adopting its energy-dependence to extrapolate between the highest energy point and $\rho(S_n)$, as was done e.g. in Ref.~\cite{Goriely2022}. 
Nevertheless, it is clear that the overall shape of our data points are significantly different from several of the level-density models.
We also remark that the slope of our level-density data points is directly linked to the slope of the $\gamma$-strength function as given in Eq.~(\ref{eq:array2}). 
If we were to renormalize our level density to the \textsf{TALYS} models, this would inevitably lead to a change in slope in the $\gamma$-strength function as well.

\begin{figure*}[tb]
\center
\includegraphics[width=2.15\columnwidth]{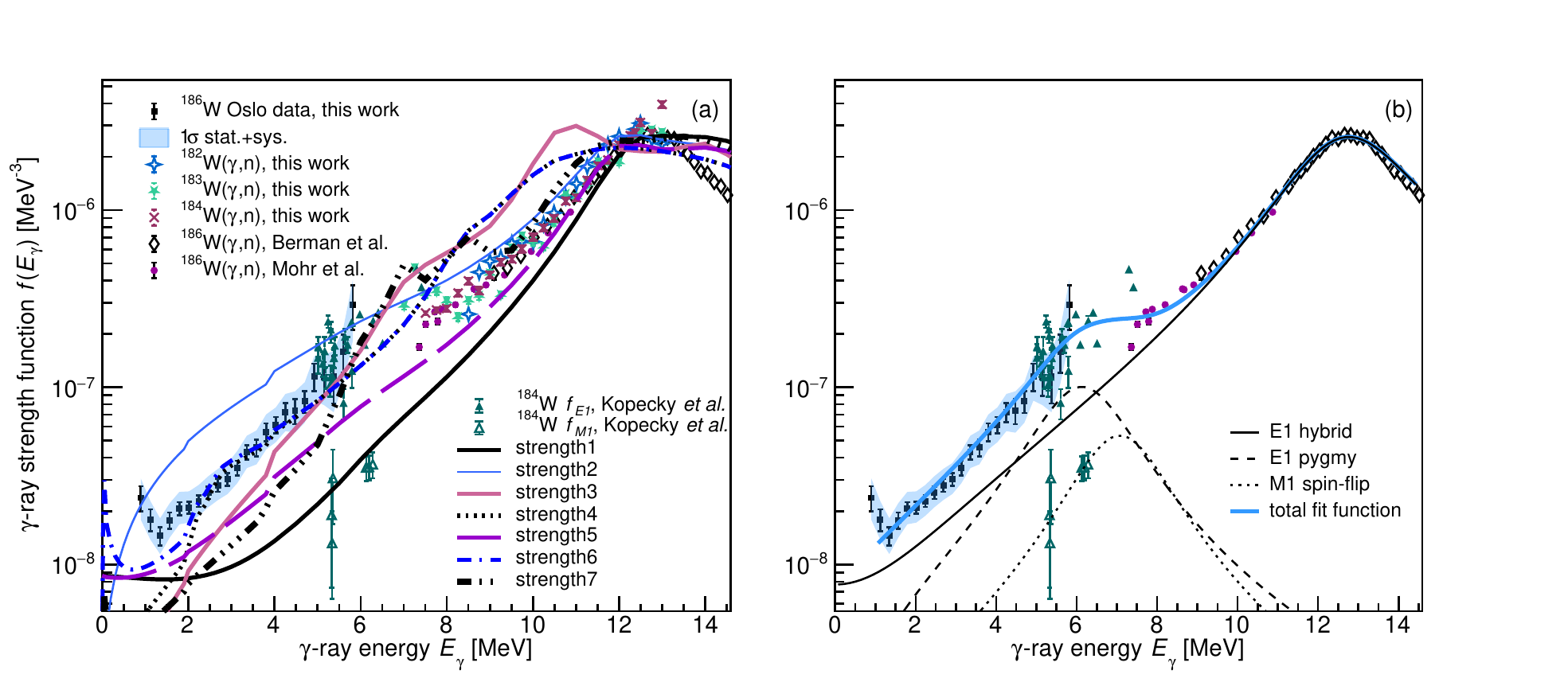}
\caption{(Color online) (a) Comparison of  $\gamma$-strength data from this work with data from the literature (Berman \textit{et al.}~\cite{Berman1969}, Mohr \textit{et al.}~\cite{mohr2004}, and Kopecky \textit{et al.}~\cite{Kopecky2017}), and to models included in the \textsf{TALYS} code (see text); (b) Fit to the $\gamma$-ray strength function data of $^{186}$W and the $^{184}$W data of Kopecky \textit{et al.}~\cite{Kopecky2017})  (see text).}
\label{fig:gammastrengthfit}
\end{figure*}

We now compare our $\gamma$-strength data from the $(\gamma,n)$ measurements and the OCL experiment to external data found in the literature, as shown in Fig.~\ref{fig:gammastrengthfit}a. 
We observe a good agreement with the $E1$ strength extracted from primary $\gamma$ rays following neutron capture by Kopecky \textit{et al}~\cite{Kopecky2017}, which brings further support to the absolute normalization procedure. 
Moreover, we compare our new photoneutron data to several data sets found in the literature, where the photoneutron cross section $\sigma_{\gamma n}$ is converted into dipole strength using the relation of Axel~\cite{Axel1968}:
\begin{equation}
    f_{\gamma n}(E_\gamma) = \frac{1}{3\pi^2\hbar^2c^2} \frac{\sigma_{\gamma n}(E_\gamma)}{E_\gamma},
\end{equation}
where $\sigma_{\gamma n}$ is in units of mb, $E_\gamma$ in MeV, and the factor $1/(3\pi^2\hbar^2c^2) = 8.674 \cdot 10^{-8}$ mb$^{-1}$MeV$^{-2}$. Overall, there is good agreement between the various data sets for the W isotopes. 

In Fig.~\ref{fig:gammastrengthfit}a, we also compare the data with available models in \textsf{TALYS}: \textit{strength 1}, the Generalized Lorentzian~\cite{Kopecky1990}; \textit{strength 2}, the Standard Lorentzian (Brink-Axel model)~\cite{brink1955,axel1962}; \textit{strength 3}, the Quasi-Particle Random Phase Approximation (QRPA) on top of a Hartree-Fock-plus-BCS calculation~\cite{Goriely2002}; \textit{strength 4},  the QRPA on top of a Hartree-Fock-Bogoliubov (HFB) calculations~\cite{Goriely2004}; \textit{strength 5}, the Hybrid model~\cite{Goriely1998} with  parameters from global systematics~\cite{TALYS}; \textit{strength 6}, QRPA as in Ref.~\cite{Goriely2004} but on top of a temperature-dependent HFB calculation~\cite{Hilaire2012}; and finally \textit{strength 7}, a relativistic mean-field calculation plus a continuum QRPA calculation~\cite{Daoutidis2012}. 
Out of these models, \textit{strength 4} and \textit{strength 6} match reasonably well the present Oslo data, but not the ($\gamma,n$) data. 
In general, the models are deviating significantly from each other and from either the Oslo data or the ($\gamma,n$) data.

\begin{table*}[bt]
\begin{center}
\caption{Parameters found from the model fits of $f_{\mathrm{tot}}$ to the $\gamma$-strength data (see text). The uncertainties given are from the fit only. Note that  $E_{M1}$ and $\Gamma_{M1}$  are fixed. }
\begin{tabular}{lcclcllclllcc}
\hline
\hline
Norm. 	& $E_{r}$ & $\Gamma_{r}$& $\sigma_{r}$  & $E_{\rm Pyg}$ & $\Gamma_{\rm Pyg}$ & $\sigma_{\rm Pyg}$ & $T_f$    &$E_{M1}$ & $\Gamma_{M1}$ & $\sigma_{M1}$ 	     \\
        & (MeV)      & (MeV)          &  (mb)            & (MeV)      & (MeV)           &  (mb)           & (MeV)    & (MeV)   & (MeV)         &(mb)     \\
\hline
Rec.   & 12.9(1)   & 4.1(1)         & 382(2)           & 6.3(1)    & 2.6(2)          & 7.2(3)         & 0.43(3)  & 7.2  & 2.5        & 4.4(4)        \\
\hline
\hline
\end{tabular}
\label{tab:gsfpar}
\end{center}
\end{table*}

To obtain a model description that can reproduce our data reasonably well over the entire energy range, we take a pragmatic approach and exploit phenomenological models for the dipole strength. 
For the main part of the $E1$ strength which is dominated by the Giant Dipole Resonance (GDR), we apply the Hybrid model of Goriely~\cite{Goriely1998}:
\begin{equation}
f_{E1}^{\rm Hyb}(E_{\gamma},T_f) = \frac{1}{3\pi^2\hbar^2c^2}\frac{E_{\gamma}\sigma_{r}\Gamma_{r}\Gamma(E_{\gamma},T_f)}{(E_\gamma^2-E_{r}^2)^2 + E_{\gamma}^2 \Gamma_{r} \Gamma(E_{\gamma},T_f)},
\end{equation}
where $\sigma_r$ is the peak cross section, $E_r$ the centroid, and $\Gamma_r$ the width of the GDR. 
Further, the $\gamma$-energy and temperature dependent width $\Gamma(E_{\gamma},T_f)$ is given by
\begin{equation}
    \Gamma(E_{\gamma},T_f) = 0.7\cdot \Gamma_{r} \frac{E_{\gamma}^2 + 4\pi^2T_f^2}{E_{\gamma}E_{r}}.
\end{equation}
The temperature of the final levels, $T_f$, is here considered as a constant, in line with the Brink-Axel hypothesis. 
We also include extra $E1$ strength (labeled ``$E1$ pygmy'' in Fig.~\ref{fig:gammastrengthfit}b) to make a smooth connection between our data and the ($\gamma,n$) data. 
Finally, we also add a magnetic-dipole component (marked ``$M1$ spin-flip'' in Fig.~\ref{fig:gammastrengthfit}b).
For both the $E1$ pygmy and the $M1$ spin-flip contributions, we apply a resonance-like form using a Standard Lorentzian: 
\begin{equation}
    f_{\rm Pyg, M1}(E_{\gamma}) = \frac{1}{3\pi^2\hbar^2 c^2}\frac{\sigma_{\rm Pyg, M1}\Gamma_{\rm Pyg, M1}^2 E_{\gamma}}{(E_{\gamma}^2 - E_{\rm Pyg, M1}^2 )^2 + \Gamma_{\rm Pyg, M1}^2 E_{\gamma}^2} 
\end{equation}
where $\sigma_{\rm Pyg, M1}$, $\Gamma_{\rm Pyg, M1}$, and $E_{\rm Pyg, M1}$ are the peak cross section, width, and centroid for the pygmy (Pyg) and the spin-flip (M1) resonance, respectively. 
The total fit function is then given by 
\begin{equation}
    f_{\rm tot}(E_\gamma) = f_{E1}^{\rm Hyb}(E_\gamma, T_f={\rm const.}) + f_{\rm Pyg}(E_{\gamma}) + f_{\rm M1}(E_{\gamma}).
\end{equation}

For the fit, we first constrain the Hybrid component by fitting only the Hybrid model to the GDR data (Mohr \textit{et al.}~\cite{mohr2004} and Berman \textit{et al.}~\cite{Berman1969}) in the range $E_\gamma = 7.7-14.5$ MeV. 
We choose to fix the $T_f$ parameter to the one used for the extrapolation of the level density (see Sec.~\ref{subsec:nldnorm}) to ease the fit, as $T_f$ is largely determined from the $\gamma$-strength function below neutron threshold. 
From this fit, we determine the GDR parameters $\sigma_r$, $E_r$, and $\Gamma_r$, to be used as start values for the next fit including the data for $\gamma$ energies  below neutron threshold as well.

For the spin-flip part, we use a fixed centroid $E_{\rm M1}$ taken from systematics~\cite{Capote2009}, and a fixed width of $\Gamma_{\rm M1}$ of 2.5 MeV.
The peak cross section $\sigma_{\rm M1}$ is then found from a fit to the $M1$ data of $^{184}$W from Kopecky~\textit{et al.}~\cite{Kopecky2017}.
Then we make a fit using the full energy range $E_\gamma = 1.0-14.5$ MeV, with only the spin-flip parameters fixed, and with the first fit of the GDR data as starting values.
In the fit, we include the present OCL data of $^{186}$W, the $E1$ data from Kopecky~\textit{et al.}~\cite{Kopecky2017} on $^{184}$W, and the GDR data from Mohr~\textit{et al.}~\cite{mohr2004} and Berman~\textit{et al.}~\cite{Berman1969}. 
The resulting fit is shown in Fig.~\ref{fig:gammastrengthfit}b, and the parameters are listed in Table~\ref{tab:gsfpar}.
As this model fit will be used to calculate the ($n,\gamma$) cross section and reactivity in the following section, we repeat the fit for all the different normalizations (varying $D_0$, $\Gamma_{\gamma 0}$, $\sigma_J$ and taking into account $\Delta f$). 
All fits are performed within the ROOT software tool~\cite{ROOT} using the Minuit package. 

The resulting fit function gives a reasonable description of the strength function data, although we note a potential issue in that the region between $E_\gamma = 6-8$ MeV contains practically no data points for $^{186}$W. 
Moreover, the $^{184}$W data points from primary transitions following neutron capture typically have large fluctuations.  
Hence, it is very difficult to assess the actual parameters for the $E1$ pygmy, and the deduced parameters given in Table~\ref{tab:gsfpar} should be used with caution.

We also remark that the data points at the lowest $\gamma$ energies, $E_\gamma < 1.5$ MeV, might indicate some low-energy increase in the $\gamma$-strength function, as first observed in iron isotopes~\cite{Voinov2004}.
However, in contrast to clear cases like $^{56}$Fe~\cite{Voinov2004,Larsen2013,Jones2018}, it is hard to conclude here as there are only a few data points that might show an increasing trend.
We therefore choose not to include an extra ``upbend'' component in the fit.

\subsection{Maxwellian-averaged cross section and reaction rate}
\label{subsec:astro}
\begin{figure}[htb]
\center
\includegraphics[width=1.0\columnwidth]{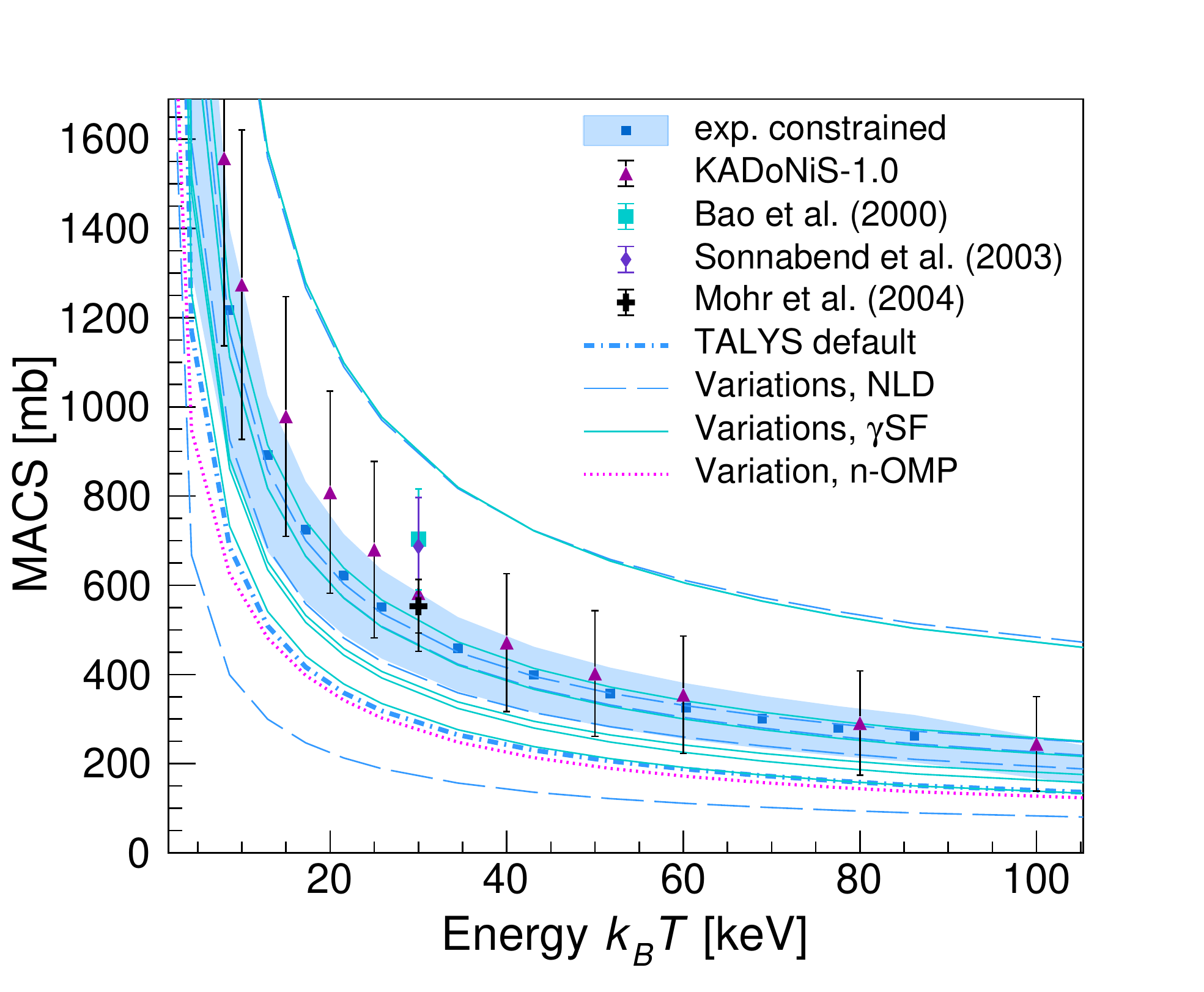}
\caption{(Color online) Maxwellian-averaged cross section for the $^{185}$W($n,\gamma$) reaction. The shaded band indicates the present data-constrained MACS. The thick, azure dashed-dotted line shows the \textsf{TALYS} result using default input, the thin, azure dashed lines show the \textsf{TALYS} MACS when varying the level-density models, and the thin, cyan lines show the variation due to different $\gamma$-strength models. The dotted line shows the deviation from the default when using the optical-model potential of Bauge \textit{et al.}~\cite{Bauge2001}.  }
\label{fig:MACS}
\end{figure}

Using our level-density data and $\gamma$-strength function data, we now calculate the Maxwellian-averaged cross section (MACS) with the \textsf{TALYS} code, which is based on the statistical model of Wolfenstein~\cite{Wolfenstein1951} and Hauser and Feshbach~\cite{HauserFeshbach1952}. 
The resulting MACS is shown in Fig.~\ref{fig:MACS}, where we also show the \textsf{TALYS} MACS with default inputs (\textit{strength 1}, \textit{ldmodel 1}, a global optical-model potential, and no upbend), and the variation of the MACS as the different level-density and $\gamma$-strength models are used. 
We have tested using the semi-microscopic optical-model potential of Bauge \textit{et al.}~\cite{Bauge2001} for comparison with the one of Koning and Delaroche~\cite{Koning2003}. 
As seen from Fig.~\ref{fig:MACS} (dashed line versus dashed-dotted line), there is only a minor difference between the two for neutron energies around $k_B T = 30$ keV, and overall the semi-microscopic potential gives a lower MACS. 
Nevertheless, the presented uncertainty band on our experimentally-constrained MACS includes the variation between the two different optical models in the lower uncertainty, in addition to uncertainties from $D_0$, $\Gamma_{\gamma 0}$, and $\sigma_J$.
\begin{figure}[tb]
\center
\includegraphics[width=1.0\columnwidth]{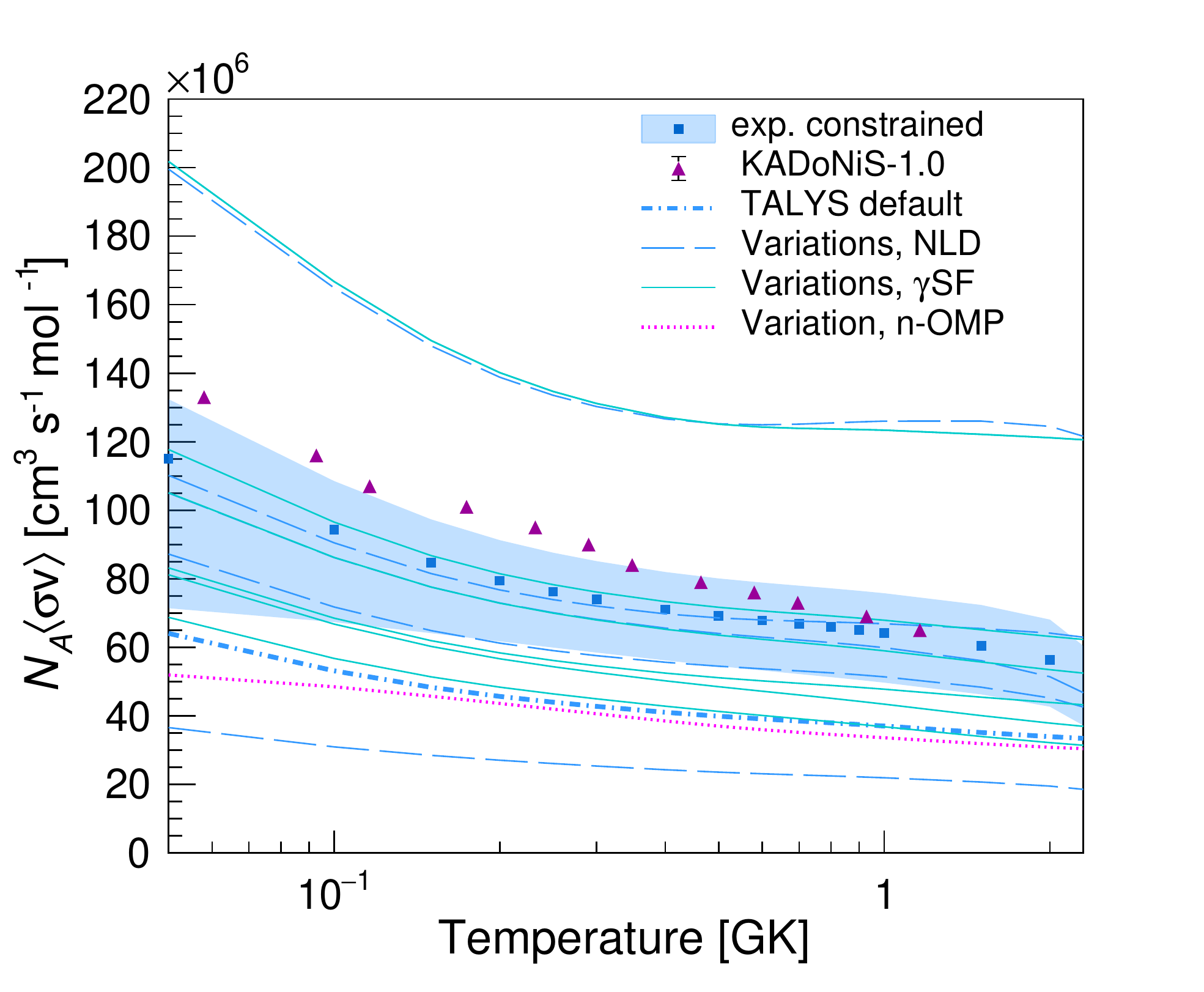}
\caption{(Color online) Reaction rate for the $^{185}$W($n,\gamma$) reaction. The shaded band indicates the present data-constrained result. See also the caption of Fig.~\ref{fig:MACS}. }
\label{fig:reactivity}
\end{figure}

In Fig.~\ref{fig:MACS}, we compare our result with the KADoNiS database~\cite{kadonis}, and find agreement within the error bars, although the KADoNiS values are overall larger than our central values. 
We remark that the KADoNiS values are from a weighted average of MACS constrained by photonuclear data above $S_n$, while our results include information on both the level density as well as the $\gamma$-strength function below $S_n$. 
We have multiplied the KADoNiS MACS values with their corresponding stellar enhancement factor (SEF) as given in Ref.~\cite{kadonis} for $^{185}$W($n,\gamma$). 
Furthermore, our estimated uncertainty band is smaller than the KADoNiS uncertainties, 
Our result at $k_BT = 30$ keV, $508^{+76}_{-106}$ mb, agrees well within error bars with the MACS from Mohr~\textit{et al.}~\cite{mohr2004}, $553(60)$ mb. 
On the other hand, the evaluation of Bao~\textit{et al.}~\cite{bao2000} of $703(113)$ mb,  and the measurement of Sonnabend~\textit{et al.}~\cite{sonnabend2003}, $687(110)$ mb, are both larger than our estimate, although still within the estimated uncertainties.
We note that none of these values are directly measured, as Bao~\textit{et al.} gives a purely theoretical prediction, while the MACS value from Sonnabend~\textit{et al.} is constrained on ($\gamma,n$) data above $S_n$. 
In comparison with the \textsf{TALYS} estimates using the default input as well as the resulting MACS when varying the level-density and $\gamma$-strength models, our deduced MACS is in between the extremes.

In Fig.~\ref{fig:reactivity}, we show the corresponding reaction rate (stellar reactivity) deduced from our data compared to the KADoNiS rate, the \textsf{TALYS} default and the variations using different model inputs.
Again we find that the KADoNiS values are overall higher than our estimated rate, in particular for temperatures below 0.3 GK. 

To address possible implications for the $s$ process and the Re/Os cosmochronometer in a reliable way, the branch points at $^{186}$Re and $^{191}$Os should also be considered in realistic stellar models for thermally-pulsing AGB stars. 
The $^{191}$Os MACS has been estimated by a similar procedure as in this work by Kullmann \textit{et al.}~\cite{Kullmann2019}. 
The $^{186}$Re MACS remains to be experimentally constrained in the same way; the $^{186}$W($\alpha,d\gamma$)$^{187}$Re data from this same experiment is currently being analyzed. 
With this experimentally-constrained MACS also at hand, we intend to perform a consistent study of the $s$ process in this mass region.

\section{Summary and outlook}
\label{sec:summary}
In this work, we have performed photoneutron cross section measurements on the $^{182,183,184}$W isotopes. 
This completes the photoneutron measurements on the stable W isotopic chain. 
Furthermore, we have presented data on the $^{186}$W($\alpha, \alpha' \gamma$) reaction, and used the extracted level density and $\gamma$-ray strength function to provide an experimentally constrained ($n,\gamma$) cross section for the branch-point nucleus $^{185}$W. 

In comparison with other data and the recommended MACS from the KADoNiS data base, we find that our estimated MACS and reaction rate are lower than most of the other available values, except for the result of Mohr \textit{et al.} 
Our reaction rate could possibly impact the $s$ process in this mass region, in particular the deduced neutron density and the calculation of the $^{186}$Os abundance. 
When the $^{186}$Re MACS also becomes available, we intend to perform a systematic study of the $s$-process conditions in the W-Re-Os region in the near future.

\acknowledgments
The authors would like to thank J.~C.~M\"{uller}, P.~A.~Sobas, and J.~C.~Wikne at the Oslo Cyclotron Laboratory for operating the cyclotron and providing excellent experimental conditions.
We sincerely thank T.~W.~Hagen, S.~J.~Rose and F. Zeiser for helping with the OCL experiment, Y.-W.~Lui for helping with the NewSUBARU experiments, and S.~N.~Liddick for inspiring discussions.
A.~C.~L. gratefully acknowledges funding of this research by the European Research Council through ERC-STG-2014 under grant agreement no. 637686, and from the Research Council of Norway, project grant no. 316116.
S.~G. acknowledges the support from the F.R.S.-FNRS.
This work was supported in part by the National Science Foundation under Grant No. OISE-1927130 (IReNA).
The photoneutron cross section measurement was performed as part of the IAEA CRP on “Updating the Photonuclear Data Library and generating a Reference Database for Photon Strength Functions” (F41032). 
A.~G., V.~W.~I., and S.~S. gratefully acknowledge financial support from the Research Council of Norway, project grant no. 325714. 
This work is in part based on the research supported partly by the National Research Foundation of South Africa (Grant Number: 118846).

\appendix
\section{Uncertainty in $\rho(S_n)$}
\label{appendix1}
To estimate the total NLD at the neutron separation energy using Eq.~(\ref{eq:d2rho}), we propagate errors from the $D_0$ value and the spin cutoff parameter $\sigma_J(S_n)$ assuming that they are independent variables, which is a justified assumption.
Thus, we get that
\begin{equation}
    \left( \frac{\delta \rho(S_n)}{\rho(S_n)} \right)^2 = \left(\frac{\delta D_0}{D_0}\right)^2 + \left(\frac{\delta \xi(\sigma_J(S_n))}{\xi(\sigma_J(S_n))}\right)^2,
\end{equation}
where $\xi$ represents the function containing the dependency on the spin cutoff parameter $\sigma_J$ at the neutron separation energy $S_n$:
\begin{equation}
    \xi(\sigma_J) = \frac{2\sigma_J^2}{I_t e^{-I_t^2/2\sigma_J^2}+ (I_t+1)e^{-(I_t+1)^2/2\sigma_J^2}}.
\end{equation}
Now we take the derivative of $\xi$ with respect to $\sigma_J$ and obtain:
\begin{widetext}
    \begin{equation}
    \label{eq:dsigma}
    \frac{\delta\xi}{\delta\sigma_J} =  
     \frac{4\sigma_J\left(I_t e^{-I_t^2/2\sigma_J^2}+ (I_t+1)e^{-(I_t+1)^2/2\sigma_J^2}\right) - \frac{2}{\sigma_J} \left(I_t^3 e^{-I_t^2/2\sigma_J^2}+ (I_t+1)^3e^{-(I_t+1)^2/2\sigma_J^2} \right) }{\left[I_t e^{-I_t^2/2\sigma_J^2}+ (I_t+1)e^{-(I_t+1)^2/2\sigma_J^2}\right]^2.}
    \end{equation}
\end{widetext}

For convenience, we now define the auxilliary functions
\[ 
    z_1 \equiv I_t^3 e^{-I_t^2/2\sigma_J^2}+ (I_t+1)^3e^{-(I_t+1)^2/2\sigma_J^2},
\]
\[ z_2 \equiv I_t e^{-I_t^2/2\sigma_J^2}+ (I_t+1)e^{-(I_t+1)^2/2\sigma_J^2}.\]
Using these and dividing Eq.~(\ref{eq:dsigma}) on the function $\xi(\sigma_J)$, we get
\begin{equation}
\frac{\delta\xi}{\xi \delta\sigma_J} = \frac{2}{\sigma_J} - \frac{z_1}{\sigma_J^3 z_2} = \frac{2}{\sigma_J}\left( 1 - \frac{1}{2\sigma_J^2} \frac{z_1}{z_2}\right). 
\end{equation}
Finally, we obtain 
\begin{equation}
    \left(\frac{\delta \xi}{\xi} \right)^2 = \left( \frac{2 \delta \sigma_J}{\sigma_J}\right)^2\left( 1 - \frac{1}{2\sigma_J^2} \frac{z_1}{z_2}\right)^2.
\end{equation}
This is what is implemented in the code \textsf{d2rho} in the Oslo software package~\cite{Oslosoftware}.

\bibliography{bibfile_186W.bib}

\providecommand{\noopsort}[1]{}\providecommand{\singleletter}[1]{#1}%
\begin{thebibliography}{92}%
\makeatletter
\providecommand \@ifxundefined [1]{%
 \@ifx{#1\undefined}
}%
\providecommand \@ifnum [1]{%
 \ifnum #1\expandafter \@firstoftwo
 \else \expandafter \@secondoftwo
 \fi
}%
\providecommand \@ifx [1]{%
 \ifx #1\expandafter \@firstoftwo
 \else \expandafter \@secondoftwo
 \fi
}%
\providecommand \natexlab [1]{#1}%
\providecommand \enquote  [1]{``#1''}%
\providecommand \bibnamefont  [1]{#1}%
\providecommand \bibfnamefont [1]{#1}%
\providecommand \citenamefont [1]{#1}%
\providecommand \href@noop [0]{\@secondoftwo}%
\providecommand \href [0]{\begingroup \@sanitize@url \@href}%
\providecommand \@href[1]{\@@startlink{#1}\@@href}%
\providecommand \@@href[1]{\endgroup#1\@@endlink}%
\providecommand \@sanitize@url [0]{\catcode `\\12\catcode `\$12\catcode
  `\&12\catcode `\#12\catcode `\^12\catcode `\_12\catcode `\%12\relax}%
\providecommand \@@startlink[1]{}%
\providecommand \@@endlink[0]{}%
\providecommand \url  [0]{\begingroup\@sanitize@url \@url }%
\providecommand \@url [1]{\endgroup\@href {#1}{\urlprefix }}%
\providecommand \urlprefix  [0]{URL }%
\providecommand \Eprint [0]{\href }%
\providecommand \doibase [0]{https://doi.org/}%
\providecommand \selectlanguage [0]{\@gobble}%
\providecommand \bibinfo  [0]{\@secondoftwo}%
\providecommand \bibfield  [0]{\@secondoftwo}%
\providecommand \translation [1]{[#1]}%
\providecommand \BibitemOpen [0]{}%
\providecommand \bibitemStop [0]{}%
\providecommand \bibitemNoStop [0]{.\EOS\space}%
\providecommand \EOS [0]{\spacefactor3000\relax}%
\providecommand \BibitemShut  [1]{\csname bibitem#1\endcsname}%
\let\auto@bib@innerbib\@empty
\bibitem [{\citenamefont {Burbidge}\ \emph {et~al.}(1957)\citenamefont
  {Burbidge}, \citenamefont {Burbidge}, \citenamefont {Fowler},\ and\
  \citenamefont {Hoyle}}]{burbidge1957}%
  \BibitemOpen
  \bibfield  {author} {\bibinfo {author} {\bibfnamefont {E.~M.}\ \bibnamefont
  {Burbidge}}, \bibinfo {author} {\bibfnamefont {G.~R.}\ \bibnamefont
  {Burbidge}}, \bibinfo {author} {\bibfnamefont {W.~A.}\ \bibnamefont
  {Fowler}},\ and\ \bibinfo {author} {\bibfnamefont {F.}~\bibnamefont
  {Hoyle}},\ }\href {https://doi.org/10.1103/RevModPhys.29.547} {\bibfield
  {journal} {\bibinfo  {journal} {Rev. Mod. Phys.}\ }\textbf {\bibinfo {volume}
  {29}},\ \bibinfo {pages} {547} (\bibinfo {year} {1957})}\BibitemShut
  {NoStop}%
\bibitem [{\citenamefont {{Cameron}}(1957)}]{cameron1957}%
  \BibitemOpen
  \bibfield  {author} {\bibinfo {author} {\bibfnamefont {A.~G.~W.}\
  \bibnamefont {{Cameron}}},\ }\href {https://doi.org/10.1086/107435}
  {\bibfield  {journal} {\bibinfo  {journal} {Astron. J.}\ }\textbf {\bibinfo
  {volume} {62}},\ \bibinfo {pages} {9} (\bibinfo {year} {1957})}\BibitemShut
  {NoStop}%
\bibitem [{\citenamefont {Arnould}\ \emph {et~al.}(2007)\citenamefont
  {Arnould}, \citenamefont {Goriely},\ and\ \citenamefont
  {Takahashi}}]{arnould2007}%
  \BibitemOpen
  \bibfield  {author} {\bibinfo {author} {\bibfnamefont {M.}~\bibnamefont
  {Arnould}}, \bibinfo {author} {\bibfnamefont {S.}~\bibnamefont {Goriely}},\
  and\ \bibinfo {author} {\bibfnamefont {K.}~\bibnamefont {Takahashi}},\ }\href
  {https://doi.org/https://doi.org/10.1016/j.physrep.2007.06.002} {\bibfield
  {journal} {\bibinfo  {journal} {Physics Reports}\ }\textbf {\bibinfo {volume}
  {450}},\ \bibinfo {pages} {97 } (\bibinfo {year} {2007})}\BibitemShut
  {NoStop}%
\bibitem [{\citenamefont {Cowan}\ \emph {et~al.}(2021)\citenamefont {Cowan},
  \citenamefont {Sneden}, \citenamefont {Lawler}, \citenamefont {Aprahamian},
  \citenamefont {Wiescher}, \citenamefont {Langanke}, \citenamefont
  {Mart\'{\i}nez-Pinedo},\ and\ \citenamefont {Thielemann}}]{cowan2021}%
  \BibitemOpen
  \bibfield  {author} {\bibinfo {author} {\bibfnamefont {J.~J.}\ \bibnamefont
  {Cowan}}, \bibinfo {author} {\bibfnamefont {C.}~\bibnamefont {Sneden}},
  \bibinfo {author} {\bibfnamefont {J.~E.}\ \bibnamefont {Lawler}}, \bibinfo
  {author} {\bibfnamefont {A.}~\bibnamefont {Aprahamian}}, \bibinfo {author}
  {\bibfnamefont {M.}~\bibnamefont {Wiescher}}, \bibinfo {author}
  {\bibfnamefont {K.}~\bibnamefont {Langanke}}, \bibinfo {author}
  {\bibfnamefont {G.}~\bibnamefont {Mart\'{\i}nez-Pinedo}},\ and\ \bibinfo
  {author} {\bibfnamefont {F.-K.}\ \bibnamefont {Thielemann}},\ }\href
  {https://doi.org/10.1103/RevModPhys.93.015002} {\bibfield  {journal}
  {\bibinfo  {journal} {Rev. Mod. Phys.}\ }\textbf {\bibinfo {volume} {93}},\
  \bibinfo {pages} {015002} (\bibinfo {year} {2021})}\BibitemShut {NoStop}%
\bibitem [{\citenamefont {K\"appeler}\ \emph {et~al.}(2011)\citenamefont
  {K\"appeler}, \citenamefont {Gallino}, \citenamefont {Bisterzo},\ and\
  \citenamefont {Aoki}}]{kappeler2011}%
  \BibitemOpen
  \bibfield  {author} {\bibinfo {author} {\bibfnamefont {F.}~\bibnamefont
  {K\"appeler}}, \bibinfo {author} {\bibfnamefont {R.}~\bibnamefont {Gallino}},
  \bibinfo {author} {\bibfnamefont {S.}~\bibnamefont {Bisterzo}},\ and\
  \bibinfo {author} {\bibfnamefont {W.}~\bibnamefont {Aoki}},\ }\bibfield
  {title} {\bibinfo {title} {The $s$ process: Nuclear physics, stellar models,
  and observations},\ }\href {https://doi.org/10.1103/RevModPhys.83.157}
  {\bibfield  {journal} {\bibinfo  {journal} {Rev. Mod. Phys.}\ }\textbf
  {\bibinfo {volume} {83}},\ \bibinfo {pages} {157} (\bibinfo {year}
  {2011})}\BibitemShut {NoStop}%
\bibitem [{\citenamefont {{Ward}}\ \emph {et~al.}(1976)\citenamefont {{Ward}},
  \citenamefont {{Newman}},\ and\ \citenamefont {{Clayton}}}]{ward1976}%
  \BibitemOpen
  \bibfield  {author} {\bibinfo {author} {\bibfnamefont {R.~A.}\ \bibnamefont
  {{Ward}}}, \bibinfo {author} {\bibfnamefont {M.~J.}\ \bibnamefont
  {{Newman}}},\ and\ \bibinfo {author} {\bibfnamefont {D.~D.}\ \bibnamefont
  {{Clayton}}},\ }\href {https://doi.org/10.1086/190373} {\bibfield  {journal}
  {\bibinfo  {journal} {Astrophys. J. Suppl.}\ }\textbf {\bibinfo {volume}
  {31}},\ \bibinfo {pages} {33} (\bibinfo {year} {1976})}\BibitemShut {NoStop}%
\bibitem [{\citenamefont {K\"{a}ppeler}\ \emph {et~al.}(1991)\citenamefont
  {K\"{a}ppeler}, \citenamefont {Jaag}, \citenamefont {Bao},\ and\
  \citenamefont {Reffo}}]{kappeler1991}%
  \BibitemOpen
  \bibfield  {author} {\bibinfo {author} {\bibfnamefont {F.}~\bibnamefont
  {K\"{a}ppeler}}, \bibinfo {author} {\bibfnamefont {S.}~\bibnamefont {Jaag}},
  \bibinfo {author} {\bibfnamefont {Z.~Y.}\ \bibnamefont {Bao}},\ and\ \bibinfo
  {author} {\bibfnamefont {G.}~\bibnamefont {Reffo}},\ }\href
  {https://ui.adsabs.harvard.edu/abs/1991ApJ...366..605K/abstract} {\bibfield
  {journal} {\bibinfo  {journal} {Astrophys. J.}\ }\textbf {\bibinfo {volume}
  {366}},\ \bibinfo {pages} {605} (\bibinfo {year} {1991})}\BibitemShut
  {NoStop}%
\bibitem [{\citenamefont {Sonnabend}\ \emph {et~al.}(2003)\citenamefont
  {Sonnabend}, \citenamefont {Mohr}, \citenamefont {Vogt}, \citenamefont
  {Zilges}, \citenamefont {Mengoni}, \citenamefont {Rauscher}, \citenamefont
  {Beer}, \citenamefont {Kappeler},\ and\ \citenamefont
  {Gallino}}]{sonnabend2003}%
  \BibitemOpen
  \bibfield  {author} {\bibinfo {author} {\bibfnamefont {K.}~\bibnamefont
  {Sonnabend}}, \bibinfo {author} {\bibfnamefont {P.}~\bibnamefont {Mohr}},
  \bibinfo {author} {\bibfnamefont {K.}~\bibnamefont {Vogt}}, \bibinfo {author}
  {\bibfnamefont {A.}~\bibnamefont {Zilges}}, \bibinfo {author} {\bibfnamefont
  {A.}~\bibnamefont {Mengoni}}, \bibinfo {author} {\bibfnamefont
  {T.}~\bibnamefont {Rauscher}}, \bibinfo {author} {\bibfnamefont
  {H.}~\bibnamefont {Beer}}, \bibinfo {author} {\bibfnamefont {F.}~\bibnamefont
  {Kappeler}},\ and\ \bibinfo {author} {\bibfnamefont {R.}~\bibnamefont
  {Gallino}},\ }\href {https://doi.org/10.1086/345086} {\bibfield  {journal}
  {\bibinfo  {journal} {The Astrophysical Journal}\ }\textbf {\bibinfo {volume}
  {583}},\ \bibinfo {pages} {506} (\bibinfo {year} {2003})}\BibitemShut
  {NoStop}%
\bibitem [{\citenamefont {Mohr}\ \emph {et~al.}(2004)\citenamefont {Mohr},
  \citenamefont {Shizuma}, \citenamefont {Ueda}, \citenamefont {Goko},
  \citenamefont {Makinaga}, \citenamefont {Hara}, \citenamefont {Hayakawa},
  \citenamefont {Lui}, \citenamefont {Ohgaki},\ and\ \citenamefont
  {Utsunomiya}}]{mohr2004}%
  \BibitemOpen
  \bibfield  {author} {\bibinfo {author} {\bibfnamefont {P.}~\bibnamefont
  {Mohr}}, \bibinfo {author} {\bibfnamefont {T.}~\bibnamefont {Shizuma}},
  \bibinfo {author} {\bibfnamefont {H.}~\bibnamefont {Ueda}}, \bibinfo {author}
  {\bibfnamefont {S.}~\bibnamefont {Goko}}, \bibinfo {author} {\bibfnamefont
  {A.}~\bibnamefont {Makinaga}}, \bibinfo {author} {\bibfnamefont {K.~Y.}\
  \bibnamefont {Hara}}, \bibinfo {author} {\bibfnamefont {T.}~\bibnamefont
  {Hayakawa}}, \bibinfo {author} {\bibfnamefont {Y.-W.}\ \bibnamefont {Lui}},
  \bibinfo {author} {\bibfnamefont {H.}~\bibnamefont {Ohgaki}},\ and\ \bibinfo
  {author} {\bibfnamefont {H.}~\bibnamefont {Utsunomiya}},\ }\href
  {https://doi.org/10.1103/PhysRevC.69.032801} {\bibfield  {journal} {\bibinfo
  {journal} {Phys. Rev. C}\ }\textbf {\bibinfo {volume} {69}},\ \bibinfo
  {pages} {032801} (\bibinfo {year} {2004})}\BibitemShut {NoStop}%
\bibitem [{\citenamefont {Emery}\ \emph {et~al.}(1972)\citenamefont {Emery},
  \citenamefont {Reynolds}, \citenamefont {Wyatt},\ and\ \citenamefont
  {Gleason}}]{Emery1972}%
  \BibitemOpen
  \bibfield  {author} {\bibinfo {author} {\bibfnamefont {J.~F.}\ \bibnamefont
  {Emery}}, \bibinfo {author} {\bibfnamefont {S.~A.}\ \bibnamefont {Reynolds}},
  \bibinfo {author} {\bibfnamefont {E.~I.}\ \bibnamefont {Wyatt}},\ and\
  \bibinfo {author} {\bibfnamefont {G.~I.}\ \bibnamefont {Gleason}},\
  }\href@noop {} {\bibfield  {journal} {\bibinfo  {journal} {Nucl. Sci. Eng.}\
  }\textbf {\bibinfo {volume} {48}},\ \bibinfo {pages} {319} (\bibinfo {year}
  {1972})}\BibitemShut {NoStop}%
\bibitem [{\citenamefont {{Clayton}}(1964)}]{Clayton1964}%
  \BibitemOpen
  \bibfield  {author} {\bibinfo {author} {\bibfnamefont {D.~D.}\ \bibnamefont
  {{Clayton}}},\ }\href {https://doi.org/10.1086/147791} {\bibfield  {journal}
  {\bibinfo  {journal} {\apj}\ }\textbf {\bibinfo {volume} {139}},\ \bibinfo
  {pages} {637} (\bibinfo {year} {1964})}\BibitemShut {NoStop}%
\bibitem [{\citenamefont {Galeazzi}\ \emph {et~al.}(2000)\citenamefont
  {Galeazzi}, \citenamefont {Fontanelli}, \citenamefont {Gatti},\ and\
  \citenamefont {Vitale}}]{galeazzi2000}%
  \BibitemOpen
  \bibfield  {author} {\bibinfo {author} {\bibfnamefont {M.}~\bibnamefont
  {Galeazzi}}, \bibinfo {author} {\bibfnamefont {F.}~\bibnamefont
  {Fontanelli}}, \bibinfo {author} {\bibfnamefont {F.}~\bibnamefont {Gatti}},\
  and\ \bibinfo {author} {\bibfnamefont {S.}~\bibnamefont {Vitale}},\ }\href
  {https://doi.org/10.1103/PhysRevC.63.014302} {\bibfield  {journal} {\bibinfo
  {journal} {Phys. Rev. C}\ }\textbf {\bibinfo {volume} {63}},\ \bibinfo
  {pages} {014302} (\bibinfo {year} {2000})}\BibitemShut {NoStop}%
\bibitem [{\citenamefont {Yokoi}\ \emph {et~al.}(1983)\citenamefont {Yokoi},
  \citenamefont {Takahashi},\ and\ \citenamefont {Arnould}}]{yokoi1983re}%
  \BibitemOpen
  \bibfield  {author} {\bibinfo {author} {\bibfnamefont {K.}~\bibnamefont
  {Yokoi}}, \bibinfo {author} {\bibfnamefont {K.}~\bibnamefont {Takahashi}},\
  and\ \bibinfo {author} {\bibfnamefont {M.}~\bibnamefont {Arnould}},\
  }\href@noop {} {\bibfield  {journal} {\bibinfo  {journal} {Astronomy and
  Astrophysics}\ }\textbf {\bibinfo {volume} {117}},\ \bibinfo {pages} {65}
  (\bibinfo {year} {1983})}\BibitemShut {NoStop}%
\bibitem [{\citenamefont {Arnould}\ \emph {et~al.}(1984)\citenamefont
  {Arnould}, \citenamefont {Takahashi},\ and\ \citenamefont
  {Yokoi}}]{arnould1984validity}%
  \BibitemOpen
  \bibfield  {author} {\bibinfo {author} {\bibfnamefont {M.}~\bibnamefont
  {Arnould}}, \bibinfo {author} {\bibfnamefont {K.}~\bibnamefont {Takahashi}},\
  and\ \bibinfo {author} {\bibfnamefont {K.}~\bibnamefont {Yokoi}},\
  }\href@noop {} {\bibfield  {journal} {\bibinfo  {journal} {Astronomy and
  Astrophysics}\ }\textbf {\bibinfo {volume} {137}},\ \bibinfo {pages} {51}
  (\bibinfo {year} {1984})}\BibitemShut {NoStop}%
\bibitem [{\citenamefont {Mosconi}\ \emph {et~al.}(2010)\citenamefont {Mosconi}
  \emph {et~al.}}]{Mosconi2010}%
  \BibitemOpen
  \bibfield  {author} {\bibinfo {author} {\bibfnamefont {M.}~\bibnamefont
  {Mosconi}} \emph {et~al.},\ }\href
  {https://doi.org/10.1103/PhysRevC.82.015802} {\bibfield  {journal} {\bibinfo
  {journal} {Phys. Rev. C}\ }\textbf {\bibinfo {volume} {82}},\ \bibinfo
  {pages} {015802} (\bibinfo {year} {2010})}\BibitemShut {NoStop}%
\bibitem [{\citenamefont {Shizuma}\ \emph {et~al.}(2005)\citenamefont
  {Shizuma}, \citenamefont {Utsunomiya}, \citenamefont {Mohr}, \citenamefont
  {Hayakawa}, \citenamefont {Goko}, \citenamefont {Makinaga}, \citenamefont
  {Akimune}, \citenamefont {Yamagata}, \citenamefont {Ohta}, \citenamefont
  {Ohgaki}, \citenamefont {Lui}, \citenamefont {Toyokawa}, \citenamefont
  {Uritani},\ and\ \citenamefont {Goriely}}]{shizuma2005}%
  \BibitemOpen
  \bibfield  {author} {\bibinfo {author} {\bibfnamefont {T.}~\bibnamefont
  {Shizuma}}, \bibinfo {author} {\bibfnamefont {H.}~\bibnamefont {Utsunomiya}},
  \bibinfo {author} {\bibfnamefont {P.}~\bibnamefont {Mohr}}, \bibinfo {author}
  {\bibfnamefont {T.}~\bibnamefont {Hayakawa}}, \bibinfo {author}
  {\bibfnamefont {S.}~\bibnamefont {Goko}}, \bibinfo {author} {\bibfnamefont
  {A.}~\bibnamefont {Makinaga}}, \bibinfo {author} {\bibfnamefont
  {H.}~\bibnamefont {Akimune}}, \bibinfo {author} {\bibfnamefont
  {T.}~\bibnamefont {Yamagata}}, \bibinfo {author} {\bibfnamefont
  {M.}~\bibnamefont {Ohta}}, \bibinfo {author} {\bibfnamefont {H.}~\bibnamefont
  {Ohgaki}}, \bibinfo {author} {\bibfnamefont {Y.-W.}\ \bibnamefont {Lui}},
  \bibinfo {author} {\bibfnamefont {H.}~\bibnamefont {Toyokawa}}, \bibinfo
  {author} {\bibfnamefont {A.}~\bibnamefont {Uritani}},\ and\ \bibinfo {author}
  {\bibfnamefont {S.}~\bibnamefont {Goriely}},\ }\href
  {https://doi.org/10.1103/PhysRevC.72.025808} {\bibfield  {journal} {\bibinfo
  {journal} {Phys. Rev. C}\ }\textbf {\bibinfo {volume} {72}},\ \bibinfo
  {pages} {025808} (\bibinfo {year} {2005})}\BibitemShut {NoStop}%
\bibitem [{\citenamefont {Humayun}\ and\ \citenamefont
  {Brandon}(2007)}]{Humayun_2007}%
  \BibitemOpen
  \bibfield  {author} {\bibinfo {author} {\bibfnamefont {M.}~\bibnamefont
  {Humayun}}\ and\ \bibinfo {author} {\bibfnamefont {A.~D.}\ \bibnamefont
  {Brandon}},\ }\href {https://doi.org/10.1086/520636} {\bibfield  {journal}
  {\bibinfo  {journal} {The Astrophysical Journal}\ }\textbf {\bibinfo {volume}
  {664}},\ \bibinfo {pages} {L59} (\bibinfo {year} {2007})}\BibitemShut
  {NoStop}%
\bibitem [{\citenamefont {Koning}\ and\ \citenamefont {Rochman}(2012)}]{TALYS}%
  \BibitemOpen
  \bibfield  {author} {\bibinfo {author} {\bibfnamefont {A.}~\bibnamefont
  {Koning}}\ and\ \bibinfo {author} {\bibfnamefont {D.}~\bibnamefont
  {Rochman}},\ }\href
  {https://doi.org/https://doi.org/10.1016/j.nds.2012.11.002} {\bibfield
  {journal} {\bibinfo  {journal} {Nuclear Data Sheets}\ }\textbf {\bibinfo
  {volume} {113}},\ \bibinfo {pages} {2841} (\bibinfo {year} {2012})},\
  \bibinfo {note} {special Issue on Nuclear Reaction Data}\BibitemShut
  {NoStop}%
\bibitem [{\citenamefont {Utsunomiya}\ \emph {et~al.}(2014)\citenamefont
  {Utsunomiya}, \citenamefont {Shima}, \citenamefont {Takahisa}, \citenamefont
  {Filipescu}, \citenamefont {Tesileanu}, \citenamefont {Gheorghe},
  \citenamefont {Nyhus}, \citenamefont {Renstr{\o}m}, \citenamefont {Lui},
  \citenamefont {Kitagawa}, \citenamefont {Amano},\ and\ \citenamefont
  {Miyamoto}}]{calofns}%
  \BibitemOpen
  \bibfield  {author} {\bibinfo {author} {\bibfnamefont {H.}~\bibnamefont
  {Utsunomiya}}, \bibinfo {author} {\bibfnamefont {T.}~\bibnamefont {Shima}},
  \bibinfo {author} {\bibfnamefont {K.}~\bibnamefont {Takahisa}}, \bibinfo
  {author} {\bibfnamefont {D.~M.}\ \bibnamefont {Filipescu}}, \bibinfo {author}
  {\bibfnamefont {O.}~\bibnamefont {Tesileanu}}, \bibinfo {author}
  {\bibfnamefont {I.}~\bibnamefont {Gheorghe}}, \bibinfo {author}
  {\bibfnamefont {H.~T.}\ \bibnamefont {Nyhus}}, \bibinfo {author}
  {\bibfnamefont {T.}~\bibnamefont {Renstr{\o}m}}, \bibinfo {author}
  {\bibfnamefont {Y.~W.}\ \bibnamefont {Lui}}, \bibinfo {author} {\bibfnamefont
  {Y.}~\bibnamefont {Kitagawa}}, \bibinfo {author} {\bibfnamefont
  {S.}~\bibnamefont {Amano}},\ and\ \bibinfo {author} {\bibfnamefont
  {S.}~\bibnamefont {Miyamoto}},\ }\href
  {https://doi.org/10.1109/TNS.2014.2312323} {\bibfield  {journal} {\bibinfo
  {journal} {IEEE Transactions on Nuclear Science}\ }\textbf {\bibinfo {volume}
  {61}},\ \bibinfo {pages} {1252} (\bibinfo {year} {2014})}\BibitemShut
  {NoStop}%
\bibitem [{\citenamefont {Gheorghe}(2017)}]{thesisioana}%
  \BibitemOpen
  \bibfield  {author} {\bibinfo {author} {\bibfnamefont {A.~I.}\ \bibnamefont
  {Gheorghe}},\ }\emph {\bibinfo {title} {Nuclear data obtained with Laser
  Compton Scattered gamma-ray beams}},\ \href@noop {} {Ph.D. thesis},\ \bibinfo
   {school} {University of Bucharest} (\bibinfo {year} {2017})\BibitemShut
  {NoStop}%
\bibitem [{\citenamefont {Agostinelli}\ \emph {et~al.}(2003)\citenamefont
  {Agostinelli} \emph {et~al.}}]{geant1}%
  \BibitemOpen
  \bibfield  {author} {\bibinfo {author} {\bibfnamefont {S.}~\bibnamefont
  {Agostinelli}} \emph {et~al.},\ }\bibfield  {title} {\bibinfo {title} {Geant4
  a simulation toolkit},\ }\href
  {https://doi.org/https://doi.org/10.1016/S0168-9002(03)01368-8} {\bibfield
  {journal} {\bibinfo  {journal} {NIM A}\ }\textbf {\bibinfo {volume} {506}},\
  \bibinfo {pages} {250 } (\bibinfo {year} {2003})}\BibitemShut {NoStop}%
\bibitem [{\citenamefont {Allison}\ \emph {et~al.}(2006)\citenamefont {Allison}
  \emph {et~al.}}]{geant2}%
  \BibitemOpen
  \bibfield  {author} {\bibinfo {author} {\bibfnamefont {J.}~\bibnamefont
  {Allison}} \emph {et~al.},\ }\href
  {https://doi.org/http://dx.doi.org/10.1109/TNS.2006.869826} {\bibfield
  {journal} {\bibinfo  {journal} {IEEE Trans. Nucl. Sci.}\ }\textbf {\bibinfo
  {volume} {53}},\ \bibinfo {pages} {270} (\bibinfo {year} {2006})}\BibitemShut
  {NoStop}%
\bibitem [{\citenamefont {Allison}\ \emph {et~al.}(2016)\citenamefont {Allison}
  \emph {et~al.}}]{geant3}%
  \BibitemOpen
  \bibfield  {author} {\bibinfo {author} {\bibfnamefont {J.}~\bibnamefont
  {Allison}} \emph {et~al.},\ }\href
  {https://doi.org/https://doi.org/10.1016/j.nima.2016.06.125} {\bibfield
  {journal} {\bibinfo  {journal} {Nuclear Instruments and Methods in Physics
  Research Section A: Accelerators, Spectrometers, Detectors and Associated
  Equipment}\ }\textbf {\bibinfo {volume} {835}},\ \bibinfo {pages} {186 }
  (\bibinfo {year} {2016})}\BibitemShut {NoStop}%
\bibitem [{\citenamefont {Shima}\ and\ \citenamefont
  {Utsunomiya}(2015)}]{Utsu15}%
  \BibitemOpen
  \bibfield  {author} {\bibinfo {author} {\bibfnamefont {T.}~\bibnamefont
  {Shima}}\ and\ \bibinfo {author} {\bibfnamefont {H.}~\bibnamefont
  {Utsunomiya}},\ }\href
  {https://doi.org/https://doi.org/10.1142/9789814635455_0018} {\bibfield
  {journal} {\bibinfo  {journal} {Nuclear Physics and Gamma-Ray Sources for
  Nuclear Security and Nonproliferation}\ ,\ \bibinfo {pages} {151}} (\bibinfo
  {year} {2015})}\BibitemShut {NoStop}%
\bibitem [{\citenamefont {Filipescu}\ \emph {et~al.}(2014)\citenamefont
  {Filipescu}, \citenamefont {Gheorghe}, \citenamefont {Utsunomiya},
  \citenamefont {Goriely}, \citenamefont {Renstr{\o}m}, \citenamefont {Nyhus},
  \citenamefont {Tesileanu}, \citenamefont {Glodariu}, \citenamefont {Shima},
  \citenamefont {Takahisa}, \citenamefont {Miyamoto}, \citenamefont {Lui},
  \citenamefont {Hilaire}, \citenamefont {Péru}, \citenamefont {Martini},\
  and\ \citenamefont {Koning}}]{Fili14}%
  \BibitemOpen
  \bibfield  {author} {\bibinfo {author} {\bibfnamefont {D.}~\bibnamefont
  {Filipescu}}, \bibinfo {author} {\bibfnamefont {I.}~\bibnamefont {Gheorghe}},
  \bibinfo {author} {\bibfnamefont {H.}~\bibnamefont {Utsunomiya}}, \bibinfo
  {author} {\bibfnamefont {S.}~\bibnamefont {Goriely}}, \bibinfo {author}
  {\bibfnamefont {T.}~\bibnamefont {Renstr{\o}m}}, \bibinfo {author}
  {\bibfnamefont {H.}~\bibnamefont {Nyhus}}, \bibinfo {author} {\bibfnamefont
  {O.}~\bibnamefont {Tesileanu}}, \bibinfo {author} {\bibfnamefont
  {T.}~\bibnamefont {Glodariu}}, \bibinfo {author} {\bibfnamefont
  {T.}~\bibnamefont {Shima}}, \bibinfo {author} {\bibfnamefont
  {K.}~\bibnamefont {Takahisa}}, \bibinfo {author} {\bibfnamefont
  {S.}~\bibnamefont {Miyamoto}}, \bibinfo {author} {\bibfnamefont {Y.-W.}\
  \bibnamefont {Lui}}, \bibinfo {author} {\bibfnamefont {S.}~\bibnamefont
  {Hilaire}}, \bibinfo {author} {\bibfnamefont {S.}~\bibnamefont {Péru}},
  \bibinfo {author} {\bibfnamefont {M.}~\bibnamefont {Martini}},\ and\ \bibinfo
  {author} {\bibfnamefont {A.}~\bibnamefont {Koning}},\ }\href
  {https://doi.org/10.1103/PhysRevC.90.064616} {\bibfield  {journal} {\bibinfo
  {journal} {Physical Review C}\ }\textbf {\bibinfo {volume} {90}} (\bibinfo
  {year} {2014})}\BibitemShut {NoStop}%
\bibitem [{\citenamefont {Itoh}\ \emph {et~al.}(2011)\citenamefont {Itoh},
  \citenamefont {Utsunomiya}, \citenamefont {Akimune}, \citenamefont {Kondo},
  \citenamefont {Kamata}, \citenamefont {Yamagata}, \citenamefont {Toyokawa},
  \citenamefont {Harada}, \citenamefont {Kitatani}, \citenamefont {Goko},
  \citenamefont {Nair},\ and\ \citenamefont {Lui}}]{itoh_2011}%
  \BibitemOpen
  \bibfield  {author} {\bibinfo {author} {\bibfnamefont {O.}~\bibnamefont
  {Itoh}}, \bibinfo {author} {\bibfnamefont {H.}~\bibnamefont {Utsunomiya}},
  \bibinfo {author} {\bibfnamefont {H.}~\bibnamefont {Akimune}}, \bibinfo
  {author} {\bibfnamefont {T.}~\bibnamefont {Kondo}}, \bibinfo {author}
  {\bibfnamefont {M.}~\bibnamefont {Kamata}}, \bibinfo {author} {\bibfnamefont
  {T.}~\bibnamefont {Yamagata}}, \bibinfo {author} {\bibfnamefont
  {H.}~\bibnamefont {Toyokawa}}, \bibinfo {author} {\bibfnamefont
  {H.}~\bibnamefont {Harada}}, \bibinfo {author} {\bibfnamefont
  {F.}~\bibnamefont {Kitatani}}, \bibinfo {author} {\bibfnamefont
  {S.}~\bibnamefont {Goko}}, \bibinfo {author} {\bibfnamefont {C.}~\bibnamefont
  {Nair}},\ and\ \bibinfo {author} {\bibfnamefont {Y.-W.}\ \bibnamefont
  {Lui}},\ }\href {https://doi.org/10.1080/18811248.2011.9711766} {\bibfield
  {journal} {\bibinfo  {journal} {Journal of Nuclear Science and Technology}\
  }\textbf {\bibinfo {volume} {48}},\ \bibinfo {pages} {834} (\bibinfo {year}
  {2011})}\BibitemShut {NoStop}%
\bibitem [{\citenamefont {Berman}\ and\ \citenamefont
  {Fultz}(1975)}]{Berman_ring_ratio}%
  \BibitemOpen
  \bibfield  {author} {\bibinfo {author} {\bibfnamefont {B.~L.}\ \bibnamefont
  {Berman}}\ and\ \bibinfo {author} {\bibfnamefont {S.~C.}\ \bibnamefont
  {Fultz}},\ }\href {https://doi.org/10.1103/RevModPhys.47.713} {\bibfield
  {journal} {\bibinfo  {journal} {Rev. Mod. Phys.}\ }\textbf {\bibinfo {volume}
  {47}},\ \bibinfo {pages} {713} (\bibinfo {year} {1975})}\BibitemShut
  {NoStop}%
\bibitem [{\citenamefont {Nyhus}\ \emph {et~al.}(2015)\citenamefont {Nyhus},
  \citenamefont {Renstr{\o}m}, \citenamefont {Utsunomiya}, \citenamefont
  {Goriely}, \citenamefont {Filipescu}, \citenamefont {Gheorghe}, \citenamefont
  {Tesileanu}, \citenamefont {Glodariu}, \citenamefont {Shima}, \citenamefont
  {Takahisa}, \citenamefont {Miyamoto}, \citenamefont {Lui}, \citenamefont
  {Hilaire}, \citenamefont {Péru}, \citenamefont {Martini}, \citenamefont
  {Siess},\ and\ \citenamefont {Koning}}]{Nyhu15}%
  \BibitemOpen
  \bibfield  {author} {\bibinfo {author} {\bibfnamefont {H.}~\bibnamefont
  {Nyhus}}, \bibinfo {author} {\bibfnamefont {T.}~\bibnamefont {Renstr{\o}m}},
  \bibinfo {author} {\bibfnamefont {H.}~\bibnamefont {Utsunomiya}}, \bibinfo
  {author} {\bibfnamefont {S.}~\bibnamefont {Goriely}}, \bibinfo {author}
  {\bibfnamefont {D.}~\bibnamefont {Filipescu}}, \bibinfo {author}
  {\bibfnamefont {I.}~\bibnamefont {Gheorghe}}, \bibinfo {author}
  {\bibfnamefont {O.}~\bibnamefont {Tesileanu}}, \bibinfo {author}
  {\bibfnamefont {T.}~\bibnamefont {Glodariu}}, \bibinfo {author}
  {\bibfnamefont {T.}~\bibnamefont {Shima}}, \bibinfo {author} {\bibfnamefont
  {K.}~\bibnamefont {Takahisa}}, \bibinfo {author} {\bibfnamefont
  {S.}~\bibnamefont {Miyamoto}}, \bibinfo {author} {\bibfnamefont {Y.-W.}\
  \bibnamefont {Lui}}, \bibinfo {author} {\bibfnamefont {S.}~\bibnamefont
  {Hilaire}}, \bibinfo {author} {\bibfnamefont {S.}~\bibnamefont {Péru}},
  \bibinfo {author} {\bibfnamefont {M.}~\bibnamefont {Martini}}, \bibinfo
  {author} {\bibfnamefont {L.}~\bibnamefont {Siess}},\ and\ \bibinfo {author}
  {\bibfnamefont {A.}~\bibnamefont {Koning}},\ }\href
  {https://doi.org/10.1103/PhysRevC.91.015808} {\bibfield  {journal} {\bibinfo
  {journal} {Physical Review C}\ }\textbf {\bibinfo {volume} {91}} (\bibinfo
  {year} {2015})}\BibitemShut {NoStop}%
\bibitem [{\citenamefont {Kondo}\ \emph {et~al.}(2011)\citenamefont {Kondo},
  \citenamefont {Utsunomiya}, \citenamefont {Akimune}, \citenamefont
  {Yamagata}, \citenamefont {Okamoto}, \citenamefont {Harada}, \citenamefont
  {Kitatani}, \citenamefont {Shima}, \citenamefont {Horikawa},\ and\
  \citenamefont {Miyamoto}}]{KONDO2011462}%
  \BibitemOpen
  \bibfield  {author} {\bibinfo {author} {\bibfnamefont {T.}~\bibnamefont
  {Kondo}}, \bibinfo {author} {\bibfnamefont {H.}~\bibnamefont {Utsunomiya}},
  \bibinfo {author} {\bibfnamefont {H.}~\bibnamefont {Akimune}}, \bibinfo
  {author} {\bibfnamefont {T.}~\bibnamefont {Yamagata}}, \bibinfo {author}
  {\bibfnamefont {A.}~\bibnamefont {Okamoto}}, \bibinfo {author} {\bibfnamefont
  {H.}~\bibnamefont {Harada}}, \bibinfo {author} {\bibfnamefont
  {F.}~\bibnamefont {Kitatani}}, \bibinfo {author} {\bibfnamefont
  {T.}~\bibnamefont {Shima}}, \bibinfo {author} {\bibfnamefont
  {K.}~\bibnamefont {Horikawa}},\ and\ \bibinfo {author} {\bibfnamefont
  {S.}~\bibnamefont {Miyamoto}},\ }\href
  {https://doi.org/https://doi.org/10.1016/j.nima.2011.08.035} {\bibfield
  {journal} {\bibinfo  {journal} {Nuclear Instruments and Methods in Physics
  Research Section A: Accelerators, Spectrometers, Detectors and Associated
  Equipment}\ }\textbf {\bibinfo {volume} {659}},\ \bibinfo {pages} {462}
  (\bibinfo {year} {2011})}\BibitemShut {NoStop}%
\bibitem [{\citenamefont {Utsunomiya}\ \emph {et~al.}(2018)\citenamefont
  {Utsunomiya}, \citenamefont {Watanabe}, \citenamefont {Ari-izumi},
  \citenamefont {Takenaka}, \citenamefont {Araki}, \citenamefont {Tsuji},
  \citenamefont {Gheorghe}, \citenamefont {Filipescu}, \citenamefont
  {Belyshev}, \citenamefont {Stopani}, \citenamefont {Symochko}, \citenamefont
  {Wang}, \citenamefont {Fan}, \citenamefont {Renstr{\o}m}, \citenamefont
  {Tveten}, \citenamefont {Lui}, \citenamefont {Sugita},\ and\ \citenamefont
  {Miyamoto}}]{UTSUNOMIYA2018103}%
  \BibitemOpen
  \bibfield  {author} {\bibinfo {author} {\bibfnamefont {H.}~\bibnamefont
  {Utsunomiya}}, \bibinfo {author} {\bibfnamefont {T.}~\bibnamefont
  {Watanabe}}, \bibinfo {author} {\bibfnamefont {T.}~\bibnamefont {Ari-izumi}},
  \bibinfo {author} {\bibfnamefont {D.}~\bibnamefont {Takenaka}}, \bibinfo
  {author} {\bibfnamefont {T.}~\bibnamefont {Araki}}, \bibinfo {author}
  {\bibfnamefont {K.}~\bibnamefont {Tsuji}}, \bibinfo {author} {\bibfnamefont
  {I.}~\bibnamefont {Gheorghe}}, \bibinfo {author} {\bibfnamefont {D.~M.}\
  \bibnamefont {Filipescu}}, \bibinfo {author} {\bibfnamefont {S.}~\bibnamefont
  {Belyshev}}, \bibinfo {author} {\bibfnamefont {K.}~\bibnamefont {Stopani}},
  \bibinfo {author} {\bibfnamefont {D.}~\bibnamefont {Symochko}}, \bibinfo
  {author} {\bibfnamefont {H.}~\bibnamefont {Wang}}, \bibinfo {author}
  {\bibfnamefont {G.}~\bibnamefont {Fan}}, \bibinfo {author} {\bibfnamefont
  {T.}~\bibnamefont {Renstr{\o}m}}, \bibinfo {author} {\bibfnamefont {G.~M.}\
  \bibnamefont {Tveten}}, \bibinfo {author} {\bibfnamefont {Y.-W.}\
  \bibnamefont {Lui}}, \bibinfo {author} {\bibfnamefont {K.}~\bibnamefont
  {Sugita}},\ and\ \bibinfo {author} {\bibfnamefont {S.}~\bibnamefont
  {Miyamoto}},\ }\href
  {https://doi.org/https://doi.org/10.1016/j.nima.2018.04.021} {\bibfield
  {journal} {\bibinfo  {journal} {Nuclear Instruments and Methods in Physics
  Research Section A: Accelerators, Spectrometers, Detectors and Associated
  Equipment}\ }\textbf {\bibinfo {volume} {896}},\ \bibinfo {pages} {103}
  (\bibinfo {year} {2018})}\BibitemShut {NoStop}%
\bibitem [{\citenamefont {Renstr\o{}m}\ \emph {et~al.}(2018)\citenamefont
  {Renstr\o{}m}, \citenamefont {Utsunomiya}, \citenamefont {Nyhus},
  \citenamefont {Larsen}, \citenamefont {Guttormsen}, \citenamefont {Tveten},
  \citenamefont {Filipescu}, \citenamefont {Gheorghe}, \citenamefont {Goriely},
  \citenamefont {Hilaire}, \citenamefont {Lui}, \citenamefont {Midtb\o{}},
  \citenamefont {P\'eru}, \citenamefont {Shima}, \citenamefont {Siem},\ and\
  \citenamefont {Tesileanu}}]{renstrom2018_Dy}%
  \BibitemOpen
  \bibfield  {author} {\bibinfo {author} {\bibfnamefont {T.}~\bibnamefont
  {Renstr\o{}m}}, \bibinfo {author} {\bibfnamefont {H.}~\bibnamefont
  {Utsunomiya}}, \bibinfo {author} {\bibfnamefont {H.~T.}\ \bibnamefont
  {Nyhus}}, \bibinfo {author} {\bibfnamefont {A.~C.}\ \bibnamefont {Larsen}},
  \bibinfo {author} {\bibfnamefont {M.}~\bibnamefont {Guttormsen}}, \bibinfo
  {author} {\bibfnamefont {G.~M.}\ \bibnamefont {Tveten}}, \bibinfo {author}
  {\bibfnamefont {D.~M.}\ \bibnamefont {Filipescu}}, \bibinfo {author}
  {\bibfnamefont {I.}~\bibnamefont {Gheorghe}}, \bibinfo {author}
  {\bibfnamefont {S.}~\bibnamefont {Goriely}}, \bibinfo {author} {\bibfnamefont
  {S.}~\bibnamefont {Hilaire}}, \bibinfo {author} {\bibfnamefont {Y.-W.}\
  \bibnamefont {Lui}}, \bibinfo {author} {\bibfnamefont {J.~E.}\ \bibnamefont
  {Midtb\o{}}}, \bibinfo {author} {\bibfnamefont {S.}~\bibnamefont {P\'eru}},
  \bibinfo {author} {\bibfnamefont {T.}~\bibnamefont {Shima}}, \bibinfo
  {author} {\bibfnamefont {S.}~\bibnamefont {Siem}},\ and\ \bibinfo {author}
  {\bibfnamefont {O.}~\bibnamefont {Tesileanu}},\ }\href
  {https://doi.org/10.1103/PhysRevC.98.054310} {\bibfield  {journal} {\bibinfo
  {journal} {Phys. Rev. C}\ }\textbf {\bibinfo {volume} {98}},\ \bibinfo
  {pages} {054310} (\bibinfo {year} {2018})}\BibitemShut {NoStop}%
\bibitem [{\citenamefont {Goryachev}\ and\ \citenamefont
  {Zalesnyy}(1978)}]{Goryachev1978}%
  \BibitemOpen
  \bibfield  {author} {\bibinfo {author} {\bibfnamefont {A.~M.}\ \bibnamefont
  {Goryachev}}\ and\ \bibinfo {author} {\bibfnamefont {G.~N.}\ \bibnamefont
  {Zalesnyy}},\ }\href@noop {} {\bibfield  {journal} {\bibinfo  {journal}
  {Izvestiya Akademii Nauk Resp.Kaz., Ser.Fiz.-Mat.}\ }\textbf {\bibinfo
  {volume} {6}},\ \bibinfo {pages} {8} (\bibinfo {year} {1978})}\BibitemShut
  {NoStop}%
\bibitem [{\citenamefont {Guttormsen}\ \emph {et~al.}(2011)\citenamefont
  {Guttormsen}, \citenamefont {B\"{u}rger}, \citenamefont {Hansen},\ and\
  \citenamefont {Lietaer}}]{Guttormsen_NIMA_2011}%
  \BibitemOpen
  \bibfield  {author} {\bibinfo {author} {\bibfnamefont {M.}~\bibnamefont
  {Guttormsen}}, \bibinfo {author} {\bibfnamefont {A.}~\bibnamefont
  {B\"{u}rger}}, \bibinfo {author} {\bibfnamefont {T.~E.}\ \bibnamefont
  {Hansen}},\ and\ \bibinfo {author} {\bibfnamefont {N.}~\bibnamefont
  {Lietaer}},\ }\href {https://doi.org/dx.doi.org/10.1016/j.nima.2011.05.055}
  {\bibfield  {journal} {\bibinfo  {journal} {Nucl. Instr. Methods Phys. Res.
  A}\ }\textbf {\bibinfo {volume} {648}},\ \bibinfo {pages} {168} (\bibinfo
  {year} {2011})}\BibitemShut {NoStop}%
\bibitem [{\citenamefont {Ingeberg}(2015)}]{Qkinz}%
  \BibitemOpen
  \bibfield  {author} {\bibinfo {author} {\bibfnamefont {V.~W.}\ \bibnamefont
  {Ingeberg}},\ }\href {https://doi.org/https://doi.org/10.5281/zenodo.1206099}
  {} (\bibinfo {year} {2015})\BibitemShut {NoStop}%
\bibitem [{\citenamefont {Guttormsen}\ \emph {et~al.}(1990)\citenamefont
  {Guttormsen}, \citenamefont {Atac}, \citenamefont {L{\o}vh{\o}iden},
  \citenamefont {Messelt}, \citenamefont {Rams{\o}y}, \citenamefont {Rekstad},
  \citenamefont {Thorsteinsen}, \citenamefont {Tveter},\ and\ \citenamefont
  {Zelazny}}]{guttormsen_CACTUS_1990}%
  \BibitemOpen
  \bibfield  {author} {\bibinfo {author} {\bibfnamefont {M.}~\bibnamefont
  {Guttormsen}}, \bibinfo {author} {\bibfnamefont {A.}~\bibnamefont {Atac}},
  \bibinfo {author} {\bibfnamefont {G.}~\bibnamefont {L{\o}vh{\o}iden}},
  \bibinfo {author} {\bibfnamefont {S.}~\bibnamefont {Messelt}}, \bibinfo
  {author} {\bibfnamefont {T.}~\bibnamefont {Rams{\o}y}}, \bibinfo {author}
  {\bibfnamefont {J.}~\bibnamefont {Rekstad}}, \bibinfo {author} {\bibfnamefont
  {T.}~\bibnamefont {Thorsteinsen}}, \bibinfo {author} {\bibfnamefont
  {T.}~\bibnamefont {Tveter}},\ and\ \bibinfo {author} {\bibfnamefont
  {Z.}~\bibnamefont {Zelazny}},\ }\href
  {https://doi.org/https://doi.org/10.1088/0031-8949/1990/T32/010} {\bibfield
  {journal} {\bibinfo  {journal} {Physica Scripta}\ }\textbf {\bibinfo {volume}
  {T32}},\ \bibinfo {pages} {54} (\bibinfo {year} {1990})}\BibitemShut
  {NoStop}%
\bibitem [{\citenamefont {Guttormsen}\ \emph {et~al.}(1996)\citenamefont
  {Guttormsen}, \citenamefont {Tveter}, \citenamefont {Bergholt}, \citenamefont
  {Ingebretsen},\ and\ \citenamefont {Rekstad}}]{guttormsen_unfolding_1996}%
  \BibitemOpen
  \bibfield  {author} {\bibinfo {author} {\bibfnamefont {M.}~\bibnamefont
  {Guttormsen}}, \bibinfo {author} {\bibfnamefont {T.}~\bibnamefont {Tveter}},
  \bibinfo {author} {\bibfnamefont {L.}~\bibnamefont {Bergholt}}, \bibinfo
  {author} {\bibfnamefont {F.}~\bibnamefont {Ingebretsen}},\ and\ \bibinfo
  {author} {\bibfnamefont {J.}~\bibnamefont {Rekstad}},\ }\href
  {https://doi.org/https://doi.org/10.1016/0168-9002(96)00197-0} {\bibfield
  {journal} {\bibinfo  {journal} {Nuclear Instruments and Methods in Physics
  Research A}\ }\textbf {\bibinfo {volume} {374}},\ \bibinfo {pages} {371}
  (\bibinfo {year} {1996})}\BibitemShut {NoStop}%
\bibitem [{\citenamefont {Guttormsen}\ \emph {et~al.}()\citenamefont
  {Guttormsen} \emph {et~al.}}]{Oslosoftware}%
  \BibitemOpen
  \bibfield  {author} {\bibinfo {author} {\bibfnamefont {M.}~\bibnamefont
  {Guttormsen}} \emph {et~al.},\ }\href
  {https://doi.org/10.5281/zenodo.6024876} {\bibinfo {title} {Oslo-method
  software package}},\ \bibinfo {howpublished} {Github, Oslo Method
  v1.1.6}\BibitemShut {NoStop}%
\bibitem [{\citenamefont {Guttormsen}\ \emph {et~al.}(1987)\citenamefont
  {Guttormsen}, \citenamefont {Rams{\o}y},\ and\ \citenamefont
  {Rekstad}}]{guttormsen_fg_1987}%
  \BibitemOpen
  \bibfield  {author} {\bibinfo {author} {\bibfnamefont {M.}~\bibnamefont
  {Guttormsen}}, \bibinfo {author} {\bibfnamefont {T.}~\bibnamefont
  {Rams{\o}y}},\ and\ \bibinfo {author} {\bibfnamefont {J.}~\bibnamefont
  {Rekstad}},\ }\href
  {https://doi.org/https://doi.org/10.1016/0168-9002(87)91221-6} {\bibfield
  {journal} {\bibinfo  {journal} {Nuclear Instruments and Methods in Physics
  Research A}\ }\textbf {\bibinfo {volume} {255}},\ \bibinfo {pages} {518}
  (\bibinfo {year} {1987})}\BibitemShut {NoStop}%
\bibitem [{\citenamefont {Larsen}\ \emph {et~al.}(2011)\citenamefont {Larsen},
  \citenamefont {Guttormsen}, \citenamefont {Krti\ifmmode~\check{c}\else
  \v{c}\fi{}ka}, \citenamefont {B\ifmmode~\check{e}\else \v{e}\fi{}t\'ak},
  \citenamefont {B\"urger}, \citenamefont {G\"orgen}, \citenamefont {Nyhus},
  \citenamefont {Rekstad}, \citenamefont {Schiller}, \citenamefont {Siem},
  \citenamefont {Toft}, \citenamefont {Tveten}, \citenamefont {Voinov},\ and\
  \citenamefont {Wikan}}]{larsen_syst_2011}%
  \BibitemOpen
  \bibfield  {author} {\bibinfo {author} {\bibfnamefont {A.~C.}\ \bibnamefont
  {Larsen}}, \bibinfo {author} {\bibfnamefont {M.}~\bibnamefont {Guttormsen}},
  \bibinfo {author} {\bibfnamefont {M.}~\bibnamefont
  {Krti\ifmmode~\check{c}\else \v{c}\fi{}ka}}, \bibinfo {author} {\bibfnamefont
  {E.}~\bibnamefont {B\ifmmode~\check{e}\else \v{e}\fi{}t\'ak}}, \bibinfo
  {author} {\bibfnamefont {A.}~\bibnamefont {B\"urger}}, \bibinfo {author}
  {\bibfnamefont {A.}~\bibnamefont {G\"orgen}}, \bibinfo {author}
  {\bibfnamefont {H.~T.}\ \bibnamefont {Nyhus}}, \bibinfo {author}
  {\bibfnamefont {J.}~\bibnamefont {Rekstad}}, \bibinfo {author} {\bibfnamefont
  {A.}~\bibnamefont {Schiller}}, \bibinfo {author} {\bibfnamefont
  {S.}~\bibnamefont {Siem}}, \bibinfo {author} {\bibfnamefont {H.~K.}\
  \bibnamefont {Toft}}, \bibinfo {author} {\bibfnamefont {G.~M.}\ \bibnamefont
  {Tveten}}, \bibinfo {author} {\bibfnamefont {A.~V.}\ \bibnamefont {Voinov}},\
  and\ \bibinfo {author} {\bibfnamefont {K.}~\bibnamefont {Wikan}},\ }\href
  {https://doi.org/10.1103/PhysRevC.83.034315} {\bibfield  {journal} {\bibinfo
  {journal} {Phys. Rev. C}\ }\textbf {\bibinfo {volume} {83}},\ \bibinfo
  {pages} {034315} (\bibinfo {year} {2011})}\BibitemShut {NoStop}%
\bibitem [{\citenamefont {Dirac}(1927)}]{Dirac1927}%
  \BibitemOpen
  \bibfield  {author} {\bibinfo {author} {\bibfnamefont {P.~A.~M.}\
  \bibnamefont {Dirac}},\ }\href
  {https://doi.org/https://doi.org/10.1098/rspa.1927.0039} {\bibfield
  {journal} {\bibinfo  {journal} {Proc. R. Soc. London A}\ }\textbf {\bibinfo
  {volume} {114}},\ \bibinfo {pages} {243} (\bibinfo {year}
  {1927})}\BibitemShut {NoStop}%
\bibitem [{\citenamefont {Fermi}(1950)}]{Fermi1950}%
  \BibitemOpen
  \bibfield  {author} {\bibinfo {author} {\bibfnamefont {E.}~\bibnamefont
  {Fermi}},\ }\href@noop {} {\emph {\bibinfo {title} {Nuclear Physics}}}\
  (\bibinfo  {publisher} {University of Chicago Press, Chicago},\ \bibinfo
  {year} {1950})\BibitemShut {NoStop}%
\bibitem [{\citenamefont {Schiller}\ \emph {et~al.}(2000)\citenamefont
  {Schiller}, \citenamefont {Bergholt}, \citenamefont {Guttormsen},
  \citenamefont {Melby}, \citenamefont {Rekstad},\ and\ \citenamefont
  {Siem}}]{schiller2000}%
  \BibitemOpen
  \bibfield  {author} {\bibinfo {author} {\bibfnamefont {A.}~\bibnamefont
  {Schiller}}, \bibinfo {author} {\bibfnamefont {L.}~\bibnamefont {Bergholt}},
  \bibinfo {author} {\bibfnamefont {M.}~\bibnamefont {Guttormsen}}, \bibinfo
  {author} {\bibfnamefont {E.}~\bibnamefont {Melby}}, \bibinfo {author}
  {\bibfnamefont {J.}~\bibnamefont {Rekstad}},\ and\ \bibinfo {author}
  {\bibfnamefont {S.}~\bibnamefont {Siem}},\ }\href
  {https://doi.org/https://doi.org/10.1016/S0168-9002(99)01187-0} {\bibfield
  {journal} {\bibinfo  {journal} {Nuclear Instruments and Methods in Physics
  Research Section A: Accelerators, Spectrometers, Detectors and Associated
  Equipment}\ }\textbf {\bibinfo {volume} {447}},\ \bibinfo {pages} {498}
  (\bibinfo {year} {2000})}\BibitemShut {NoStop}%
\bibitem [{\citenamefont {Brink}(1955)}]{brink1955}%
  \BibitemOpen
  \bibfield  {author} {\bibinfo {author} {\bibfnamefont {D.~M.}\ \bibnamefont
  {Brink}},\ }\emph {\bibinfo {title} {Some aspects of the interaction of
  fields with matter}},\ \href@noop {} {Ph.D. thesis},\ \bibinfo  {school}
  {Oxford University} (\bibinfo {year} {1955})\BibitemShut {NoStop}%
\bibitem [{\citenamefont {Axel}(1962)}]{axel1962}%
  \BibitemOpen
  \bibfield  {author} {\bibinfo {author} {\bibfnamefont {P.}~\bibnamefont
  {Axel}},\ }\href {https://doi.org/10.1103/PhysRev.126.671} {\bibfield
  {journal} {\bibinfo  {journal} {Phys. Rev.}\ }\textbf {\bibinfo {volume}
  {126}},\ \bibinfo {pages} {671} (\bibinfo {year} {1962})}\BibitemShut
  {NoStop}%
\bibitem [{\citenamefont {Misch}\ \emph {et~al.}(2014)\citenamefont {Misch},
  \citenamefont {Fuller},\ and\ \citenamefont
  {Brown}}]{Misch_PhysRevC.90.065808}%
  \BibitemOpen
  \bibfield  {author} {\bibinfo {author} {\bibfnamefont {G.~W.}\ \bibnamefont
  {Misch}}, \bibinfo {author} {\bibfnamefont {G.~M.}\ \bibnamefont {Fuller}},\
  and\ \bibinfo {author} {\bibfnamefont {B.~A.}\ \bibnamefont {Brown}},\ }\href
  {https://doi.org/10.1103/PhysRevC.90.065808} {\bibfield  {journal} {\bibinfo
  {journal} {Phys. Rev. C}\ }\textbf {\bibinfo {volume} {90}},\ \bibinfo
  {pages} {065808} (\bibinfo {year} {2014})}\BibitemShut {NoStop}%
\bibitem [{\citenamefont {Johnson}(2015)}]{JOHNSON201572}%
  \BibitemOpen
  \bibfield  {author} {\bibinfo {author} {\bibfnamefont {C.~W.}\ \bibnamefont
  {Johnson}},\ }\href
  {https://doi.org/https://doi.org/10.1016/j.physletb.2015.08.054} {\bibfield
  {journal} {\bibinfo  {journal} {Physics Letters B}\ }\textbf {\bibinfo
  {volume} {750}},\ \bibinfo {pages} {72} (\bibinfo {year} {2015})}\BibitemShut
  {NoStop}%
\bibitem [{\citenamefont {Hung}\ \emph {et~al.}(2017)\citenamefont {Hung},
  \citenamefont {Dang},\ and\ \citenamefont
  {Huong}}]{Hung_PhysRevLett.118.022502}%
  \BibitemOpen
  \bibfield  {author} {\bibinfo {author} {\bibfnamefont {N.~Q.}\ \bibnamefont
  {Hung}}, \bibinfo {author} {\bibfnamefont {N.~D.}\ \bibnamefont {Dang}},\
  and\ \bibinfo {author} {\bibfnamefont {L.~T.~Q.}\ \bibnamefont {Huong}},\
  }\href {https://doi.org/10.1103/PhysRevLett.118.022502} {\bibfield  {journal}
  {\bibinfo  {journal} {Phys. Rev. Lett.}\ }\textbf {\bibinfo {volume} {118}},\
  \bibinfo {pages} {022502} (\bibinfo {year} {2017})}\BibitemShut {NoStop}%
\bibitem [{\citenamefont {Guttormsen}\ \emph {et~al.}(2016)\citenamefont
  {Guttormsen}, \citenamefont {Larsen}, \citenamefont {G\"orgen}, \citenamefont
  {Renstr\o{}m}, \citenamefont {Siem}, \citenamefont {Tornyi},\ and\
  \citenamefont {Tveten}}]{Guttormsen_PhysRevLett.116.012502}%
  \BibitemOpen
  \bibfield  {author} {\bibinfo {author} {\bibfnamefont {M.}~\bibnamefont
  {Guttormsen}}, \bibinfo {author} {\bibfnamefont {A.~C.}\ \bibnamefont
  {Larsen}}, \bibinfo {author} {\bibfnamefont {A.}~\bibnamefont {G\"orgen}},
  \bibinfo {author} {\bibfnamefont {T.}~\bibnamefont {Renstr\o{}m}}, \bibinfo
  {author} {\bibfnamefont {S.}~\bibnamefont {Siem}}, \bibinfo {author}
  {\bibfnamefont {T.~G.}\ \bibnamefont {Tornyi}},\ and\ \bibinfo {author}
  {\bibfnamefont {G.~M.}\ \bibnamefont {Tveten}},\ }\href
  {https://doi.org/10.1103/PhysRevLett.116.012502} {\bibfield  {journal}
  {\bibinfo  {journal} {Phys. Rev. Lett.}\ }\textbf {\bibinfo {volume} {116}},\
  \bibinfo {pages} {012502} (\bibinfo {year} {2016})}\BibitemShut {NoStop}%
\bibitem [{\citenamefont {Martin}\ \emph {et~al.}(2017)\citenamefont {Martin},
  \citenamefont {von Neumann-Cosel}, \citenamefont {Tamii}, \citenamefont
  {Aoi}, \citenamefont {Bassauer}, \citenamefont {Bertulani}, \citenamefont
  {Carter}, \citenamefont {Donaldson}, \citenamefont {Fujita}, \citenamefont
  {Fujita}, \citenamefont {Hashimoto}, \citenamefont {Hatanaka}, \citenamefont
  {Ito}, \citenamefont {Krugmann}, \citenamefont {Liu}, \citenamefont {Maeda},
  \citenamefont {Miki}, \citenamefont {Neveling}, \citenamefont {Pietralla},
  \citenamefont {Poltoratska}, \citenamefont {Ponomarev}, \citenamefont
  {Richter}, \citenamefont {Shima}, \citenamefont {Yamamoto},\ and\
  \citenamefont {Zweidinger}}]{Martin_PhysRevLett.119.182503}%
  \BibitemOpen
  \bibfield  {author} {\bibinfo {author} {\bibfnamefont {D.}~\bibnamefont
  {Martin}}, \bibinfo {author} {\bibfnamefont {P.}~\bibnamefont {von
  Neumann-Cosel}}, \bibinfo {author} {\bibfnamefont {A.}~\bibnamefont {Tamii}},
  \bibinfo {author} {\bibfnamefont {N.}~\bibnamefont {Aoi}}, \bibinfo {author}
  {\bibfnamefont {S.}~\bibnamefont {Bassauer}}, \bibinfo {author}
  {\bibfnamefont {C.~A.}\ \bibnamefont {Bertulani}}, \bibinfo {author}
  {\bibfnamefont {J.}~\bibnamefont {Carter}}, \bibinfo {author} {\bibfnamefont
  {L.}~\bibnamefont {Donaldson}}, \bibinfo {author} {\bibfnamefont
  {H.}~\bibnamefont {Fujita}}, \bibinfo {author} {\bibfnamefont
  {Y.}~\bibnamefont {Fujita}}, \bibinfo {author} {\bibfnamefont
  {T.}~\bibnamefont {Hashimoto}}, \bibinfo {author} {\bibfnamefont
  {K.}~\bibnamefont {Hatanaka}}, \bibinfo {author} {\bibfnamefont
  {T.}~\bibnamefont {Ito}}, \bibinfo {author} {\bibfnamefont {A.}~\bibnamefont
  {Krugmann}}, \bibinfo {author} {\bibfnamefont {B.}~\bibnamefont {Liu}},
  \bibinfo {author} {\bibfnamefont {Y.}~\bibnamefont {Maeda}}, \bibinfo
  {author} {\bibfnamefont {K.}~\bibnamefont {Miki}}, \bibinfo {author}
  {\bibfnamefont {R.}~\bibnamefont {Neveling}}, \bibinfo {author}
  {\bibfnamefont {N.}~\bibnamefont {Pietralla}}, \bibinfo {author}
  {\bibfnamefont {I.}~\bibnamefont {Poltoratska}}, \bibinfo {author}
  {\bibfnamefont {V.~Y.}\ \bibnamefont {Ponomarev}}, \bibinfo {author}
  {\bibfnamefont {A.}~\bibnamefont {Richter}}, \bibinfo {author} {\bibfnamefont
  {T.}~\bibnamefont {Shima}}, \bibinfo {author} {\bibfnamefont
  {T.}~\bibnamefont {Yamamoto}},\ and\ \bibinfo {author} {\bibfnamefont
  {M.}~\bibnamefont {Zweidinger}},\ }\href
  {https://doi.org/10.1103/PhysRevLett.119.182503} {\bibfield  {journal}
  {\bibinfo  {journal} {Phys. Rev. Lett.}\ }\textbf {\bibinfo {volume} {119}},\
  \bibinfo {pages} {182503} (\bibinfo {year} {2017})}\BibitemShut {NoStop}%
\bibitem [{\citenamefont {Campo}\ \emph {et~al.}(2018)\citenamefont {Campo},
  \citenamefont {Guttormsen}, \citenamefont {Garrote}, \citenamefont {Eriksen},
  \citenamefont {Giacoppo}, \citenamefont {G\"orgen}, \citenamefont
  {Hadynska-Klek}, \citenamefont {Klintefjord}, \citenamefont {Larsen},
  \citenamefont {Renstr\o{}m}, \citenamefont {Sahin}, \citenamefont {Siem},
  \citenamefont {Springer}, \citenamefont {Tornyi},\ and\ \citenamefont
  {Tveten}}]{CrespoCampo_PhysRevC.98.054303}%
  \BibitemOpen
  \bibfield  {author} {\bibinfo {author} {\bibfnamefont {L.~C.}\ \bibnamefont
  {Campo}}, \bibinfo {author} {\bibfnamefont {M.}~\bibnamefont {Guttormsen}},
  \bibinfo {author} {\bibfnamefont {F.~L.~B.}\ \bibnamefont {Garrote}},
  \bibinfo {author} {\bibfnamefont {T.~K.}\ \bibnamefont {Eriksen}}, \bibinfo
  {author} {\bibfnamefont {F.}~\bibnamefont {Giacoppo}}, \bibinfo {author}
  {\bibfnamefont {A.}~\bibnamefont {G\"orgen}}, \bibinfo {author}
  {\bibfnamefont {K.}~\bibnamefont {Hadynska-Klek}}, \bibinfo {author}
  {\bibfnamefont {M.}~\bibnamefont {Klintefjord}}, \bibinfo {author}
  {\bibfnamefont {A.~C.}\ \bibnamefont {Larsen}}, \bibinfo {author}
  {\bibfnamefont {T.}~\bibnamefont {Renstr\o{}m}}, \bibinfo {author}
  {\bibfnamefont {E.}~\bibnamefont {Sahin}}, \bibinfo {author} {\bibfnamefont
  {S.}~\bibnamefont {Siem}}, \bibinfo {author} {\bibfnamefont {A.}~\bibnamefont
  {Springer}}, \bibinfo {author} {\bibfnamefont {T.~G.}\ \bibnamefont
  {Tornyi}},\ and\ \bibinfo {author} {\bibfnamefont {G.~M.}\ \bibnamefont
  {Tveten}},\ }\href {https://doi.org/10.1103/PhysRevC.98.054303} {\bibfield
  {journal} {\bibinfo  {journal} {Phys. Rev. C}\ }\textbf {\bibinfo {volume}
  {98}},\ \bibinfo {pages} {054303} (\bibinfo {year} {2018})}\BibitemShut
  {NoStop}%
\bibitem [{\citenamefont {Scholz}\ \emph {et~al.}(2020)\citenamefont {Scholz},
  \citenamefont {Guttormsen}, \citenamefont {Heim}, \citenamefont {Larsen},
  \citenamefont {Mayer}, \citenamefont {Savran}, \citenamefont {Spieker},
  \citenamefont {Tveten}, \citenamefont {Voinov}, \citenamefont {Wilhelmy},
  \citenamefont {Zeiser},\ and\ \citenamefont
  {Zilges}}]{Scholz_PhysRevC.101.045806}%
  \BibitemOpen
  \bibfield  {author} {\bibinfo {author} {\bibfnamefont {P.}~\bibnamefont
  {Scholz}}, \bibinfo {author} {\bibfnamefont {M.}~\bibnamefont {Guttormsen}},
  \bibinfo {author} {\bibfnamefont {F.}~\bibnamefont {Heim}}, \bibinfo {author}
  {\bibfnamefont {A.~C.}\ \bibnamefont {Larsen}}, \bibinfo {author}
  {\bibfnamefont {J.}~\bibnamefont {Mayer}}, \bibinfo {author} {\bibfnamefont
  {D.}~\bibnamefont {Savran}}, \bibinfo {author} {\bibfnamefont
  {M.}~\bibnamefont {Spieker}}, \bibinfo {author} {\bibfnamefont {G.~M.}\
  \bibnamefont {Tveten}}, \bibinfo {author} {\bibfnamefont {A.~V.}\
  \bibnamefont {Voinov}}, \bibinfo {author} {\bibfnamefont {J.}~\bibnamefont
  {Wilhelmy}}, \bibinfo {author} {\bibfnamefont {F.}~\bibnamefont {Zeiser}},\
  and\ \bibinfo {author} {\bibfnamefont {A.}~\bibnamefont {Zilges}},\ }\href
  {https://doi.org/10.1103/PhysRevC.101.045806} {\bibfield  {journal} {\bibinfo
   {journal} {Phys. Rev. C}\ }\textbf {\bibinfo {volume} {101}},\ \bibinfo
  {pages} {045806} (\bibinfo {year} {2020})}\BibitemShut {NoStop}%
\bibitem [{\citenamefont {Angell}\ \emph {et~al.}(2012)\citenamefont {Angell},
  \citenamefont {Hammond}, \citenamefont {Karwowski}, \citenamefont {Kelley},
  \citenamefont {Krti\ifmmode~\check{c}\else \v{c}\fi{}ka}, \citenamefont
  {Kwan}, \citenamefont {Makinaga},\ and\ \citenamefont
  {Rusev}}]{Angell_PhysRevC.86.051302}%
  \BibitemOpen
  \bibfield  {author} {\bibinfo {author} {\bibfnamefont {C.~T.}\ \bibnamefont
  {Angell}}, \bibinfo {author} {\bibfnamefont {S.~L.}\ \bibnamefont {Hammond}},
  \bibinfo {author} {\bibfnamefont {H.~J.}\ \bibnamefont {Karwowski}}, \bibinfo
  {author} {\bibfnamefont {J.~H.}\ \bibnamefont {Kelley}}, \bibinfo {author}
  {\bibfnamefont {M.}~\bibnamefont {Krti\ifmmode~\check{c}\else \v{c}\fi{}ka}},
  \bibinfo {author} {\bibfnamefont {E.}~\bibnamefont {Kwan}}, \bibinfo {author}
  {\bibfnamefont {A.}~\bibnamefont {Makinaga}},\ and\ \bibinfo {author}
  {\bibfnamefont {G.}~\bibnamefont {Rusev}},\ }\href
  {https://doi.org/10.1103/PhysRevC.86.051302} {\bibfield  {journal} {\bibinfo
  {journal} {Phys. Rev. C}\ }\textbf {\bibinfo {volume} {86}},\ \bibinfo
  {pages} {051302} (\bibinfo {year} {2012})}\BibitemShut {NoStop}%
\bibitem [{\citenamefont {Isaak}\ \emph {et~al.}(2019)\citenamefont {Isaak},
  \citenamefont {Savran}, \citenamefont {Löher}, \citenamefont {Beck},
  \citenamefont {Bhike}, \citenamefont {Gayer}, \citenamefont {Krishichayan},
  \citenamefont {Pietralla}, \citenamefont {Scheck}, \citenamefont {Tornow},
  \citenamefont {Werner}, \citenamefont {Zilges},\ and\ \citenamefont
  {Zweidinger}}]{ISAAK2019225}%
  \BibitemOpen
  \bibfield  {author} {\bibinfo {author} {\bibfnamefont {J.}~\bibnamefont
  {Isaak}}, \bibinfo {author} {\bibfnamefont {D.}~\bibnamefont {Savran}},
  \bibinfo {author} {\bibfnamefont {B.}~\bibnamefont {Löher}}, \bibinfo
  {author} {\bibfnamefont {T.}~\bibnamefont {Beck}}, \bibinfo {author}
  {\bibfnamefont {M.}~\bibnamefont {Bhike}}, \bibinfo {author} {\bibfnamefont
  {U.}~\bibnamefont {Gayer}}, \bibinfo {author} {\bibnamefont {Krishichayan}},
  \bibinfo {author} {\bibfnamefont {N.}~\bibnamefont {Pietralla}}, \bibinfo
  {author} {\bibfnamefont {M.}~\bibnamefont {Scheck}}, \bibinfo {author}
  {\bibfnamefont {W.}~\bibnamefont {Tornow}}, \bibinfo {author} {\bibfnamefont
  {V.}~\bibnamefont {Werner}}, \bibinfo {author} {\bibfnamefont
  {A.}~\bibnamefont {Zilges}},\ and\ \bibinfo {author} {\bibfnamefont
  {M.}~\bibnamefont {Zweidinger}},\ }\href
  {https://doi.org/https://doi.org/10.1016/j.physletb.2018.11.038} {\bibfield
  {journal} {\bibinfo  {journal} {Physics Letters B}\ }\textbf {\bibinfo
  {volume} {788}},\ \bibinfo {pages} {225} (\bibinfo {year}
  {2019})}\BibitemShut {NoStop}%
\bibitem [{\citenamefont {{National Nuclear Data Center}}(2022)}]{NNDC}%
  \BibitemOpen
  \bibfield  {author} {\bibinfo {author} {\bibnamefont {{National Nuclear Data
  Center}}},\ }\href {https://www.nndc.bnl.gov/nudat3/} {\bibinfo {title}
  {Nudat 3 database}} (\bibinfo {year} {2022})\BibitemShut {NoStop}%
\bibitem [{\citenamefont {Mughabghab}(2018)}]{Mughabghab2018}%
  \BibitemOpen
  \bibfield  {author} {\bibinfo {author} {\bibfnamefont {S.~F.}\ \bibnamefont
  {Mughabghab}},\ }\href@noop {} {\emph {\bibinfo {title} {Atlas of Neutron
  Resonances, 6th ed.}}}\ (\bibinfo  {publisher} {Elsevier Science,
  Amsterdam},\ \bibinfo {year} {2018})\BibitemShut {NoStop}%
\bibitem [{\citenamefont {von Egidy}\ and\ \citenamefont
  {Bucurescu}(2005)}]{Egidy_PhysRevC.72.044311}%
  \BibitemOpen
  \bibfield  {author} {\bibinfo {author} {\bibfnamefont {T.}~\bibnamefont {von
  Egidy}}\ and\ \bibinfo {author} {\bibfnamefont {D.}~\bibnamefont
  {Bucurescu}},\ }\href {https://doi.org/10.1103/PhysRevC.72.044311} {\bibfield
   {journal} {\bibinfo  {journal} {Phys. Rev. C}\ }\textbf {\bibinfo {volume}
  {72}},\ \bibinfo {pages} {044311} (\bibinfo {year} {2005})}\BibitemShut
  {NoStop}%
\bibitem [{\citenamefont {von Egidy}\ and\ \citenamefont
  {Bucurescu}(2006)}]{Egidy_PhysRevC.73.049901}%
  \BibitemOpen
  \bibfield  {author} {\bibinfo {author} {\bibfnamefont {T.}~\bibnamefont {von
  Egidy}}\ and\ \bibinfo {author} {\bibfnamefont {D.}~\bibnamefont
  {Bucurescu}},\ }\href {https://doi.org/10.1103/PhysRevC.73.049901} {\bibfield
   {journal} {\bibinfo  {journal} {Phys. Rev. C}\ }\textbf {\bibinfo {volume}
  {73}},\ \bibinfo {pages} {049901} (\bibinfo {year} {2006})}\BibitemShut
  {NoStop}%
\bibitem [{\citenamefont {Uhrenholt}\ \emph {et~al.}(2013)\citenamefont
  {Uhrenholt}, \citenamefont {{\AA}berg}, \citenamefont {Dobrowolski},
  \citenamefont {D{\o}ssing}, \citenamefont {Ichikawa},\ and\ \citenamefont
  {M{\"{o}}ller}}]{UHRENHOLT2013127}%
  \BibitemOpen
  \bibfield  {author} {\bibinfo {author} {\bibfnamefont {H.}~\bibnamefont
  {Uhrenholt}}, \bibinfo {author} {\bibfnamefont {S.}~\bibnamefont
  {{\AA}berg}}, \bibinfo {author} {\bibfnamefont {A.}~\bibnamefont
  {Dobrowolski}}, \bibinfo {author} {\bibfnamefont {T.}~\bibnamefont
  {D{\o}ssing}}, \bibinfo {author} {\bibfnamefont {T.}~\bibnamefont
  {Ichikawa}},\ and\ \bibinfo {author} {\bibfnamefont {P.}~\bibnamefont
  {M{\"{o}}ller}},\ }\href
  {https://doi.org/https://doi.org/10.1016/j.nuclphysa.2013.06.002} {\bibfield
  {journal} {\bibinfo  {journal} {Nuclear Physics A}\ }\textbf {\bibinfo
  {volume} {913}},\ \bibinfo {pages} {127} (\bibinfo {year}
  {2013})}\BibitemShut {NoStop}%
\bibitem [{\citenamefont {Ericson}(1959)}]{Ericson1959}%
  \BibitemOpen
  \bibfield  {author} {\bibinfo {author} {\bibfnamefont {T.}~\bibnamefont
  {Ericson}},\ }\href
  {https://doi.org/https://doi.org/10.1016/0029-5582(59)90291-3} {\bibfield
  {journal} {\bibinfo  {journal} {Nuclear Physics}\ }\textbf {\bibinfo {volume}
  {11}},\ \bibinfo {pages} {481} (\bibinfo {year} {1959})}\BibitemShut
  {NoStop}%
\bibitem [{\citenamefont {Voinov}\ \emph {et~al.}(2001)\citenamefont {Voinov},
  \citenamefont {Guttormsen}, \citenamefont {Melby}, \citenamefont {Rekstad},
  \citenamefont {Schiller},\ and\ \citenamefont {Siem}}]{Voinov2001}%
  \BibitemOpen
  \bibfield  {author} {\bibinfo {author} {\bibfnamefont {A.}~\bibnamefont
  {Voinov}}, \bibinfo {author} {\bibfnamefont {M.}~\bibnamefont {Guttormsen}},
  \bibinfo {author} {\bibfnamefont {E.}~\bibnamefont {Melby}}, \bibinfo
  {author} {\bibfnamefont {J.}~\bibnamefont {Rekstad}}, \bibinfo {author}
  {\bibfnamefont {A.}~\bibnamefont {Schiller}},\ and\ \bibinfo {author}
  {\bibfnamefont {S.}~\bibnamefont {Siem}},\ }\href
  {https://doi.org/10.1103/PhysRevC.63.044313} {\bibfield  {journal} {\bibinfo
  {journal} {Phys. Rev. C}\ }\textbf {\bibinfo {volume} {63}},\ \bibinfo
  {pages} {044313} (\bibinfo {year} {2001})}\BibitemShut {NoStop}%
\bibitem [{\citenamefont {Bethe}(1936)}]{Bethe1936}%
  \BibitemOpen
  \bibfield  {author} {\bibinfo {author} {\bibfnamefont {H.~A.}\ \bibnamefont
  {Bethe}},\ }\href {https://doi.org/10.1103/PhysRev.50.332} {\bibfield
  {journal} {\bibinfo  {journal} {Phys. Rev.}\ }\textbf {\bibinfo {volume}
  {50}},\ \bibinfo {pages} {332} (\bibinfo {year} {1936})}\BibitemShut
  {NoStop}%
\bibitem [{\citenamefont {Ericson}\ and\ \citenamefont
  {Strutinski}(1958)}]{Ericson1958}%
  \BibitemOpen
  \bibfield  {author} {\bibinfo {author} {\bibfnamefont {T.}~\bibnamefont
  {Ericson}}\ and\ \bibinfo {author} {\bibfnamefont {V.}~\bibnamefont
  {Strutinski}},\ }\href
  {https://doi.org/https://doi.org/10.1016/0029-5582(58)90156-1} {\bibfield
  {journal} {\bibinfo  {journal} {Nuclear Physics}\ }\textbf {\bibinfo {volume}
  {8}},\ \bibinfo {pages} {284} (\bibinfo {year} {1958})}\BibitemShut {NoStop}%
\bibitem [{\citenamefont {Capote}\ \emph {et~al.}(2009)\citenamefont {Capote},
  \citenamefont {Herman}, \citenamefont {Obložinský}, \citenamefont {Young},
  \citenamefont {Goriely}, \citenamefont {Belgya}, \citenamefont {Ignatyuk},
  \citenamefont {Koning}, \citenamefont {Hilaire}, \citenamefont {Plujko},
  \citenamefont {Avrigeanu}, \citenamefont {Bersillon}, \citenamefont
  {Chadwick}, \citenamefont {Fukahori}, \citenamefont {Ge}, \citenamefont
  {Han}, \citenamefont {Kailas}, \citenamefont {Kopecky}, \citenamefont
  {Maslov}, \citenamefont {Reffo}, \citenamefont {Sin}, \citenamefont
  {Soukhovitskii},\ and\ \citenamefont {Talou}}]{Capote2009}%
  \BibitemOpen
  \bibfield  {author} {\bibinfo {author} {\bibfnamefont {R.}~\bibnamefont
  {Capote}}, \bibinfo {author} {\bibfnamefont {M.}~\bibnamefont {Herman}},
  \bibinfo {author} {\bibfnamefont {P.}~\bibnamefont {Obložinský}}, \bibinfo
  {author} {\bibfnamefont {P.}~\bibnamefont {Young}}, \bibinfo {author}
  {\bibfnamefont {S.}~\bibnamefont {Goriely}}, \bibinfo {author} {\bibfnamefont
  {T.}~\bibnamefont {Belgya}}, \bibinfo {author} {\bibfnamefont
  {A.}~\bibnamefont {Ignatyuk}}, \bibinfo {author} {\bibfnamefont
  {A.}~\bibnamefont {Koning}}, \bibinfo {author} {\bibfnamefont
  {S.}~\bibnamefont {Hilaire}}, \bibinfo {author} {\bibfnamefont
  {V.}~\bibnamefont {Plujko}}, \bibinfo {author} {\bibfnamefont
  {M.}~\bibnamefont {Avrigeanu}}, \bibinfo {author} {\bibfnamefont
  {O.}~\bibnamefont {Bersillon}}, \bibinfo {author} {\bibfnamefont
  {M.}~\bibnamefont {Chadwick}}, \bibinfo {author} {\bibfnamefont
  {T.}~\bibnamefont {Fukahori}}, \bibinfo {author} {\bibfnamefont
  {Z.}~\bibnamefont {Ge}}, \bibinfo {author} {\bibfnamefont {Y.}~\bibnamefont
  {Han}}, \bibinfo {author} {\bibfnamefont {S.}~\bibnamefont {Kailas}},
  \bibinfo {author} {\bibfnamefont {J.}~\bibnamefont {Kopecky}}, \bibinfo
  {author} {\bibfnamefont {V.}~\bibnamefont {Maslov}}, \bibinfo {author}
  {\bibfnamefont {G.}~\bibnamefont {Reffo}}, \bibinfo {author} {\bibfnamefont
  {M.}~\bibnamefont {Sin}}, \bibinfo {author} {\bibfnamefont {E.}~\bibnamefont
  {Soukhovitskii}},\ and\ \bibinfo {author} {\bibfnamefont {P.}~\bibnamefont
  {Talou}},\ }\href {https://doi.org/https://doi.org/10.1016/j.nds.2009.10.004}
  {\bibfield  {journal} {\bibinfo  {journal} {Nuclear Data Sheets}\ }\textbf
  {\bibinfo {volume} {110}},\ \bibinfo {pages} {3107} (\bibinfo {year}
  {2009})},\ \bibinfo {note} {special Issue on Nuclear Reaction
  Data}\BibitemShut {NoStop}%
\bibitem [{\citenamefont {Renstr{\o}m}\ \emph {et~al.}(2018)\citenamefont
  {Renstr{\o}m}, \citenamefont {Tveten}, \citenamefont {Midtb{\o}},
  \citenamefont {Utsunomiya}, \citenamefont {Achakovskiy}, \citenamefont
  {Kamerdzhiev}, \citenamefont {Brown}, \citenamefont {Avdeenkov},
  \citenamefont {Ari-izumi}, \citenamefont {G\''{o}rgen}, \citenamefont
  {Grimes}, \citenamefont {Guttormsen}, \citenamefont {Hagen}, \citenamefont
  {Ingeberg}, \citenamefont {Katayama}, \citenamefont {Kheswa}, \citenamefont
  {Larsen}, \citenamefont {Lui}, \citenamefont {Nyhus}, \citenamefont {Siem},
  \citenamefont {Symochko}, \citenamefont {Takenaka},\ and\ \citenamefont
  {Voinov}}]{renstrom2018_Ni}%
  \BibitemOpen
  \bibfield  {author} {\bibinfo {author} {\bibfnamefont {T.}~\bibnamefont
  {Renstr{\o}m}}, \bibinfo {author} {\bibfnamefont {G.~M.}\ \bibnamefont
  {Tveten}}, \bibinfo {author} {\bibfnamefont {J.~E.}\ \bibnamefont
  {Midtb{\o}}}, \bibinfo {author} {\bibfnamefont {H.}~\bibnamefont
  {Utsunomiya}}, \bibinfo {author} {\bibfnamefont {O.}~\bibnamefont
  {Achakovskiy}}, \bibinfo {author} {\bibfnamefont {S.}~\bibnamefont
  {Kamerdzhiev}}, \bibinfo {author} {\bibfnamefont {B.~A.}\ \bibnamefont
  {Brown}}, \bibinfo {author} {\bibfnamefont {A.}~\bibnamefont {Avdeenkov}},
  \bibinfo {author} {\bibfnamefont {T.}~\bibnamefont {Ari-izumi}}, \bibinfo
  {author} {\bibfnamefont {A.}~\bibnamefont {G\''{o}rgen}}, \bibinfo {author}
  {\bibfnamefont {S.~M.}\ \bibnamefont {Grimes}}, \bibinfo {author}
  {\bibfnamefont {M.}~\bibnamefont {Guttormsen}}, \bibinfo {author}
  {\bibfnamefont {T.~W.}\ \bibnamefont {Hagen}}, \bibinfo {author}
  {\bibfnamefont {V.~W.}\ \bibnamefont {Ingeberg}}, \bibinfo {author}
  {\bibfnamefont {S.}~\bibnamefont {Katayama}}, \bibinfo {author}
  {\bibfnamefont {B.~V.}\ \bibnamefont {Kheswa}}, \bibinfo {author}
  {\bibfnamefont {A.~C.}\ \bibnamefont {Larsen}}, \bibinfo {author}
  {\bibfnamefont {Y.-W.}\ \bibnamefont {Lui}}, \bibinfo {author} {\bibfnamefont
  {H.-T.}\ \bibnamefont {Nyhus}}, \bibinfo {author} {\bibfnamefont
  {S.}~\bibnamefont {Siem}}, \bibinfo {author} {\bibfnamefont {D.}~\bibnamefont
  {Symochko}}, \bibinfo {author} {\bibfnamefont {D.}~\bibnamefont {Takenaka}},\
  and\ \bibinfo {author} {\bibfnamefont {A.~V.}\ \bibnamefont {Voinov}},\
  }\href {https://arxiv.org/abs/1804.08086} {\bibfield  {journal} {\bibinfo
  {journal} {arxiv}\ } (\bibinfo {year} {2018})}\BibitemShut {NoStop}%
\bibitem [{\citenamefont {Schwengner}\ \emph {et~al.}(2013)\citenamefont
  {Schwengner}, \citenamefont {Frauendorf},\ and\ \citenamefont
  {Larsen}}]{Schwengner2013}%
  \BibitemOpen
  \bibfield  {author} {\bibinfo {author} {\bibfnamefont {R.}~\bibnamefont
  {Schwengner}}, \bibinfo {author} {\bibfnamefont {S.}~\bibnamefont
  {Frauendorf}},\ and\ \bibinfo {author} {\bibfnamefont {A.~C.}\ \bibnamefont
  {Larsen}},\ }\href {https://doi.org/10.1103/PhysRevLett.111.232504}
  {\bibfield  {journal} {\bibinfo  {journal} {Phys. Rev. Lett.}\ }\textbf
  {\bibinfo {volume} {111}},\ \bibinfo {pages} {232504} (\bibinfo {year}
  {2013})}\BibitemShut {NoStop}%
\bibitem [{\citenamefont {Brown}\ and\ \citenamefont
  {Larsen}(2014)}]{Brown2014}%
  \BibitemOpen
  \bibfield  {author} {\bibinfo {author} {\bibfnamefont {B.~A.}\ \bibnamefont
  {Brown}}\ and\ \bibinfo {author} {\bibfnamefont {A.~C.}\ \bibnamefont
  {Larsen}},\ }\href {https://doi.org/10.1103/PhysRevLett.113.252502}
  {\bibfield  {journal} {\bibinfo  {journal} {Phys. Rev. Lett.}\ }\textbf
  {\bibinfo {volume} {113}},\ \bibinfo {pages} {252502} (\bibinfo {year}
  {2014})}\BibitemShut {NoStop}%
\bibitem [{\citenamefont {Kopecky}\ and\ \citenamefont
  {Uhl}(1990)}]{Kopecky1990}%
  \BibitemOpen
  \bibfield  {author} {\bibinfo {author} {\bibfnamefont {J.}~\bibnamefont
  {Kopecky}}\ and\ \bibinfo {author} {\bibfnamefont {M.}~\bibnamefont {Uhl}},\
  }\href {https://doi.org/10.1103/PhysRevC.41.1941} {\bibfield  {journal}
  {\bibinfo  {journal} {Phys. Rev. C}\ }\textbf {\bibinfo {volume} {41}},\
  \bibinfo {pages} {1941} (\bibinfo {year} {1990})}\BibitemShut {NoStop}%
\bibitem [{\citenamefont {Larsen}\ \emph {et~al.}(2013)\citenamefont {Larsen},
  \citenamefont {Blasi}, \citenamefont {Bracco}, \citenamefont {Camera},
  \citenamefont {Eriksen}, \citenamefont {G\"orgen}, \citenamefont
  {Guttormsen}, \citenamefont {Hagen}, \citenamefont {Leoni}, \citenamefont
  {Million}, \citenamefont {Nyhus}, \citenamefont {Renstr\o{}m}, \citenamefont
  {Rose}, \citenamefont {Ruud}, \citenamefont {Siem}, \citenamefont {Tornyi},
  \citenamefont {Tveten}, \citenamefont {Voinov},\ and\ \citenamefont
  {Wiedeking}}]{Larsen2013}%
  \BibitemOpen
  \bibfield  {author} {\bibinfo {author} {\bibfnamefont {A.~C.}\ \bibnamefont
  {Larsen}}, \bibinfo {author} {\bibfnamefont {N.}~\bibnamefont {Blasi}},
  \bibinfo {author} {\bibfnamefont {A.}~\bibnamefont {Bracco}}, \bibinfo
  {author} {\bibfnamefont {F.}~\bibnamefont {Camera}}, \bibinfo {author}
  {\bibfnamefont {T.~K.}\ \bibnamefont {Eriksen}}, \bibinfo {author}
  {\bibfnamefont {A.}~\bibnamefont {G\"orgen}}, \bibinfo {author}
  {\bibfnamefont {M.}~\bibnamefont {Guttormsen}}, \bibinfo {author}
  {\bibfnamefont {T.~W.}\ \bibnamefont {Hagen}}, \bibinfo {author}
  {\bibfnamefont {S.}~\bibnamefont {Leoni}}, \bibinfo {author} {\bibfnamefont
  {B.}~\bibnamefont {Million}}, \bibinfo {author} {\bibfnamefont {H.~T.}\
  \bibnamefont {Nyhus}}, \bibinfo {author} {\bibfnamefont {T.}~\bibnamefont
  {Renstr\o{}m}}, \bibinfo {author} {\bibfnamefont {S.~J.}\ \bibnamefont
  {Rose}}, \bibinfo {author} {\bibfnamefont {I.~E.}\ \bibnamefont {Ruud}},
  \bibinfo {author} {\bibfnamefont {S.}~\bibnamefont {Siem}}, \bibinfo {author}
  {\bibfnamefont {T.}~\bibnamefont {Tornyi}}, \bibinfo {author} {\bibfnamefont
  {G.~M.}\ \bibnamefont {Tveten}}, \bibinfo {author} {\bibfnamefont {A.~V.}\
  \bibnamefont {Voinov}},\ and\ \bibinfo {author} {\bibfnamefont
  {M.}~\bibnamefont {Wiedeking}},\ }\href
  {https://doi.org/10.1103/PhysRevLett.111.242504} {\bibfield  {journal}
  {\bibinfo  {journal} {Phys. Rev. Lett.}\ }\textbf {\bibinfo {volume} {111}},\
  \bibinfo {pages} {242504} (\bibinfo {year} {2013})}\BibitemShut {NoStop}%
\bibitem [{\citenamefont {Jones}\ \emph {et~al.}(2018)\citenamefont {Jones},
  \citenamefont {Macchiavelli}, \citenamefont {Wiedeking}, \citenamefont
  {Bernstein}, \citenamefont {Crawford}, \citenamefont {Campbell},
  \citenamefont {Clark}, \citenamefont {Cromaz}, \citenamefont {Fallon},
  \citenamefont {Lee}, \citenamefont {Salathe}, \citenamefont {Wiens},
  \citenamefont {Ayangeakaa}, \citenamefont {Bleuel}, \citenamefont {Bottoni},
  \citenamefont {Carpenter}, \citenamefont {Davids}, \citenamefont {Elson},
  \citenamefont {G\"orgen}, \citenamefont {Guttormsen}, \citenamefont
  {Janssens}, \citenamefont {Kinnison}, \citenamefont {Kirsch}, \citenamefont
  {Larsen}, \citenamefont {Lauritsen}, \citenamefont {Reviol}, \citenamefont
  {Sarantites}, \citenamefont {Siem}, \citenamefont {Voinov},\ and\
  \citenamefont {Zhu}}]{Jones2018}%
  \BibitemOpen
  \bibfield  {author} {\bibinfo {author} {\bibfnamefont {M.~D.}\ \bibnamefont
  {Jones}}, \bibinfo {author} {\bibfnamefont {A.~O.}\ \bibnamefont
  {Macchiavelli}}, \bibinfo {author} {\bibfnamefont {M.}~\bibnamefont
  {Wiedeking}}, \bibinfo {author} {\bibfnamefont {L.~A.}\ \bibnamefont
  {Bernstein}}, \bibinfo {author} {\bibfnamefont {H.~L.}\ \bibnamefont
  {Crawford}}, \bibinfo {author} {\bibfnamefont {C.~M.}\ \bibnamefont
  {Campbell}}, \bibinfo {author} {\bibfnamefont {R.~M.}\ \bibnamefont {Clark}},
  \bibinfo {author} {\bibfnamefont {M.}~\bibnamefont {Cromaz}}, \bibinfo
  {author} {\bibfnamefont {P.}~\bibnamefont {Fallon}}, \bibinfo {author}
  {\bibfnamefont {I.~Y.}\ \bibnamefont {Lee}}, \bibinfo {author} {\bibfnamefont
  {M.}~\bibnamefont {Salathe}}, \bibinfo {author} {\bibfnamefont
  {A.}~\bibnamefont {Wiens}}, \bibinfo {author} {\bibfnamefont {A.~D.}\
  \bibnamefont {Ayangeakaa}}, \bibinfo {author} {\bibfnamefont {D.~L.}\
  \bibnamefont {Bleuel}}, \bibinfo {author} {\bibfnamefont {S.}~\bibnamefont
  {Bottoni}}, \bibinfo {author} {\bibfnamefont {M.~P.}\ \bibnamefont
  {Carpenter}}, \bibinfo {author} {\bibfnamefont {H.~M.}\ \bibnamefont
  {Davids}}, \bibinfo {author} {\bibfnamefont {J.}~\bibnamefont {Elson}},
  \bibinfo {author} {\bibfnamefont {A.}~\bibnamefont {G\"orgen}}, \bibinfo
  {author} {\bibfnamefont {M.}~\bibnamefont {Guttormsen}}, \bibinfo {author}
  {\bibfnamefont {R.~V.~F.}\ \bibnamefont {Janssens}}, \bibinfo {author}
  {\bibfnamefont {J.~E.}\ \bibnamefont {Kinnison}}, \bibinfo {author}
  {\bibfnamefont {L.}~\bibnamefont {Kirsch}}, \bibinfo {author} {\bibfnamefont
  {A.~C.}\ \bibnamefont {Larsen}}, \bibinfo {author} {\bibfnamefont
  {T.}~\bibnamefont {Lauritsen}}, \bibinfo {author} {\bibfnamefont
  {W.}~\bibnamefont {Reviol}}, \bibinfo {author} {\bibfnamefont {D.~G.}\
  \bibnamefont {Sarantites}}, \bibinfo {author} {\bibfnamefont
  {S.}~\bibnamefont {Siem}}, \bibinfo {author} {\bibfnamefont {A.~V.}\
  \bibnamefont {Voinov}},\ and\ \bibinfo {author} {\bibfnamefont
  {S.}~\bibnamefont {Zhu}},\ }\href
  {https://doi.org/10.1103/PhysRevC.97.024327} {\bibfield  {journal} {\bibinfo
  {journal} {Phys. Rev. C}\ }\textbf {\bibinfo {volume} {97}},\ \bibinfo
  {pages} {024327} (\bibinfo {year} {2018})}\BibitemShut {NoStop}%
\bibitem [{\citenamefont {Gilbert}\ and\ \citenamefont
  {Cameron}(1965)}]{gilbert1965}%
  \BibitemOpen
  \bibfield  {author} {\bibinfo {author} {\bibfnamefont {A.}~\bibnamefont
  {Gilbert}}\ and\ \bibinfo {author} {\bibfnamefont {A.~G.~W.}\ \bibnamefont
  {Cameron}},\ }\href {https://doi.org/10.1139/p65-139} {\bibfield  {journal}
  {\bibinfo  {journal} {Canadian Journal of Physics}\ }\textbf {\bibinfo
  {volume} {43}},\ \bibinfo {pages} {1446} (\bibinfo {year} {1965})},\ \Eprint
  {https://arxiv.org/abs/https://doi.org/10.1139/p65-139}
  {https://doi.org/10.1139/p65-139} \BibitemShut {NoStop}%
\bibitem [{\citenamefont {Huizenga}\ \emph {et~al.}(1969)\citenamefont
  {Huizenga}, \citenamefont {Vonach}, \citenamefont {Katsanos}, \citenamefont
  {Gorski},\ and\ \citenamefont {Stephan}}]{Huizenga1969}%
  \BibitemOpen
  \bibfield  {author} {\bibinfo {author} {\bibfnamefont {J.~R.}\ \bibnamefont
  {Huizenga}}, \bibinfo {author} {\bibfnamefont {H.~K.}\ \bibnamefont
  {Vonach}}, \bibinfo {author} {\bibfnamefont {A.~A.}\ \bibnamefont
  {Katsanos}}, \bibinfo {author} {\bibfnamefont {A.~J.}\ \bibnamefont
  {Gorski}},\ and\ \bibinfo {author} {\bibfnamefont {C.~J.}\ \bibnamefont
  {Stephan}},\ }\href {https://doi.org/10.1103/PhysRev.182.1149} {\bibfield
  {journal} {\bibinfo  {journal} {Phys. Rev.}\ }\textbf {\bibinfo {volume}
  {182}},\ \bibinfo {pages} {1149} (\bibinfo {year} {1969})}\BibitemShut
  {NoStop}%
\bibitem [{\citenamefont {Ignatyuk}\ \emph {et~al.}(1993)\citenamefont
  {Ignatyuk}, \citenamefont {Weil}, \citenamefont {Raman},\ and\ \citenamefont
  {Kahane}}]{Ignatyuk1993}%
  \BibitemOpen
  \bibfield  {author} {\bibinfo {author} {\bibfnamefont {A.~V.}\ \bibnamefont
  {Ignatyuk}}, \bibinfo {author} {\bibfnamefont {J.~L.}\ \bibnamefont {Weil}},
  \bibinfo {author} {\bibfnamefont {S.}~\bibnamefont {Raman}},\ and\ \bibinfo
  {author} {\bibfnamefont {S.}~\bibnamefont {Kahane}},\ }\href
  {https://doi.org/10.1103/PhysRevC.47.1504} {\bibfield  {journal} {\bibinfo
  {journal} {Phys. Rev. C}\ }\textbf {\bibinfo {volume} {47}},\ \bibinfo
  {pages} {1504} (\bibinfo {year} {1993})}\BibitemShut {NoStop}%
\bibitem [{\citenamefont {Demetriou}\ and\ \citenamefont
  {Goriely}(2001)}]{Demetriou2001}%
  \BibitemOpen
  \bibfield  {author} {\bibinfo {author} {\bibfnamefont {P.}~\bibnamefont
  {Demetriou}}\ and\ \bibinfo {author} {\bibfnamefont {S.}~\bibnamefont
  {Goriely}},\ }\href
  {https://doi.org/https://doi.org/10.1016/S0375-9474(01)01095-8} {\bibfield
  {journal} {\bibinfo  {journal} {Nuclear Physics A}\ }\textbf {\bibinfo
  {volume} {695}},\ \bibinfo {pages} {95} (\bibinfo {year} {2001})}\BibitemShut
  {NoStop}%
\bibitem [{\citenamefont {Goriely}\ \emph {et~al.}(2008)\citenamefont
  {Goriely}, \citenamefont {Hilaire},\ and\ \citenamefont
  {Koning}}]{Goriely2008}%
  \BibitemOpen
  \bibfield  {author} {\bibinfo {author} {\bibfnamefont {S.}~\bibnamefont
  {Goriely}}, \bibinfo {author} {\bibfnamefont {S.}~\bibnamefont {Hilaire}},\
  and\ \bibinfo {author} {\bibfnamefont {A.~J.}\ \bibnamefont {Koning}},\
  }\href {https://doi.org/10.1103/PhysRevC.78.064307} {\bibfield  {journal}
  {\bibinfo  {journal} {Phys. Rev. C}\ }\textbf {\bibinfo {volume} {78}},\
  \bibinfo {pages} {064307} (\bibinfo {year} {2008})}\BibitemShut {NoStop}%
\bibitem [{\citenamefont {Hilaire}\ \emph {et~al.}(2012)\citenamefont
  {Hilaire}, \citenamefont {Girod}, \citenamefont {Goriely},\ and\
  \citenamefont {Koning}}]{Hilaire2012}%
  \BibitemOpen
  \bibfield  {author} {\bibinfo {author} {\bibfnamefont {S.}~\bibnamefont
  {Hilaire}}, \bibinfo {author} {\bibfnamefont {M.}~\bibnamefont {Girod}},
  \bibinfo {author} {\bibfnamefont {S.}~\bibnamefont {Goriely}},\ and\ \bibinfo
  {author} {\bibfnamefont {A.~J.}\ \bibnamefont {Koning}},\ }\href
  {https://doi.org/10.1103/PhysRevC.86.064317} {\bibfield  {journal} {\bibinfo
  {journal} {Phys. Rev. C}\ }\textbf {\bibinfo {volume} {86}},\ \bibinfo
  {pages} {064317} (\bibinfo {year} {2012})}\BibitemShut {NoStop}%
\bibitem [{\citenamefont {Goriely}\ \emph {et~al.}(2022)\citenamefont
  {Goriely}, \citenamefont {Larsen},\ and\ \citenamefont
  {M\"ucher}}]{Goriely2022}%
  \BibitemOpen
  \bibfield  {author} {\bibinfo {author} {\bibfnamefont {S.}~\bibnamefont
  {Goriely}}, \bibinfo {author} {\bibfnamefont {A.-C.}\ \bibnamefont
  {Larsen}},\ and\ \bibinfo {author} {\bibfnamefont {D.}~\bibnamefont
  {M\"ucher}},\ }\href {https://doi.org/10.1103/PhysRevC.106.044315} {\bibfield
   {journal} {\bibinfo  {journal} {Phys. Rev. C}\ }\textbf {\bibinfo {volume}
  {106}},\ \bibinfo {pages} {044315} (\bibinfo {year} {2022})}\BibitemShut
  {NoStop}%
\bibitem [{\citenamefont {Berman}\ \emph {et~al.}(1969)\citenamefont {Berman},
  \citenamefont {Kelly}, \citenamefont {Bramblett}, \citenamefont {Caldwell},
  \citenamefont {Davis},\ and\ \citenamefont {Fultz}}]{Berman1969}%
  \BibitemOpen
  \bibfield  {author} {\bibinfo {author} {\bibfnamefont {B.~L.}\ \bibnamefont
  {Berman}}, \bibinfo {author} {\bibfnamefont {M.~A.}\ \bibnamefont {Kelly}},
  \bibinfo {author} {\bibfnamefont {R.~L.}\ \bibnamefont {Bramblett}}, \bibinfo
  {author} {\bibfnamefont {J.~T.}\ \bibnamefont {Caldwell}}, \bibinfo {author}
  {\bibfnamefont {H.~S.}\ \bibnamefont {Davis}},\ and\ \bibinfo {author}
  {\bibfnamefont {S.~C.}\ \bibnamefont {Fultz}},\ }\href
  {https://doi.org/10.1103/PhysRev.185.1576} {\bibfield  {journal} {\bibinfo
  {journal} {Phys. Rev.}\ }\textbf {\bibinfo {volume} {185}},\ \bibinfo {pages}
  {1576} (\bibinfo {year} {1969})}\BibitemShut {NoStop}%
\bibitem [{\citenamefont {Kopecky}\ \emph {et~al.}(2017)\citenamefont
  {Kopecky}, \citenamefont {Goriely}, \citenamefont {P\'eru}, \citenamefont
  {Hilaire},\ and\ \citenamefont {Martini}}]{Kopecky2017}%
  \BibitemOpen
  \bibfield  {author} {\bibinfo {author} {\bibfnamefont {J.}~\bibnamefont
  {Kopecky}}, \bibinfo {author} {\bibfnamefont {S.}~\bibnamefont {Goriely}},
  \bibinfo {author} {\bibfnamefont {S.}~\bibnamefont {P\'eru}}, \bibinfo
  {author} {\bibfnamefont {S.}~\bibnamefont {Hilaire}},\ and\ \bibinfo {author}
  {\bibfnamefont {M.}~\bibnamefont {Martini}},\ }\href
  {https://doi.org/10.1103/PhysRevC.95.054317} {\bibfield  {journal} {\bibinfo
  {journal} {Phys. Rev. C}\ }\textbf {\bibinfo {volume} {95}},\ \bibinfo
  {pages} {054317} (\bibinfo {year} {2017})}\BibitemShut {NoStop}%
\bibitem [{\citenamefont {Axel}(1968)}]{Axel1968}%
  \BibitemOpen
  \bibfield  {author} {\bibinfo {author} {\bibfnamefont {P.}~\bibnamefont
  {Axel}},\ }\href@noop {} {\bibfield  {journal} {\bibinfo  {journal}
  {Proceedings of the International Symposium on Nuclear Structure}\ ,\
  \bibinfo {pages} {p.299}} (\bibinfo {year} {1968})}\BibitemShut {NoStop}%
\bibitem [{\citenamefont {Goriely}\ and\ \citenamefont
  {Khan}(2002)}]{Goriely2002}%
  \BibitemOpen
  \bibfield  {author} {\bibinfo {author} {\bibfnamefont {S.}~\bibnamefont
  {Goriely}}\ and\ \bibinfo {author} {\bibfnamefont {E.}~\bibnamefont {Khan}},\
  }\href {https://doi.org/https://doi.org/10.1016/S0375-9474(02)00860-6}
  {\bibfield  {journal} {\bibinfo  {journal} {Nuclear Physics A}\ }\textbf
  {\bibinfo {volume} {706}},\ \bibinfo {pages} {217} (\bibinfo {year}
  {2002})}\BibitemShut {NoStop}%
\bibitem [{\citenamefont {Goriely}\ \emph {et~al.}(2004)\citenamefont
  {Goriely}, \citenamefont {Khan},\ and\ \citenamefont {Samyn}}]{Goriely2004}%
  \BibitemOpen
  \bibfield  {author} {\bibinfo {author} {\bibfnamefont {S.}~\bibnamefont
  {Goriely}}, \bibinfo {author} {\bibfnamefont {E.}~\bibnamefont {Khan}},\ and\
  \bibinfo {author} {\bibfnamefont {M.}~\bibnamefont {Samyn}},\ }\href
  {https://doi.org/https://doi.org/10.1016/j.nuclphysa.2004.04.105} {\bibfield
  {journal} {\bibinfo  {journal} {Nuclear Physics A}\ }\textbf {\bibinfo
  {volume} {739}},\ \bibinfo {pages} {331} (\bibinfo {year}
  {2004})}\BibitemShut {NoStop}%
\bibitem [{\citenamefont {Goriely}(1998)}]{Goriely1998}%
  \BibitemOpen
  \bibfield  {author} {\bibinfo {author} {\bibfnamefont {S.}~\bibnamefont
  {Goriely}},\ }\href
  {https://doi.org/https://doi.org/10.1016/S0370-2693(98)00907-1} {\bibfield
  {journal} {\bibinfo  {journal} {Physics Letters B}\ }\textbf {\bibinfo
  {volume} {436}},\ \bibinfo {pages} {10} (\bibinfo {year} {1998})}\BibitemShut
  {NoStop}%
\bibitem [{\citenamefont {Daoutidis}\ and\ \citenamefont
  {Goriely}(2012)}]{Daoutidis2012}%
  \BibitemOpen
  \bibfield  {author} {\bibinfo {author} {\bibfnamefont {I.}~\bibnamefont
  {Daoutidis}}\ and\ \bibinfo {author} {\bibfnamefont {S.}~\bibnamefont
  {Goriely}},\ }\href {https://doi.org/10.1103/PhysRevC.86.034328} {\bibfield
  {journal} {\bibinfo  {journal} {Phys. Rev. C}\ }\textbf {\bibinfo {volume}
  {86}},\ \bibinfo {pages} {034328} (\bibinfo {year} {2012})}\BibitemShut
  {NoStop}%
\bibitem [{\citenamefont {Brun}\ and\ \citenamefont {Rademakers}(1997)}]{ROOT}%
  \BibitemOpen
  \bibfield  {author} {\bibinfo {author} {\bibfnamefont {R.}~\bibnamefont
  {Brun}}\ and\ \bibinfo {author} {\bibfnamefont {F.}~\bibnamefont
  {Rademakers}},\ }\href {https://root.cern} {\bibinfo {title} {Root - {An
  Object Oriented Data Analysis Framework, Proceedings AIHENP'96 Workshop,
  Lausanne, Sep. 1996,}}} (\bibinfo {year} {1997})\BibitemShut {NoStop}%
\bibitem [{\citenamefont {Voinov}\ \emph {et~al.}(2004)\citenamefont {Voinov},
  \citenamefont {Algin}, \citenamefont {Agvaanluvsan}, \citenamefont {Belgya},
  \citenamefont {Chankova}, \citenamefont {Guttormsen}, \citenamefont
  {Mitchell}, \citenamefont {Rekstad}, \citenamefont {Schiller},\ and\
  \citenamefont {Siem}}]{Voinov2004}%
  \BibitemOpen
  \bibfield  {author} {\bibinfo {author} {\bibfnamefont {A.}~\bibnamefont
  {Voinov}}, \bibinfo {author} {\bibfnamefont {E.}~\bibnamefont {Algin}},
  \bibinfo {author} {\bibfnamefont {U.}~\bibnamefont {Agvaanluvsan}}, \bibinfo
  {author} {\bibfnamefont {T.}~\bibnamefont {Belgya}}, \bibinfo {author}
  {\bibfnamefont {R.}~\bibnamefont {Chankova}}, \bibinfo {author}
  {\bibfnamefont {M.}~\bibnamefont {Guttormsen}}, \bibinfo {author}
  {\bibfnamefont {G.~E.}\ \bibnamefont {Mitchell}}, \bibinfo {author}
  {\bibfnamefont {J.}~\bibnamefont {Rekstad}}, \bibinfo {author} {\bibfnamefont
  {A.}~\bibnamefont {Schiller}},\ and\ \bibinfo {author} {\bibfnamefont
  {S.}~\bibnamefont {Siem}},\ }\href
  {https://doi.org/10.1103/PhysRevLett.93.142504} {\bibfield  {journal}
  {\bibinfo  {journal} {Phys. Rev. Lett.}\ }\textbf {\bibinfo {volume} {93}},\
  \bibinfo {pages} {142504} (\bibinfo {year} {2004})}\BibitemShut {NoStop}%
\bibitem [{\citenamefont {Wolfenstein}(1951)}]{Wolfenstein1951}%
  \BibitemOpen
  \bibfield  {author} {\bibinfo {author} {\bibfnamefont {L.}~\bibnamefont
  {Wolfenstein}},\ }\href {https://doi.org/10.1103/PhysRev.82.690} {\bibfield
  {journal} {\bibinfo  {journal} {Phys. Rev.}\ }\textbf {\bibinfo {volume}
  {82}},\ \bibinfo {pages} {690} (\bibinfo {year} {1951})}\BibitemShut
  {NoStop}%
\bibitem [{\citenamefont {Hauser}\ and\ \citenamefont
  {Feshbach}(1952)}]{HauserFeshbach1952}%
  \BibitemOpen
  \bibfield  {author} {\bibinfo {author} {\bibfnamefont {W.}~\bibnamefont
  {Hauser}}\ and\ \bibinfo {author} {\bibfnamefont {H.}~\bibnamefont
  {Feshbach}},\ }\href {https://doi.org/10.1103/PhysRev.87.366} {\bibfield
  {journal} {\bibinfo  {journal} {Phys. Rev.}\ }\textbf {\bibinfo {volume}
  {87}},\ \bibinfo {pages} {366} (\bibinfo {year} {1952})}\BibitemShut
  {NoStop}%
\bibitem [{\citenamefont {Bauge}\ \emph {et~al.}(2001)\citenamefont {Bauge},
  \citenamefont {Delaroche},\ and\ \citenamefont {Girod}}]{Bauge2001}%
  \BibitemOpen
  \bibfield  {author} {\bibinfo {author} {\bibfnamefont {E.}~\bibnamefont
  {Bauge}}, \bibinfo {author} {\bibfnamefont {J.~P.}\ \bibnamefont
  {Delaroche}},\ and\ \bibinfo {author} {\bibfnamefont {M.}~\bibnamefont
  {Girod}},\ }\href {https://doi.org/10.1103/PhysRevC.63.024607} {\bibfield
  {journal} {\bibinfo  {journal} {Phys. Rev. C}\ }\textbf {\bibinfo {volume}
  {63}},\ \bibinfo {pages} {024607} (\bibinfo {year} {2001})}\BibitemShut
  {NoStop}%
\bibitem [{\citenamefont {Koning}\ and\ \citenamefont
  {Delaroche}(2003)}]{Koning2003}%
  \BibitemOpen
  \bibfield  {author} {\bibinfo {author} {\bibfnamefont {A.}~\bibnamefont
  {Koning}}\ and\ \bibinfo {author} {\bibfnamefont {J.}~\bibnamefont
  {Delaroche}},\ }\href
  {https://doi.org/https://doi.org/10.1016/S0375-9474(02)01321-0} {\bibfield
  {journal} {\bibinfo  {journal} {Nuclear Physics A}\ }\textbf {\bibinfo
  {volume} {713}},\ \bibinfo {pages} {231} (\bibinfo {year}
  {2003})}\BibitemShut {NoStop}%
\bibitem [{\citenamefont {Dillmann}\ \emph {et~al.}()\citenamefont {Dillmann},
  \citenamefont {Plag}, \citenamefont {K\"{a}ppeler},\ and\ \citenamefont
  {Rauscher}}]{kadonis}%
  \BibitemOpen
  \bibfield  {author} {\bibinfo {author} {\bibfnamefont {I.}~\bibnamefont
  {Dillmann}}, \bibinfo {author} {\bibfnamefont {R.}~\bibnamefont {Plag}},
  \bibinfo {author} {\bibfnamefont {F.}~\bibnamefont {K\"{a}ppeler}},\ and\
  \bibinfo {author} {\bibfnamefont {T.}~\bibnamefont {Rauscher}},\ }\href
  {https://exp-astro.de/kadonis1.0/index.php} {\bibinfo {title} {Data extracted
  using the {KADoNiS} {On-Line Data Service}}}\BibitemShut {NoStop}%
\bibitem [{\citenamefont {Bao}\ \emph {et~al.}(2000)\citenamefont {Bao},
  \citenamefont {Beer}, \citenamefont {K\"{a}ppeler}, \citenamefont {Voss},
  \citenamefont {Wisshak},\ and\ \citenamefont {Rauscher}}]{bao2000}%
  \BibitemOpen
  \bibfield  {author} {\bibinfo {author} {\bibfnamefont {Z.}~\bibnamefont
  {Bao}}, \bibinfo {author} {\bibfnamefont {H.}~\bibnamefont {Beer}}, \bibinfo
  {author} {\bibfnamefont {F.}~\bibnamefont {K\"{a}ppeler}}, \bibinfo {author}
  {\bibfnamefont {F.}~\bibnamefont {Voss}}, \bibinfo {author} {\bibfnamefont
  {K.}~\bibnamefont {Wisshak}},\ and\ \bibinfo {author} {\bibfnamefont
  {T.}~\bibnamefont {Rauscher}},\ }\href
  {https://doi.org/https://doi.org/10.1006/adnd.2000.0838} {\bibfield
  {journal} {\bibinfo  {journal} {Atomic Data and Nuclear Data Tables}\
  }\textbf {\bibinfo {volume} {76}},\ \bibinfo {pages} {70} (\bibinfo {year}
  {2000})}\BibitemShut {NoStop}%
\bibitem [{\citenamefont {Kullmann}\ \emph {et~al.}(2019)\citenamefont
  {Kullmann}, \citenamefont {Larsen}, \citenamefont {Renstr\o{}m},
  \citenamefont {Beckmann}, \citenamefont {Garrote}, \citenamefont {Campo},
  \citenamefont {G\"orgen}, \citenamefont {Guttormsen}, \citenamefont
  {Midtb\o{}}, \citenamefont {Sahin}, \citenamefont {Siem}, \citenamefont
  {Tveten},\ and\ \citenamefont {Zeiser}}]{Kullmann2019}%
  \BibitemOpen
  \bibfield  {author} {\bibinfo {author} {\bibfnamefont {I.~K.~B.}\
  \bibnamefont {Kullmann}}, \bibinfo {author} {\bibfnamefont {A.~C.}\
  \bibnamefont {Larsen}}, \bibinfo {author} {\bibfnamefont {T.}~\bibnamefont
  {Renstr\o{}m}}, \bibinfo {author} {\bibfnamefont {K.~S.}\ \bibnamefont
  {Beckmann}}, \bibinfo {author} {\bibfnamefont {F.~L.~B.}\ \bibnamefont
  {Garrote}}, \bibinfo {author} {\bibfnamefont {L.~C.}\ \bibnamefont {Campo}},
  \bibinfo {author} {\bibfnamefont {A.}~\bibnamefont {G\"orgen}}, \bibinfo
  {author} {\bibfnamefont {M.}~\bibnamefont {Guttormsen}}, \bibinfo {author}
  {\bibfnamefont {J.~E.}\ \bibnamefont {Midtb\o{}}}, \bibinfo {author}
  {\bibfnamefont {E.}~\bibnamefont {Sahin}}, \bibinfo {author} {\bibfnamefont
  {S.}~\bibnamefont {Siem}}, \bibinfo {author} {\bibfnamefont {G.~M.}\
  \bibnamefont {Tveten}},\ and\ \bibinfo {author} {\bibfnamefont
  {F.}~\bibnamefont {Zeiser}},\ }\href
  {https://doi.org/10.1103/PhysRevC.99.065806} {\bibfield  {journal} {\bibinfo
  {journal} {Phys. Rev. C}\ }\textbf {\bibinfo {volume} {99}},\ \bibinfo
  {pages} {065806} (\bibinfo {year} {2019})}\BibitemShut {NoStop}%
\end{thebibliography}%

\end{document}